\documentclass[11pt]{article}

\usepackage{arxiv}
\usepackage[utf8]{inputenc} 
\usepackage[T1]{fontenc}    
\usepackage{hyperref}       
\usepackage{url}            
\usepackage{booktabs}       
\usepackage{tabularx}
\usepackage{amsfonts}       
\usepackage{nicefrac}       
\usepackage{microtype}      
\usepackage{nomencl}
\usepackage{amssymb}
\usepackage{amsmath}
\usepackage{tikz}
\usetikzlibrary{positioning, matrix, shapes.geometric, arrows.meta, backgrounds, fit, calc}
\usepackage{subcaption}
\usepackage[normalem]{ulem}
\hypersetup{colorlinks,urlcolor=blue}
\usepackage[bbgreekl]{mathbbol}
\usepackage{etoolbox}
\usepackage{letltxmacro}
\usepackage{mathtools}
\usepackage{algpseudocode}
\usepackage{color}
\usepackage{placeins}
\usepackage{xspace}
\usepackage[linesnumbered,ruled]{algorithm2e}
\usepackage[style=numeric-comp,backend=biber, giveninits=true, sorting=none,maxbibnames=999, maxcitenames=999]{biblatex}
\usepackage{ntheorem}
\theoremstyle{plain}
\newtheorem{remark}{Remark}[section]
\DeclareSymbolFontAlphabet{\mathbb}{AMSb}
\DeclareSymbolFontAlphabet{\mathbbl}{bbold}
\DeclarePairedDelimiterX{\infdivx}[2]{}{}{%
  {#1}\;\big\|\;{#2}%
}

\newcommand{\eg}{e.\,g.,\xspace}
\newcommand{\ie}{i.\,e.,\xspace}

\newcommand{\wrt}{w.\,r.\,t.\xspace}
\newcommand{\rhs}{r.\,h.\,s.\xspace}
\newcommand{\iid}{i.\,i.\,d.\xspace}

\newcommand{\dd}{\,\mathrm{d}} 

\newcommand{\bs}{\boldsymbol}
\newcommand{\bx}{{\bs{x}}}
\newcommand{\bxele}{{\bs{x}_{\mathrm{ele}}}}
\newcommand{\by}{{\bs{y}}}

\newcommand{\bd}{\boldsymbol{d}}

\newcommand{\Mod}{\mathfrak{M}}

\newcommand{\nbatch}{n_{\mathrm{batch}}}

\newcommand{\Ex}[2]{{\mathbb{E}}_{#1}\left[#2\right]}

\newcommand{\DKL}[2]{\mathrm{D}_{\text{KL}}\left[#1\, ||\, #2\right]}

\newcommand{\be}{\begin{equation}}
\newcommand{\ee}{\end{equation}}
\newcommand{\ed}{\end{description}}
\newcommand{\bi}{\begin{itemize}}
\newcommand{\ei}{\end{itemize}}

\newcommand{\bc}{\boldsymbol{c}}

\newcommand{\bmu}{\boldsymbol{\mu}}
\newcommand{\bv}{\boldsymbol{v}}

\newcommand{\bu}{\boldsymbol{u}}
\newcommand{\bp}{\boldsymbol{p}}
\newcommand{\bn}{\boldsymbol{n}}
\newcommand{\bnil}{\boldsymbol{0}}

\newcommand{\ndofs}{n_{\text{dofs}}}

\newcommand{\dimx}{\dim(\bx)}

\newcommand{\bz}{\boldsymbol{z}}

\newcommand{\bxgt}{\boldsymbol{x}_{\mathrm{gt}}}
\newcommand{\txgt}{\tilde{x}_{\mathrm{gt}}}
\newcommand{\kgt}{k_{\mathrm{gt}}}
\newcommand{\bphi}{\boldsymbol{\phi}}
\newcommand{\logl}{\mathcal{L}}
\newcommand{\br}{\boldsymbol{r}}
\newcommand{\trp}{\boldsymbol{t}_{\text{rp}}}
\newcommand{\nsamples}{n_{\mathrm{samples}}}

\newcommand{\Adj}{\mathfrak{A}}

\newcommand{\ELBO}{\mathrm{ELBO}}
\newcommand{\softplus}{\mathrm{softplus}}
\newcommand{\sigmoid}{\mathrm{sigmoid}}
\newcommand{\bphimu}{\bphi_{\bmu}}
\newcommand{\bphiL}{\bphi_{L}}
\newcommand{\bphiLdiag}{\bphi_{L,\text{diag.}}}
\newcommand{\bphiLoffdiag}{\bphi_{L,\text{off-diag.}}}
\newcommand{\LQ}{L_{Q}}
\newcommand{\byobs}{\boldsymbol{y}_{\mathrm{obs}}}
\newcommand{\blambda}{\boldsymbol{\lambda}}

\newcommand{\bw}{\boldsymbol{w}}
\newcommand{\bh}{\boldsymbol{h}}

\newcommand{\Assembly}{\underset{\mathrm{ele},ij}{\bigwedge}}
\newcommand{\AssemblyVec}{\underset{\mathrm{ele},i}{\bigwedge}}
\newcommand{\AL}{A_{L}}

\newcommand{\tx}{\tilde{x}}

\newsavebox{\blueline}
\savebox{\blueline}{\tikz[baseline=-0.5ex]  \draw[blue, thick] (0,0) -- (0.3,0);}

\definecolor{mydarkgreen}{RGB}{0,100,0}
\newsavebox{\darkgreenline}
\savebox{\darkgreenline}{\tikz[baseline=-0.5ex]  \draw[mydarkgreen, thick] (0,0) -- (0.3,0);}

\newsavebox{\bluedashed}
\savebox{\bluedashed}{\tikz[baseline=-0.5ex]  \draw[blue, thick, dashed] (0,0) -- (0.3,0);}

\newsavebox{\greydot}
\savebox{\greydot}{\tikz[baseline=-0.5ex] \fill[gray] (0,0) circle (0.2ex);} 

\definecolor{customLightGrey}{rgb}{0.6, 0.6, 0.6} 

\newsavebox{\lightgreybox}
\savebox{\lightgreybox}{\tikz\fill[customLightGrey] (0,0) rectangle (0.4,0.2);} 

\newsavebox{\reddot}
\savebox{\reddot}{\tikz[baseline=-0.5ex] \fill[red] (0,0) circle (0.3ex);}

\tikzset{
  proc/.style   = {rectangle, rounded corners=3pt,
                   draw=black, very thick,
                   inner sep=4pt,
                   text width=7.4em, align=center,
                   font=\small\linespread{0.96}\selectfont},
  arrow/.style  = {-{Stealth[length=2.2mm,width=1.6mm]}, thick},
}

\DeclareRobustCommand{\cnum}[1]{\tikz[baseline=(c.base)]{%
  \node[circle,draw,inner sep=0.4pt,minimum size=1.25em,line width=0.4pt,font=\scriptsize](c){#1};}}
\newlength{\commentindent}
\makeatletter

\LetLtxMacro{\oldalgorithmic}{\algorithmic}
\renewcommand{\algorithmic}[1][0]{%
  \oldalgorithmic[#1]%
}
\makeatother 
\allowdisplaybreaks 
\title{Scalable High-Dimensional Bayesian Field Reconstruction with Finite Elements: Application to 3D Porous Media Flow}

\author{
  Jonas Nitzler\\
  Institute for Computational Mechanics\\
  Technical University of Munich\\
  D-85748 Garching b. München\\
  \texttt{jonas.nitzler@tum.de} \\
  \And
   Maximilian Bergbauer\\
  Institute for Computational Mechanics\\
  Technical University of Munich\\
  D-85748 Garching b. München\\
  \texttt{maximilian.bergbauer@tum.de} \\
  \And
  Phaedon-Stelios Koutsourelakis \\
 Professorship of Data-driven Materials Modeling\\
   Technical University of Munich\\
  D-85748 Garching b. München\\
  \texttt{p.s.koutsourelakis@tum.de} \\
  \And
  Wolfgang A. Wall\\
 Institute for Computational Mechanics\\
  Technical University of Munich\\
  D-85748 Garching b. München\\
  \texttt{wolfgang.a.wall@tum.de} \\
}

\begin{document}

\maketitle

\begin{abstract}
We present a unified, finite-element-native variational inference framework for very high-dimensional Bayesian spatial field reconstruction in physics-based problems governed by partial differential equations (PDEs) that are nonlinear in the inferred parameters. The framework delivers a full-covariance Gaussian variational posterior (a dense covariance represented implicitly through a sparse precision Cholesky factor), with a probabilistic treatment of the prior and likelihood variance/precision hyperparameters, on a three-dimensional curved finite-element discretization at a stochastic field dimension exceeding $4\cdot 10^5$. To our knowledge, this is the first full-covariance variational reconstruction at this scale, complementing the low-rank Hessian-Laplace approaches that dominate extreme-scale Bayesian inversion. The following contributions enable this:                                  
The spatial prior is derived from the stochastic PDE (SPDE) connection and formulated natively in terms of finite-element (FE) operators. The sparse Gaussian variational distribution is parameterized via its precision Cholesky factor, with the sparsity pattern inherited from the domain's Laplacian. 
Unlike covariance-based sparse parameterizations, which encode only short-range correlations, the sparse precision implicitly represents dense posterior covariances through its sparse inverse, yielding smooth, physically plausible samples at $\mathcal{O}(n)$ memory cost and enabling direct evidence-lower-bound (ELBO) gradients via the path-derivative (sticking-the-landing) estimator, without reparameterization Jacobians. 
Prior, variational distribution, hyperparameter updates, and adjoint-based log-posterior gradients are all expressed in terms of the same FE operators assembled from a single discretization shared with the forward problem. 
Natural gradient strategies account for the ELBO curvature and stabilize convergence. All hyperparameters are marginalized analytically within a variational Bayes expectation-maximization (VB-EM) loop, which additionally induces an automatic coarse-to-fine continuation that accelerates convergence.
The framework is demonstrated on Bayesian permeability field reconstruction for a porous-media flow problem on this domain, recovering all major spatial features with high fidelity. A detailed quantitative study, comprising an algorithmic ablation of every component and a comparison with alternative inference methods, provides numerical evidence of the improvements achieved by our developments relative to state-of-the-art baselines. The code for the forward problem and the inference algorithm is openly available.
\end{abstract}


\section{Introduction}
\label{sec:introduction}

Bayesian inverse problems for spatially varying physical fields arise throughout engineering and science, with applications ranging from subsurface flow characterization \cite{fox1997sampling, lee2002markov, chen2007reservoir}, constitutive tissue property inference \cite{Moireau_2010, kehl2017calibration, nitzler2024bmfia} to material identification in structural and geotechnical engineering \cite{kaipio2007statistical, Rappel_2019, Papaioannou_2016}.
The mathematical foundations of spatial Bayesian inversion in infinite-dimensional function spaces are well established \cite{stuart2010inverse, dashti2017bayesian}, and the finite element method (FEM) provides a natural discretization framework for both the underlying forward models and the spatial fields to be inferred. However, when the field is resolved on a three-dimensional triangulation, the stochastic dimension, \ie the number of unknown parameters $\bx$ that describe the latter, routinely exceeds $\mathcal{O}(10^5)$. At this scale, inference becomes a formidable computational challenge: each likelihood evaluation requires an expensive partial differential equation (PDE) solve (using FEM), and convergence slows dramatically in high-dimensional parameter spaces, becomes unstable, or may fail altogether.

Markov chain Monte Carlo methods, whether gradient-based (Hamiltonian Monte Carlo (HMC) \cite{betancourt2017conceptual}, Riemann manifold variants \cite{girolami2011riemann}, the No-U-Turn sampler \cite{hoffman2014no}) or designed for dimension-robust mixing in function-space Bayesian inversion (preconditioned Crank-Nicolson, pCN \cite{cotter2013mcmc}), require many sequential and expensive forward solves and remain impractical at this scale; likelihood-informed subspace methods \cite{cui2014likelihood, cui2016dimension} and multiscale strategies \cite{wan2011bayesian, xia2022bayesian} alleviate this cost through a-posteriori dimensionality reductions to subspaces of $\mathcal{O}(10^3)$ parameters, at the cost of discarding fine spatial features. Hessian-based and Laplace approximations \cite{bui2008model, tan2013, villa2021hippylib}, whose scalability to high dimensions rests on low-rank approximations of the prior-preconditioned data-misfit Hessian \cite{flath2011fast, bui2013computational}, are effective for linear or mildly nonlinear forwards; a recent linear-forward demonstration reaches $10^9$ parameters for tsunami forecasting \cite{henneking2025gordonbell}, and sequential Monte Carlo built on a Laplace or mean-field variational proposal can refine accuracy for nonlinear forwards at non-negligible additional cost \cite{kehl2017thesis}. The resulting Gaussian is a local quadratic approximation at the maximum a posteriori (MAP) estimate, with its covariance structured as the prior plus a rank-$r$, data-informed correction along the dominant informed directions. Ensemble Kalman inversion (EKI) \cite{iglesias2013ensemble, schillings2017analysis} offers derivative-free parallelism at the cost of restrictive Gaussian assumptions, a larger number of forward-solver evaluations, and ensemble degeneration in high stochastic dimensions. 

Bayesian physics-informed neural networks (B-PINNs) \cite{yang2021b} take a different route: the field is represented as the output of a neural network, the PDE operator is applied to the network output pointwise via automatic differentiation at a set of collocation points, and the resulting strong-form residual enters the likelihood as a Gaussian term with prescribed noise scale. This removes the need for an adjoint solver and reduces the stochastic dimension to the network weight count, at the price of additional hyperparameters that balance data fit against PDE residual, the practical challenges of training deep networks (non-convex optimization, sensitivity to architecture and prior), and a less mature convergence theory than that of finite-element discretization. Recent work on accelerating posterior sampling \cite{pensoneault2024efficient, thiagarajan2025accelerating} has broadened the toolkit, but demonstrations remain largely confined to one- and two-dimensional benchmarks. Adjacent neural strategies instead replace either the forward map with a trained surrogate \cite{zhu2019physics, chatzopoulos2024physics} or the prior over fields with a deep generative model \cite{patel2022solution, kaltenbach2023semi, scholz2024weak}, shifting the computational burden from FE solves to offline network training. The present work takes a complementary approach: rather than replacing the forward model, we operate directly with the finite element (FE) solver and its adjoint, exploiting the algebraic structure of the FE discretization within the Bayesian inference scheme.

Stochastic variational inference (SVI) \cite{hoffman2013stochastic, blei2017variational} offers a different strategy by recasting posterior inference as deterministic optimization of the evidence lower bound (ELBO) using noisy gradient estimates. Combined with adjoint-based model gradients, which provide the log-likelihood gradient at the cost of a single additional PDE solve regardless of the parameter dimension, SVI has been applied to physics-based Bayesian inverse (reconstruction) problems with relatively high-dimensional spatial parameterizations \cite{franck2016sparse, koutsourelakis2016variational, bruder2018beyond, koutsourelakis2012novel}. Related black-box variational methods \cite{ranganath2014black, kucukelbir2017automatic, rei2025blackbox} have broadened the applicability of this paradigm, while normalizing flows \cite{rezende2015variational} have been explored for richer variational families, though their memory overhead and complexity limit applicability at very high dimensions. However, existing adjoint-based SVI approaches rely on covariance-based variational parameterizations and prior structures defined on structured rectangular grids \cite{franck2016sparse, koutsourelakis2016variational, povala2022variational, bruder2018beyond}, which become inefficient and unstable for complex three-dimensional geometries. 

A central challenge in very high-dimensional Bayesian inversion is constructing appropriate priors that provide the necessary regularization. Dense covariance matrices from standard kernel-based Gaussian process (GP) priors \cite{rasmussen2003gaussian} are infeasible beyond $10^4$ parameters, motivating sparse formulations such as Total Variation priors \cite{lassas2004can, franck2016sparse}, Cauchy Markov random fields \cite{suuronen2022cauchy}, Besov priors \cite{dashti2011besov}, and sparse Gaussian processes \cite{hensman2013gaussian}. Among these, Gaussian Markov Random Field (GMRF) priors \cite{markov_priors, rue2005gaussian} are uniquely suited to our setting: their precision matrices arise naturally from FE operators \cite{lindgren2011explicit, roininen2014whittle}, admit conjugate hyperprior updates within a variational Bayes expectation maximization (VB-EM) scheme \cite{dempster1977maximum}, and have found broad use in Bayesian inverse problems \cite{kaipio2011bayesian, koutsourelakis2012novel, wang2006markov, lee2002}. The seminal stochastic partial differential equation (SPDE) approach of \cite{lindgren2011explicit} established the connection between GMRFs and Mat\'ern-class Gaussian fields, showing that samples of such fields can be obtained as solutions to a sparse linear SPDE which, upon FE discretization, directly yields a GMRF with sparse precision matrix. This bypasses the dense $\mathcal{O}(n^2)$ Mat\'ern covariance, which is infeasible to construct or store at the scales targeted here.

Integrated nested Laplace approximations (INLA) \cite{rue2009approximate} exploit GMRF structure for fast approximate Bayesian inference, but the established literature primarily targets uniform, structured grids. The present work, by contrast, formulates both the prior and the variational precision directly in terms of FE operators, namely the discrete Laplacian and the mass matrix, assembled from the same FE mesh used by the forward solver, so that the construction generalizes to arbitrary geometries without manual specification of neighborhood structures. Moreover, the Markov sparsity of the prior precision provides a natural template for the variational distribution: by parameterizing the approximate posterior via its precision Cholesky factor with the \emph{same} sparsity pattern, long-range posterior correlations are captured implicitly through its sparse inverse, in contrast to covariance-based parameterizations \cite{franck2016sparse, koutsourelakis2016variational, bruder2018beyond} that encode only short-range correlations.

This paper combines sparse SVI, adjoint-based gradients, GMRF/SPDE priors, and variational Bayes expectation-maximization (VB-EM) on a single FE discretization for Bayesian spatial field reconstruction at genuinely high parameter dimensions. To our knowledge, the present work is the first to combine (i)~a PDE forward model nonlinear in the inferred parameters, (ii)~a full-covariance Gaussian variational posterior represented implicitly through a sparse precision Cholesky factor, (iii)~a fully Bayesian treatment of the prior and likelihood variance/precision hyperparameters ($\delta$ and $\tau$), with the SPDE shift parameter $\kappa^2$ held fixed as a small nugget (Section~\ref{sec: Prior}), and (iv)~a three-dimensional finite-element discretization on a curved domain, at a stochastic dimension exceeding $4\cdot 10^5$. Prior work has matched individual elements of this combination, notably the million-parameter, nonlinear-in-parameters Hessian-Laplace inversion of \cite{Bui_Thanh_2012} on a 3D curved domain, the recent linear-forward, billion-parameter Hessian-Laplace tsunami inversion of \cite{henneking2025gordonbell} that exploits Toeplitz structure for real-time inference at extreme scale, and the sparse-precision variational approximations of \cite{povala2022variational} on low-dimensional structured-grid problems, but not their integration at this scale. Both our framework and the Hessian-Laplace approaches yield Gaussian approximations of the posterior with comparable mode estimates, but differ in how they capture the covariance. Hessian-Laplace fits the covariance locally at the MAP via the Hessian curvature, with a prior-plus-low-rank structure. Such a local approximation can underestimate posterior dispersion when the log-posterior departs from its quadratic approximation near the mode, which is critical in safety-relevant applications that rely on calibrated uncertainty. Our framework instead fits the full covariance globally via the ELBO and encodes prior-mediated long-range spatial correlations implicitly through a sparse precision matrix whose inverse is dense, yielding smooth, physically plausible posterior samples. We elaborate this comparison in Section~\ref{sec:positioning}. While the individual algorithmic components, namely SVI, adjoint gradients, sparse priors, and variational expectation-maximization \cite{dempster1977maximum}, are each established, integrating them on three-dimensional FE discretizations required two levels of development: genuinely new methodological choices and the careful joint deployment of established techniques to address the practical scale challenges that arise.

\paragraph{Methodological developments:}
\begin{enumerate}
\item A precision-matrix-parameterized sparse Gaussian variational family whose sparsity pattern is inherited from the SPDE prior (Section~\ref{sec: appendix_svi}). Sparse-precision variational inference has been explored in the statistics literature \cite{tan2018gaussian} and for low-dimensional PDE inverse problems \cite{povala2022variational}; the distinguishing choice here is to inherit the sparsity pattern directly from the FE Laplacian $\AL$ of the forward problem's discretization, which scales to three-dimensional FE discretizations. Unlike covariance-based sparse parameterizations, which encode only short-range correlations and produce rough, noisy samples, the precision parameterization implicitly represents dense posterior covariances via its sparse precision matrix (the inverse of the covariance matrix), yielding smooth, physically plausible samples that are essential for stable forward solves in high dimensions.
\item A direct ELBO gradient for this precision-Cholesky parameterization via the path-derivative (\emph{sticking-the-landing}) estimator \cite{roeder2017sticking}, avoiding the reparameterization Jacobian through the inverse Cholesky factor that limits the closely related scheme of \cite{povala2022variational} to low-dimensional structured-grid problems (Section~\ref{sec: appendix_svi}).
\item A unified FE-native assembly in which the SPDE/GMRF prior, the variational precision sparsity pattern, the VB-EM hyperparameter updates, and the adjoint-based log-posterior gradient are all expressed through the same sparse FE operators, namely the discrete Laplacian $\AL$ and the mass matrix $M$, that the forward problem already provides, eliminating bespoke data structures for the inference (Sections~\ref{sec: Prior}, \ref{sec:adjoint}).
\end{enumerate}

\paragraph{Integrated deployment at three-dimensional scale:}
\begin{enumerate}
\setcounter{enumi}{3}
\item A joint VB-EM treatment of the prior and noise precision hyperparameters that, beyond providing fully Bayesian uncertainty quantification, induces an automatic coarse-to-fine continuation, substantially accelerating SVI convergence (Section~\ref{sec:latent_variables}).
\item Natural gradient strategies for both the mean and the precision Cholesky parameters that account for the spatially varying posterior curvature without requiring inversion of the full Fisher information matrix, significantly stabilizing convergence at high dimensions (Section~\ref{sec:natural_gradient_mean}).
\item Practical guidelines, \eg initialization strategies, element-level gradient assembly, and other implementation aspects for stable convergence in three-dimensional FE-based SVI.
\end{enumerate}

We further validate the unified framework through an algorithmic ablation and a comparison with alternative inference methods (Section~\ref{sec:convergence_analysis}). The precision parameterization outperforms both the diagonal mean-field and the sparse banded-covariance variants by approximately $20$ percentage points in posterior-mean discrepancy. The code is available at \url{https://github.com/jnitzler/high_dim_bia_queens} and \url{https://github.com/jnitzler/porous_media_flow_3d}.
The remainder of this paper is organized as follows: Section~\ref{sec:methodology} develops the methodology, Section~\ref{sec:demonstration} presents the numerical demonstration, and Section~\ref{sec:conclusion} concludes with a summary and outlook.

\section{Methodology: High-dimensional Bayesian reconstruction problems with finite elements}
\label{sec:methodology}

We consider Bayesian inverse problems for inferring the posterior distribution of spatial (random) fields from noisy point-wise observations $\byobs$ of a physical system \cite{stuart2010inverse, dashti2017bayesian}. Such problems frequently arise in biomechanics, materials science, and industrial engineering, where spatially varying fields (\eg material properties) are inferred from limited observational data. The spatial field is defined on a physical domain $\Omega \subset \mathbb{R}^3$ with spatial coordinates $\bc \in \Omega$. It is discretized via finite elements, and the vector $\bx \in \mathbb{R}^{\ndofs}$ collects the global FE degrees of freedom (DoFs) of this field. Note that $\bx$ refers to the DoFs of the inferred field of interest, not to those of the forward problem's solution variables such as velocity or pressure (see Section~\ref{sec:fe_representation}). Before observing any data, one typically starts with a prior distribution $p(\bx)$ that reflects the beliefs about the field's correlation structure, smoothness, or overall variability. The prior belief is updated via Bayes' rule:
\begin{equation}
\label{eqn:bayes_rule}
\underbrace{p(\bx|\byobs)}_{\text{posterior}} = \frac{\overbrace{p(\byobs|\Mod(\bx))}^{\text{likelihood}}\cdot \overbrace{p(\bx)}^{\text{prior}}}{\underbrace{\int p(\byobs|\Mod(\bx))\cdot p(\bx)\dd \bx}_{=p(\byobs), \text{ evidence}}}\propto \underbrace{p(\byobs|\Mod(\bx))\cdot p(\bx)}_{\substack{=p(\Mod(\bx),\byobs),\\ \text{unnormalized posterior}}},
\end{equation}
given the observational data $\byobs$ and using a physics-based forward model $\by=\Mod(\bx)$ (in the form of an FE model) within the likelihood $p(\byobs|\Mod(\bx))$, or shorter: $p(\byobs|\bx)$. We abbreviate the log-likelihood function as $\logl(\bx):= \log p(\byobs|\Mod(\bx))$. The computational model links the spatial field parameterization $\bx$ to the model outputs $\by$, which reside in the same observational space as the noisy and potentially sparse observations $\byobs$. The inference objective is to characterize the full posterior distribution $p(\bx|\byobs)$, which represents the updated belief over the spatial field $\bx$ after conditioning on the observations $\byobs$, and quantifies all remaining uncertainty consistent with both the prior and the data. As surveyed in Section~\ref{sec:introduction}, inference methods scalable to $\dim(\bx) > 10^{5}$ predominantly produce Gaussian approximations of $p(\bx|\byobs)$, even though variational inference admits richer parametric families in principle (\eg log-Gaussian or normalizing-flow variants). The remainder of this section develops the sparse-precision variational inference framework, and Section~\ref{sec:positioning} situates it relative to low-rank Hessian-Laplace and ensemble Kalman alternatives.

\subsection{Sparse spatial priors on finite element discretizations}
\label{sec: Prior}
\label{sec:fe_representation}
We discretize the spatial field $\tx(\bc,\bx)$ on arbitrary domains $\Omega$ using a FE approach with shape functions of degree $p$, spatial coordinates $\bc$, and nodal field values $\bx$. The FE interpolation vector $\bs{s}_{\mathrm{ele}}(\bc)$ yields the parameterization:
\begin{subequations}
\begin{align}
    \label{eqn: rf_fe}
    \underbrace{\tx(\bc,\bx)}_{\substack{\text{continuous}\\ \text{(scalar) field}}} &= \underbrace{\bs{s}_\mathrm{ele}(\bc)}_{\substack{\text{local element}\\ \text{interpolation}\\ \text{vector}}}\cdot\underbrace{\bxele}_{\substack{\text{local}\\ \text{DoF}\\ \text{vector}}}\ ,\quad \mathrm{for}\ \bc\in\Omega_{\mathrm{ele}}\in \Omega\\
    \underbrace{\bxele}_{\substack{\text{local}\\ \text{DoF}\\ \text{vector}}} &= \underbrace{\mathfrak{R}}_{\text{restrictor}} [\underbrace{\bx}_{\substack{\text{global}\\ \text{DoF}\\ \text{vector}}}]
\end{align}
\end{subequations}
For \emph{random} fields, $\bx$ is a random vector whose posterior density $p(\bx|\byobs)$ we seek to infer. Constrained fields, \eg strictly positive quantities, are handled through nonlinear transformations such as $\exp(\tx(\bc,\bx))$, enabling inference in an unconstrained space.

Scaling to three-dimensional problems with millions of DoFs requires \emph{sparse prior formulations}, \ie formulations that only involve sparse matrices, since dense covariance matrices are infeasible at this scale. Among sparse priors, Gaussian Markov Random Fields (GMRFs) \cite{markov_priors, rue2005gaussian}, Total Variation priors \cite{lassas2004can,franck2016sparse,koutsourelakis2016variational}, Cauchy MRFs \cite{suuronen2022cauchy}, Besov priors \cite{dashti2011besov}, and sparse Gaussian Processes \cite{hensman2013gaussian, snelson2007local} are the most established.
While Total Variation and Cauchy MRF priors are non-Gaussian and promote piecewise-constant or heavy-tailed spatial variations, which are physically inappropriate for the smooth fields considered here, and Besov priors are defined in the wavelet domain and do not connect naturally to general FE discretizations, GMRFs are uniquely well suited to the present framework: they are Gaussian, which admits conjugate VB-EM hyperparameter updates and direct integration with the Gaussian SVI variational family; their Mat\'ern covariance captures the spatial smoothness of the fields of interest; and, unlike inducing-point sparse GP approximations, their exact sparse precision matrices arise directly from the FE Laplacian and mass matrices already assembled for the forward problem.
We employ a GMRF prior whose precision matrix is constructed via the stochastic partial differential equation (SPDE) approach \cite{lindgren2011explicit}, which establishes an explicit link between Gaussian random fields with Mat\'ern covariance and sparse GMRFs. The resulting precision matrix can be assembled from the same FE matrices -- the Laplacian and mass matrices -- that the forward problem already provides, yielding a principled prior for high-dimensional spatial reconstruction on general FE meshes.

\paragraph{The Whittle-Mat\'ern SPDE.}
The foundational observation, due to Whittle \cite{whittle1954stationary} and formalized in a FE context by Lindgren et al.\ \cite{lindgren2011explicit}, is that a Gaussian random field with Mat\'ern covariance can be represented as the solution of a SPDE, which reads in its simplest form as:
\begin{equation}
\label{eqn:spde}
(\kappa^2 - \Delta)\,\tx(\bc) = \mathcal{W}(\bc), \qquad \bc \in \Omega,
\end{equation}
where $\mathcal{W}$ denotes Gaussian white noise on $\Omega$ and $\kappa > 0$ controls the correlation length of the resulting field (smaller $\kappa$ yields longer-range correlations). More generally, the operator $(\kappa^2 - \Delta)$ can be raised to a power $\alpha/2$ to produce smoother fields with longer-range correlations, but at increased computational cost. We adopt the first-order formulation ($\alpha = 2$, Equation~\eqref{eqn:spde}) throughout this work, which is the standard choice for GMRF priors in inverse problems \cite{markov_priors} and provides sufficient spatial regularity for the reconstruction while avoiding excessive smoothing that can suppress the signal from the likelihood\footnote{On unbounded domains, the solution of Equation~\eqref{eqn:spde} is a Gaussian random field with Mat\'ern covariance of smoothness $\nu = \alpha - d/2$, where $d$ is the spatial dimension \cite{lindgren2011explicit, whittle1954stationary}. For $\alpha = 2$ in three dimensions, this gives $\nu = 1/2$ (exponential covariance). The Mat\'ern smoothness parameter $\nu$ does not appear elsewhere in this work and is just used for reference to the Mat\'ern kernel theory; only $\kappa$ enters the discretized formulation.}.

\paragraph{Finite element discretization of the SPDE.}
The key advantage of the SPDE formulation is that it admits a direct FE discretization on the same mesh. We discretize $\tx(\bc) = \sum_{j=1}^{\ndofs} s_j(\bc) x_j$ using the FE basis from Equation~\eqref{eqn: rf_fe}, with $s_j(\bc)$ being Lagrange finite element basis or shape functions of degree $p$, and apply a Bubnov-Galerkin projection of Equation~\eqref{eqn:spde}. Multiplying by test functions $s_i(\bc)$, integrating over $\Omega$, and integrating the Laplacian term by parts yields:
\begin{equation}
\label{eqn:spde_discrete}
\underbrace{(\AL + \kappa^2\,M)}_{\text{SPDE operator } \AL^*}\cdot\bx = \boldsymbol{\eta},
\end{equation}
where $\AL$ is the symmetric positive semi-definite Laplace matrix and $M$ is the mass matrix, assembled from element-level contributions. 
The right-hand side $\boldsymbol{\eta}$ is the $L^2$ projection of the white noise $\mathcal{W}$ onto the FE basis, with components ${\eta_i = \int_{\Omega} s_i(\bc)\,\mathcal{W}(\bc)\,\dd\Omega}$. A crucial consequence of this projection is that the covariance of $\boldsymbol{\eta}$ yields the FE mass matrix\footnote{White noise is spatially uncorrelated: $\mathbb{E}[\mathcal{W}(\bc)\,\mathcal{W}(\bc')] = \delta(\bc - \bc')$. Computing $\mathrm{Cov}(\eta_i,\eta_j)$ requires a double integral over $\Omega$ since each $\eta_i$ is itself an integral over $\Omega$; the Dirac delta then collapses this to the single integral $\int_\Omega s_i\,s_j\,\dd\Omega = M_{ij}$.}:
\begin{equation}
\label{eqn:noise_cov}
\mathrm{Cov}(\boldsymbol{\eta}) = M.
\end{equation}
This is the central distinction from finite-difference discretizations (as done in \cite{markov_priors}), where the mass matrix reduces to a scaled identity and can be absorbed into the scalar precision parameter. For general FE discretizations, $M$ is non-diagonal, and correctly accounting for its structure is essential.

\paragraph{Boundary conditions.}
The Galerkin discretization of the SPDE implicitly imposes homogeneous Neumann boundary conditions on the random field, \ie $\nabla \tx \cdot \boldsymbol{n} = 0$ on $\partial\Omega$. Integrating the Laplacian term by parts produces the boundary integral $\int_{\partial\Omega} s_i\,(\nabla \tx \cdot \boldsymbol{n})\,\dd\Gamma$, and dropping this term is equivalent to enforcing that natural homogeneous Neumann condition. The choice of boundary conditions for the prior is independent of those for the forward problem.

\paragraph{SPDE-based GMRF precision.}
We denote the SPDE operator from Equation~\eqref{eqn:spde_discrete} by $\AL^* := \AL + \kappa^2\,M$, where $\kappa^2 > 0$ controls the correlation length of the prior field (smaller values yield longer-range correlations). We fix $\kappa^2 = 10^{-4}$ throughout\footnote{At this value, the term $\kappa^2 M$ primarily serves as a regularization that ensures positive definiteness of $\AL^*$ (since $\AL$ alone is positive semi-definite with a null space spanned by constant functions), while permitting long-range spatial correlations. Small values ($\kappa^2 \ll 1$) impose minimal prior bias on the correlation length and let the data determine the spatial structure through the likelihood; the sensitivity of the reconstruction to $\kappa^2$ is examined in Section~\ref{sec:convergence_analysis}. In principle, $\kappa$ could also be learned from the data, but since it enters $\AL^*$ nonlinearly, it cannot be treated within the closed-form hyperparameter framework of Section~\ref{sec:latent_variables}; fixing it is the standard approach \cite{markov_priors}. An \emph{approximate} Bayesian treatment (for instance a gradient or grid step on $\log\kappa^2$ in the outer VB-EM loop, reusing the same sparse $\AL^*$ operators and thus preserving scalability) is a natural direction for future work.}. In addition, we introduce a scalar precision parameter $\delta > 0$ that controls the marginal variance of the prior field independently of its correlation structure. Formally, $\delta$ scales the SPDE noise amplitude: $\boldsymbol{\eta}\sim\mathcal{N}(\boldsymbol{0},\delta^{-1} M)$ instead of $\mathcal{N}(\boldsymbol{0}, M)$. The resulting covariance and precision of the prior field are:
\begin{equation}
\label{eqn:gmrf_precision}
\mathrm{Cov}(\bx|\delta) = \delta^{-1}\,(\AL^*)^{-1}\,M\,(\AL^*)^{-1}, \qquad Q(\delta) = \delta\cdot\AL^*\,M^{-1}\,\AL^*.
\end{equation}
The $M^{-1}$ factor in the precision is a direct consequence of $\mathrm{Cov}(\boldsymbol{\eta}) = M$ (Equation~\eqref{eqn:noise_cov}): on finite-difference grids, $M$ reduces to a scaled identity and absorbs into $\delta$, but on general FE meshes it must be retained.
The GMRF prior is thus:
\begin{equation}
\label{eqn:gmrf_full}
p(\bx|\delta) = \mathcal{N}\left(\bx\,\big|\bmu_{\bx},\; Q(\delta)^{-1}\right),
\end{equation}
where $\bmu_{\bx}$ is the prior mean, set to a constant reflecting the expected field value in the absence of data. The parameter $\kappa^2$ determines the correlation structure (fixed), while $\delta$ controls the overall variance scale. Rather than fixing $\delta$ manually, we treat it fully Bayesian by treating it as a latent variable and assigning a conjugate Gamma hyperprior $p(\delta) =\Gamma\left[\delta|a_0,b_0\right]$ with $a_0=b_0=1\cdot10^{-9}$ \cite{koutsourelakis2012novel} and marginalizing $\delta$ out analytically. The latent variable framework and the VB-EM procedure that enable this marginalization are developed in Section~\ref{sec:latent_variables}, where we also present the resulting closed-form expressions for the marginalized log-prior and its gradient (Equations~\eqref{eqn:latent_var_model_markov} - \eqref{eqn:markov_delta_params}).

\paragraph{Computational aspects.}
Although $\AL^*\,M^{-1}\,\AL^*$ involves the inverse of the mass matrix and is therefore formally dense, it is \emph{never computed explicitly}. All operations decompose into sparse primitives:
\begin{enumerate}
\item \emph{Matrix-vector product} $Q\bv$: compute $\br = \AL^*\bv$, solve $M\bw = \br$, return $\AL^*\bw$.
\item \emph{Quadratic form} $\bv^T Q\bv$: compute $Q\bv$ as above, then form the dot product with $\bv$.
\item \emph{Prior sampling}: solve $\AL^*\bx = \boldsymbol{\eta}$ with $\boldsymbol{\eta} \sim \mathcal{N}(\boldsymbol{0}, M)$, yielding the correct SPDE covariance.
\end{enumerate}
The mass-matrix solve (step~1) is very well-conditioned and typically converges in $\sim$10 conjugate gradient iterations with a Jacobi preconditioner, adding negligible cost. The individual matrices $\AL^*$ and $M$ are sparse with $\mathcal{O}(\ndofs)$ nonzero entries, so the total cost per gradient evaluation remains $\mathcal{O}(\ndofs)$. Yet the implied covariance $Q^{-1}$ is dense, encoding smooth, long-range spatial correlations without ever being formed. The sparsity pattern of $\AL^*$ also provides a natural template for the variational approximation developed in Section~\ref{sec: appendix_svi}: although the full precision $Q = \AL^*\,M^{-1}\,\AL^*$ has a wider bandwidth due to the mass-matrix inverse, the sparsity of $\AL^*$ captures the dominant conditional dependencies dictated by the mesh connectivity.

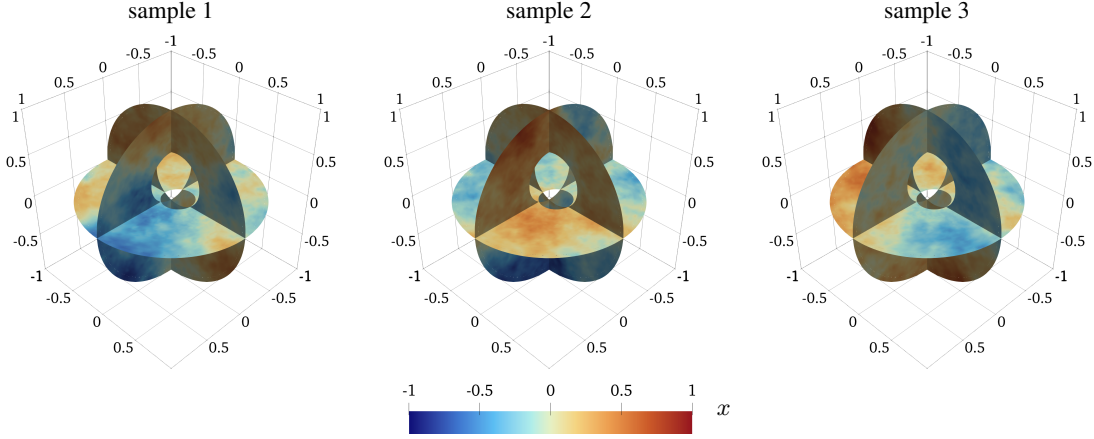
\begin{figure}[htbp]
    \centering
    \begin{tikzpicture}
    \def\imgscale{0.09}
    \def\dx{5.0}  

    \node[inner sep=0pt, anchor=north] (img1) at (0,0)
      {\includegraphics[scale=\imgscale]{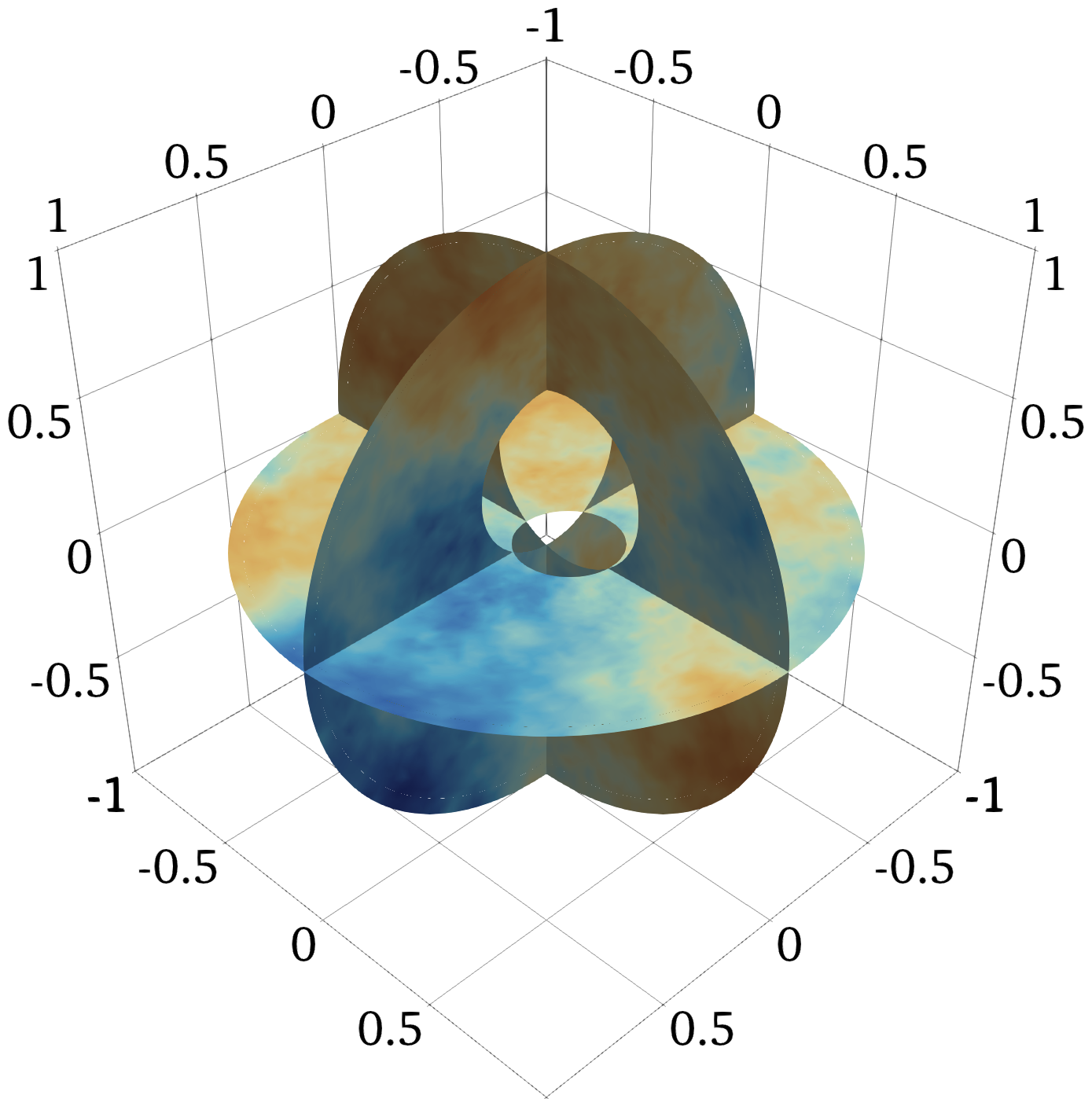}};
    \node[anchor=south, font=\small] at (img1.north) {sample 1};

    \node[inner sep=0pt, anchor=north] (img2) at (\dx,0)
      {\includegraphics[scale=\imgscale]{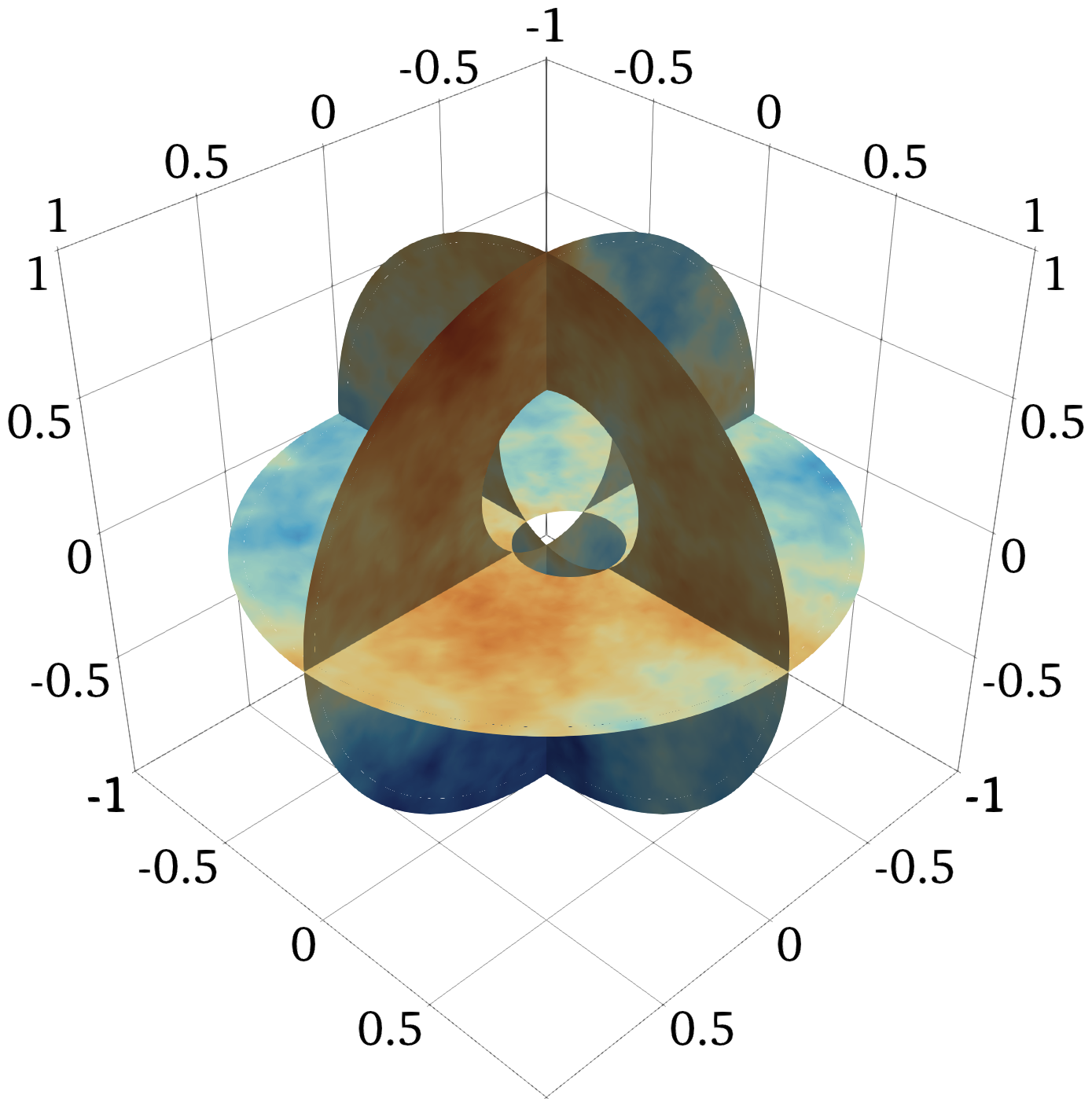}};
    \node[anchor=south, font=\small] at (img2.north) {sample 2};

    \node[inner sep=0pt, anchor=north] (img3) at ({2*\dx},0)
      {\includegraphics[scale=\imgscale]{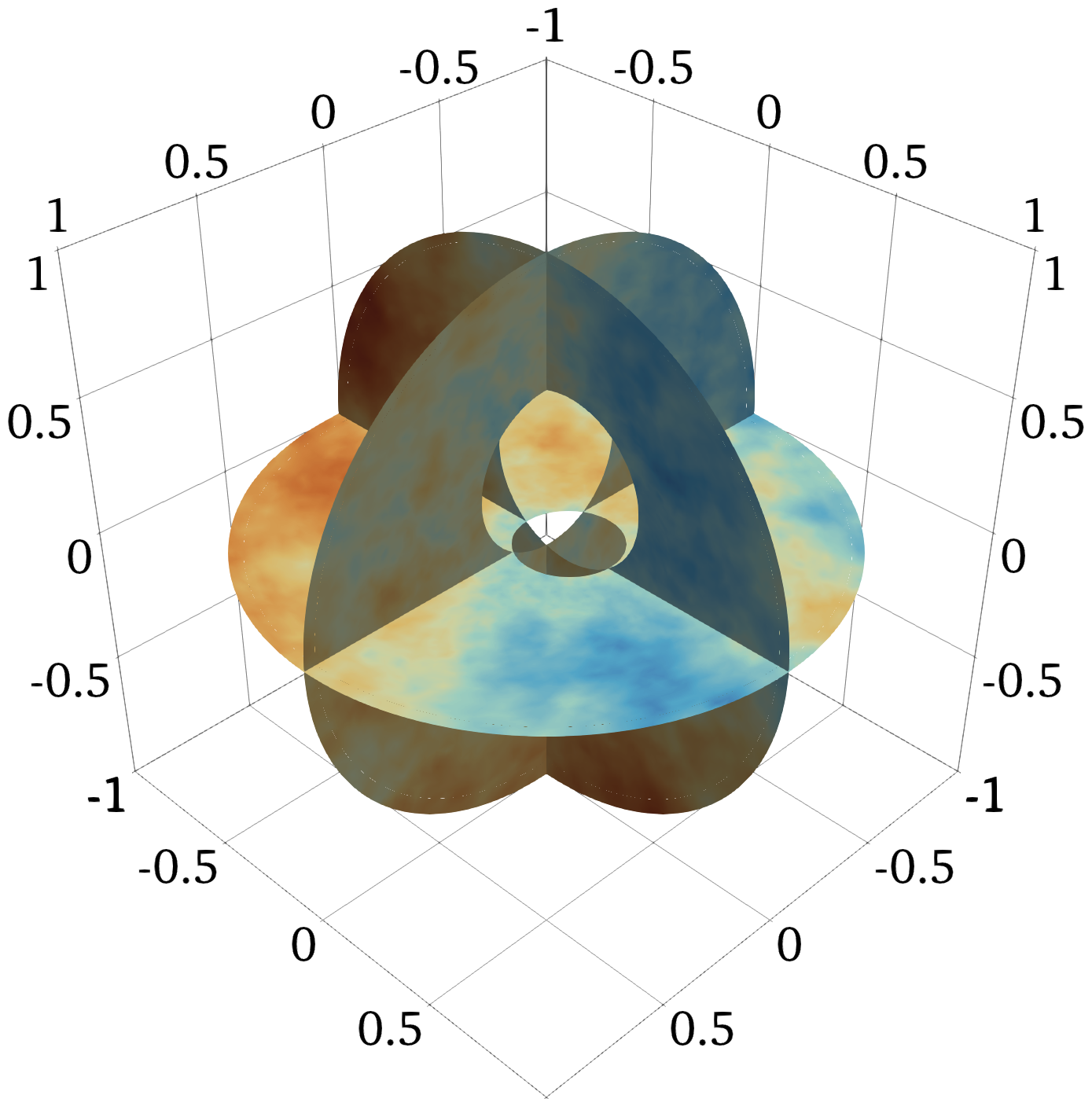}};
    \node[anchor=south, font=\small] at (img3.north) {sample 3};

    \node[inner sep=0pt, anchor=north] (cbar) at (\dx,-4.6)
      {\includegraphics[scale=0.11]{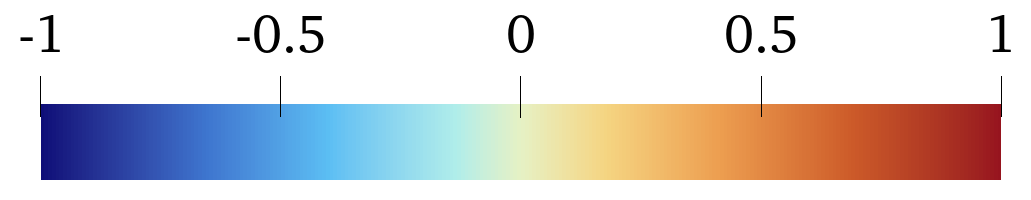}};
    \node[anchor=west, font=\small] at (cbar.east) {$\;x$};

    \end{tikzpicture}
    \caption{Three independent samples from the conditional SPDE-based GMRF prior $p(\bx|\delta{=}1)$ on the three-dimensional eccentric hyper-shell domain (see Section~\ref{sec:porous_media_flow} for the geometry). Samples are obtained by solving $\AL^*\,\bx = \boldsymbol{\eta}$ with $\boldsymbol{\eta} \sim \mathcal{N}(\boldsymbol{0}, M)$, where $\AL^* = \AL + \kappa^2\,M$ is the SPDE operator (Equation~\eqref{eqn:spde_discrete}, $\kappa^2 = 10^{-4}$), yielding the SPDE covariance $\mathrm{Cov}(\bx|\delta) = \delta^{-1}\,(\AL^*)^{-1}\,M\,(\AL^*)^{-1}$. The colorbar indicates the value of the individual field samples $\tilde{x}_i(\bc)$.}
    \label{fig:prior_samples}
\end{figure}

\paragraph{Related work.}
GMRFs are a special case of Markov Random Fields, formalized by the Hammersley-Clifford theorem \cite{hammersley1971markov} and extended to spatial statistics by Besag \cite{Besag_1974}. The SPDE approach of Lindgren et al.\ \cite{lindgren2011explicit} established the explicit link between GMRFs and Mat\'ern-class Gaussian fields on FE meshes, and Roininen et al.\ \cite{roininen2014whittle} applied this connection to Bayesian inverse problems. Bardsley \cite{markov_priors} developed GMRF priors based on discrete Laplacian operators for inverse problems on uniform grids, where the mass matrix reduces to a scaled identity; the present formulation retains the non-diagonal mass matrix required on general FE meshes. For further background on GMRFs, see \cite{rue2005gaussian, perdikaris2015multi}.

\paragraph{Applicability of the SPDE-GMRF prior.}
The SPDE-GMRF prior defined above is appropriate for reconstructing spatial fields that (i)~vary continuously without a priori known sharp discontinuities, (ii)~have no strongly preferred direction or location (approximate isotropy and second-order stationarity), and (iii)~have variability well described by a Gaussian distribution. These conditions apply to many material-property fields, like, for example, tissue permeability and elasticity parameters in biomechanics, and subsurface porosity and thermal conductivity of continuously heterogeneous media in engineering. Gaussianity additionally enables the conjugate VB-EM updates of Section~\ref{sec:latent_variables} and the gradient-penalty interpretation of Remark~\ref{rem:gradient_penalty} that underpin the framework's scalability.

Other sparse priors apply to different regimes: Total Variation \cite{lassas2004can, franck2016sparse, koutsourelakis2016variational} and Cauchy MRF \cite{suuronen2022cauchy} for piecewise-constant or sharply-bounded fields (multi-phase composites, lithology), at the cost of non-Gaussianity; Besov priors \cite{dashti2011besov} for fields with heavy-tailed marginals or sparse localized features; hierarchical SPDE formulations \cite{roininen2014whittle} for known non-stationarity or multiscale behavior. This paper infers the scalar log-permeability field $\tx(\bc)$, yielding the isotropic permeability tensor $K(\bc) = \exp(\tx(\bc))\cdot I$; a fully tensor-valued $K(\bc) \in \mathbb{R}^{3\times 3}$ follows as a straightforward extension with its independent scalar components modeled as cross-correlated SPDE fields.

\begin{remark}[Gradient penalty interpretation and the Laplacian prior]
\label{rem:gradient_penalty}
The SPDE-based precision admits a complementary interpretation through the gradient penalty functional. Minimizing the roughness of the field leads, after FE discretization, to the quadratic form
\begin{equation}
\label{eqn: disc_quadratic_lap}
\min_{\tx}\int_{\Omega}\|\nabla\tx(\bc)\|^2\,\dd\Omega \quad\longrightarrow\quad \min_{\bx}\;\bx^T\cdot\AL\cdot\bx,
\end{equation}
which directly motivates interpreting $\AL$ as the precision of a Gaussian prior, \ie setting $Q_{\mathrm{simple}}(\delta)=\delta\cdot\AL$ instead of the SPDE-based $Q(\delta) = \delta\cdot\AL^*\,M^{-1}\,\AL^*$. This bypasses the SPDE derivation entirely: no $\kappa^2$ regularization, no mass-matrix correction for the white-noise projection, and no mass-matrix solve in the precision evaluation. Positive definiteness can be ensured by a small diagonal perturbation or by eliminating the constant-function null space through boundary conditions. In practice, this simpler formulation works surprisingly well, but it yields fields with shorter correlation lengths and weaker smoothness constraints, leading to less accurate reconstructions. The quantitative comparison in Section~\ref{sec:convergence_analysis} confirms this: the SPDE-based formulation adopted throughout this work yields a lower posterior-mean discrepancy, at the cost of the additional mass-matrix solve documented in the computational aspects above.
\end{remark}

\subsection{Modeling the data log-likelihood}
\label{sec: data_log_lik}
The data log-likelihood connects the simulation output $\by=\Mod(\bx)$ to the noisy observations $\byobs$. We adopt an \iid Gaussian noise model, which is the maximum-entropy distribution for continuous variables with known first and second moments \cite{jaynes1957information} and therefore encodes the least informative assumption about the noise structure beyond its variance. Our noise assumption relates the simulation output and the observations through additive noise $\bs{\epsilon}$:
\begin{subequations}
\begin{align}
    \label{eqn:gaussian_noise}
    \byobs&=\by + \frac{1}{\sqrt{\tau}}\cdot \boldsymbol{\epsilon}\\
    \text{with }\; \bs{\epsilon}&\sim \mathcal{N}\left(\bs{\epsilon}\big|\bs{0},I\right)
\end{align}
\end{subequations}
The noise precision $\tau$ is treated as a latent variable with conjugate Gamma hyperprior $\tau\sim\Gamma(\tau|a_0,b_0)$, $a_0=b_0=10^{-9}$, giving rise to the conditional Gaussian likelihood:
\begin{align}
    \label{eqn:gauss_lik}
    \logl(\bx,\tau)=\log p(\byobs|\Mod(\bx),\tau)=\log \mathcal{N}\left(\byobs\big|\Mod(\bx),\;\tau^{-1}\cdot I\right)
\end{align}
Rather than fixing the noise precision $\tau$ manually, we treat it fully Bayesian by analytically marginalizing it out. The latent variable framework and VB-EM procedure developed in Section~\ref{sec:latent_variables} yield a closed-form marginalized log-likelihood gradient (Equations~\eqref{eqn:lik_marg_1}--\eqref{eqn:lik_marg_2}) that acts as an automatic coarse-to-fine continuation scheme during the SVI optimization. This continuation effect has been observed in earlier work on variational inference for inverse problems \cite{koutsourelakis2012novel, bruder2018beyond}, but becomes particularly critical in the high-dimensional regime we target, where a fixed, potentially misspecified noise precision can easily destabilize gradient-based optimization.

\subsection{Sparse stochastic variational inference for high-dimensional spatial Bayesian inverse problems}
\label{sec: appendix_svi}

Due to the high-dimensional parameter space ($\dim(\bx) \gg 10^3$) and the computational expense of the forward model $\Mod(\cdot)$, sampling-based methods such as Markov chain Monte Carlo (MCMC), sequential Monte Carlo (SMC), and even their gradient-informed variants (such as Stochastic Gradient Hamiltonian Monte Carlo) become impractical. We therefore employ \emph{Stochastic Variational Inference} (SVI) \cite{hoffman2013stochastic,blei2017variational}, building on its successful application to physics-based Bayesian inverse problems in moderately high stochastic dimensions \cite{bruder2018beyond, koutsourelakis2012novel}. The local support of the FE basis functions (Section~\ref{sec:fe_representation}) aligns naturally with the sparse variational distribution developed below, and the high parametric dimension resulting from the FE discretization is less challenging for SVI than the complex posterior structures that arise from global series expansions such as Fourier or Karhunen-Lo\`eve bases.

\paragraph{Basic Concepts of Stochastic Variational Inference (SVI)}
We approximate the posterior $p(\bx|\byobs)$ by a variational distribution $q(\bx|\bphi)$, parameterized by $\bphi$, and measure the approximation quality by the Kullback-Leibler divergence (KLD):
\begin{equation}
\label{eqn:KLD}
\DKL{q(\bx|\bphi)}{p(\bx|\byobs)} = \Ex{q(\bx|\bphi)}{\log\left(\frac{q(\bx|\bphi)}{p(\bx|\byobs)}\right)}
\end{equation}
Since the posterior is unknown, we instead maximize the evidence lower bound (ELBO), which is equivalent to minimizing the KLD \cite{blei2017variational}:
\begin{equation}
\label{eqn:elbo}
\ELBO(\bphi) = \underbrace{\Ex{q(\bx|\bphi)}{\log p(\bx,\byobs)}}_{\substack{\text{expected log-unnormalized posterior}}} \underbrace{- \Ex{q(\bx|\bphi)}{\log q(\bx|\bphi)}}_{\substack{:= \mathbb{H}\left[q(\bx|\bphi)\right]\\ \rightarrow \text{entropy of }q(\bx|\bphi)}},
\end{equation}
with $\log p(\bx,\byobs)=\log p(\byobs|\bx) + \log p(\bx)$. The log-joint expectation is approximated by Monte Carlo with a few samples per iteration, while the entropy $\mathbb{H}[q(\bx|\bphi)]$ is typically available in closed form.

We employ the reparameterization trick \cite{Kingma_Welling_2022}, expressing each sample as a deterministic transformation $\bx=\trp(\br, \bphi)$ of a draw $\br\sim \mathcal{N}(\boldsymbol{0},I)$, which enables differentiation through the sampling process. The stochastic estimate of the ELBO gradient is given by:
\begin{equation}
\label{eqn: svi_rp}
\nabla_{\bphi}\ELBO(\bphi) \approx \frac{1}{\nsamples}\sum_{i=1}^{\nsamples} \nabla_{\bphi}\left[\log p\!\left(\trp(\br_i,\bphi),\byobs\right) - \log q\!\left(\trp(\br_i,\bphi)\big|\bphi\right)\right], \quad \br_i\sim \mathcal{N}(\boldsymbol{0}, I).
\end{equation}
Computing this gradient requires the derivative of the log-unnormalized posterior \wrt the input variable~$\bx$, which decomposes as:
\begin{equation}
\label{eqn: chain_rule_grad}
\nabla_{\bx}\log p(\bx_i,\byobs)=\underbrace{\nabla_{\bx}\logl(\bx_i)}_{\substack{\text{log-likelihood gradient}\\ \text{(adjoint, Sec.~\ref{sec:adjoint})}}} + \underbrace{\nabla_{\bx}\log p(\bx_i)}_{\substack{\text{log-prior gradient}\\ \text{(Sec.~\ref{sec: Prior}, computed in adjoint)}}}.
\end{equation}
The log-likelihood gradient $\nabla_{\bx}\logl(\bx)$ requires differentiation through the forward model $\Mod(\bx)$, which we resolve via the adjoint method \cite{bui2023adjoint, givoli2021tutorial} (see Section~\ref{sec:adjoint}). Each sample $\br_i$ implies a distinct simulation input $\bx_i=\trp(\br_i, \bphi)$, requiring a separate forward and adjoint solve.
The variational parameters are then updated by stochastic gradient ascent:
\begin{equation}
    \label{eqn: stochastic_ascent}
    \bphi_{i+1} = \bphi_i + \boldsymbol{s}_i \cdot \nabla_{\bphi}\ELBO(\bphi_i),
\end{equation}
with step sizes $\boldsymbol{s}_i$ chosen by the Adam optimizer \cite{kingma2014adam}. The specific computation of $\nabla_{\bphi}\ELBO$ for our precision-parameterized variational family, which avoids the explicit reparameterization Jacobian, is detailed below.

\paragraph{Precision-parameterized sparse Gaussian variational family}
The key idea of our variational approach is to exploit the GMRF/SPDE prior's Markov sparsity (Section~\ref{sec: Prior}) for the approximate posterior: by parameterizing the variational distribution via its \emph{precision} Cholesky factor rather than a covariance Cholesky factor, we inherit the prior's sparse precision structure and represent the full, dense posterior covariance \emph{implicitly}, without ever forming or storing it. All required operations, including sampling, ELBO gradient evaluation, and entropy computation, are carried out entirely in terms of sparse triangular solves with the precision's lower Cholesky factor $\LQ$.

We thus define the variational distribution as:
\begin{align}
\label{eqn:var_family}
\begin{split}
q(\bx|\bphi) &= \mathcal{N}\left(\bx\,\big|\,\bmu(\bphimu),\, Q^{-1}(\bphiL)\right),\\
\text{with } Q(\bphiL)&=\LQ(\bphiL)\cdot \LQ^T(\bphiL),\\
\text{and } \bphi &= [\bphimu,\bphiL]^T,\\
\text{with } \bphiL& =[\bphiLdiag, \bphiLoffdiag]^T.
\end{split}
\end{align}
The sparsity pattern of the lower triangular matrix $\LQ$ is chosen to match the sparsity of the SPDE operator $\AL^*$: entry $(\LQ)_{ij}$ is nonzero only if nodes $i$ and $j$ are neighbors on the FE mesh and $i>j$, or if $i=j$. While the full SPDE precision $\AL^*\,M^{-1}\,\AL^*$ has a wider bandwidth, the sparsity of $\AL^*$ captures the dominant Markov neighborhood structure (see Section~\ref{sec: Prior}). This is a natural choice because the posterior precision is expected to share the Markov structure of the prior, with the observational data adding localized corrections. In plain terms, the inferred field is modeled as a Gaussian Mat\'ern-type random field: the prior favors smooth, spatially correlated variation, so sharp features are not forbidden but are smoothed in the reconstruction, while the precision encodes only \emph{local} conditional dependence (given its immediate mesh neighbors, a node is approximately independent of the rest). Crucially, the long-range correlations are not lost: this sparse precision has a \emph{dense} inverse, exactly as for the Mat\'ern Gaussian process it represents. The chosen pattern only approximates the prior's conditional structure (the exact precision $\AL^*\,M^{-1}\,\AL^*$ couples a slightly wider neighborhood), which a broader stencil would refine and is left to future work. In contrast, a covariance-based parameterization $\Sigma = LL^T$ would require $L$ to be dense (or approximated by a banded matrix with substantial fill-in) to represent the long-range correlations that are compactly encoded in the sparse precision. For the three-dimensional problems we target, with $\dim(\bx)>4\cdot 10^5$, a dense covariance factor is infeasible, while the sparse precision Cholesky $\LQ$ stores only the nonzero entries dictated by the mesh connectivity. It is efficiently implemented in our software framework QUEENS \cite{queens} using sparse matrix representations. No additional discretization or sparsity assumptions are required beyond those already present in the GMRF/SPDE prior of Section~\ref{sec: Prior}.

The variational parameters $\bphiL$ are mapped to entries of $\LQ$ through (under slight abuse of notation):
\begin{subequations}
\label{eqn:param_transforms}
\begin{align}
\label{eqn:param_diag}
(\LQ)_{ii}(\bphiLdiag) &= \softplus(\bphiLdiag) = \log\left(1+e^{\bphiLdiag}\right), \quad\text{for diagonal entries},\\
\label{eqn:param_offdiag}
(\LQ)_{ij}(\bphiLoffdiag) &= \bphiLoffdiag,\quad \text{for off-diagonal entries, } i\neq j.
\end{align}
\end{subequations}
The softplus transformation ensures positivity of the diagonal, and therefore positive definiteness of $Q$, while providing bounded gradients $\frac{\dd}{\dd\bphiLdiag}\softplus(\bphiLdiag) = \sigmoid(\bphiLdiag) = \frac{e^{\bphiLdiag}}{1+e^{\bphiLdiag}} \in (0,1)$, where $\sigmoid$ denotes the logistic sigmoid function. Since this derivative is bounded above by one, it precludes exploding gradients; it does not, however, prevent vanishing gradients, since it tends to zero for highly negative $\bphiLdiag$. The off-diagonal entries use an identity mapping, which is sufficient as the Markov sparsity pattern already restricts the set of learnable entries; a small nugget $\epsilon_{\LQ}$ added to the diagonal of $\LQ$ further ensures well-conditioning.

The reparameterization from Equation~\eqref{eqn: svi_rp} takes the form of a sparse triangular solve:
\begin{subequations}
\begin{align}
    \label{eqn:rep_1}
   \trp(\br, \bphi)&= \bphimu + \LQ^{-T}(\bphiL)\cdot\br,\\
   \text{with }\; \br&\sim \mathcal{N}\left(\br|\bs{0},I\right), \label{eqn:rep_2}
\end{align}
\end{subequations}
\ie drawing a sample requires solving the sparse upper-triangular system $\LQ^T\bz = \br$ for $\bz$, followed by the shift $\bx = \bphimu + \bz$, yielding a sample with covariance $Q^{-1}=(\LQ\LQ^T)^{-1}$.
\begin{remark}[Non-Gaussian effective fields]
\label{rem:non_gaussian_fields}
The resulting random fields do not need to follow a Gaussian distribution, as we typically apply a nonlinear transformation $g(\bx)$, \eg to enforce positivity constraints (see Section~\ref{sec:fe_representation}), such that the effective random field follows the distribution implied by the translation process \cite{grigoriu1998simulation}. Richer variational families -- such as normalizing flows \cite{rezende2015variational, papamakarios2021normalizing} or mixture models \cite{lin2019fast} -- could additionally capture non-Gaussian structure directly in $\bx$-space, though at substantially higher memory and computational cost at the scale $\dim(\bx) > 4\cdot 10^5$ considered here.
\end{remark}

\paragraph{Direct ELBO gradient via the \emph{path derivative} estimator}
\label{sec:direct_elbo_gradient}
For a covariance-parameterized variational distribution with $\trp(\br,\bphi)= \bmu + L\cdot\br$, the ELBO gradient in Equation~\eqref{eqn: svi_rp} is typically computed via the reparameterization Jacobian $\nabla_{\bphi}\trp$, which has a known sparse structure. For the precision parameterization with $\trp(\br,\bphi)= \bmu + \LQ^{-T}\cdot\br$, the Jacobian \wrt the entries of $\LQ$ involves derivatives of a matrix inverse, which are complex and expensive to form. We circumvent this by deriving the ELBO gradient \emph{directly}, using the path derivative gradient estimator \cite{roeder2017sticking}. This estimator eliminates the high-variance score-function term $\nabla_{\bphi}\log q(\bx|\bphi)$ from the ELBO gradient, ensuring that the gradient estimator approaches zero variance as the variational posterior approaches the true posterior, a property referred to as \emph{sticking the landing}.

For each sample $\bx_i$, we subtract the score function $\nabla_{\bx}\log q(\bx_i|\bphi)$ from the log-joint gradient to obtain the variance-reduced per-sample gradient:
\begin{equation}
\label{eqn:stl_control_variate}
\boldsymbol{g}_i = \nabla_{\bx}\log p(\bx_i,\byobs) - \nabla_{\bx}\log q(\bx_i|\bphi),
\end{equation}
where $\nabla_{\bx}\log q(\bx_i|\bphi) = -Q(\bx_i - \bmu) = -\LQ\cdot\br_i$ is computed efficiently from the cached base samples $\br_i$, without forming $Q$ explicitly. The score function $\nabla_{\bx}\log q(\bx_i|\bphi)$ serves as a \emph{control variate} \cite{roeder2017sticking}: since $\Ex{q}{\nabla_{\bx}\log q} = \boldsymbol{0}$ (the expected score vanishes), its subtraction does not introduce bias but significantly reduces the variance of the gradient estimator.

The ELBO gradient decomposes into contributions for the mean and the lower precision Cholesky factor:
\begin{subequations}
\label{eqn:direct_elbo_grad}
\begin{align}
\label{eqn:grad_mean}
\nabla_{\bphimu}\ELBO &= \frac{1}{\nsamples}\sum_{i=1}^{\nsamples} \boldsymbol{g}_i,\\
\label{eqn:grad_chol}
\nabla_{(\LQ)_{kl}}\ELBO &= -\frac{1}{\nsamples}\sum_{i=1}^{\nsamples} (\bx_i - \bmu)_k\cdot (\boldsymbol{z}_i)_l,\quad \text{with }\boldsymbol{z}_i = \LQ^{-1}\boldsymbol{g}_i.
\end{align}
\end{subequations}
\begin{subequations}
\label{eqn:chain_rule_transforms}
\begin{align}
\nabla_{\bphiLdiag}\ELBO &= \nabla_{(\LQ)_{kk}}\ELBO \cdot \sigmoid(\bphiLdiag),\\
\nabla_{\bphiLoffdiag}\ELBO &= \nabla_{(\LQ)_{kl}}\ELBO.
\end{align}
\end{subequations}

The idea of sparse precision parameterizations for variational inference has been explored in the statistics literature \cite{tan2018gaussian} and applied to low-dimensional PDE inverse problems with simple forward models \cite{povala2022variational}. However, these works rely on either coordinate-ascent updates with closed-form expressions \cite{tan2018gaussian} or the standard reparameterization Jacobian $\nabla_{\bphi}\trp$ \cite{povala2022variational}, which requires differentiating through the inverse of $\LQ$ and becomes prohibitively expensive at scale. The path derivative gradient derivation presented above (Equations~\eqref{eqn:stl_control_variate} --\eqref{eqn:direct_elbo_grad}) avoids this Jacobian entirely, reducing the per-sample cost to sparse triangular solves with $\LQ$, and is, to our knowledge, new for precision-parameterized variational families.

\paragraph{Natural gradient for the mean parameters}
\label{sec:natural_gradient_mean}
Stochastic gradient ascent on the ELBO treats the variational parameter space as Euclidean, ignoring the Riemannian geometry of the underlying probability distribution. The natural gradient \cite{amari1998natural} corrects for this by premultiplying the Euclidean gradient with the inverse Fisher information matrix (FIM) of the variational distribution, yielding updates that are invariant to the parameterization and typically converge faster \cite{khan2017conjugate, lin2019fast, hoffman2013stochastic}.

For the Gaussian variational family in Equation~\eqref{eqn:var_family}, the FIM with respect to the mean parameters $\bphimu$ is exactly the precision matrix $Q$ \cite{lin2019fast}. The natural gradient update for the mean is therefore:
\begin{equation}
\label{eqn:natural_grad_mean}
\widetilde{\nabla}_{\bphimu}\ELBO = Q^{-1}\nabla_{\bphimu}\ELBO = Q^{-1}\cdot\frac{1}{\nsamples}\sum_{i=1}^{\nsamples}\boldsymbol{g}_i,
\end{equation}
which requires solving the sparse system $Q\widetilde{\nabla}_{\bphimu} = \nabla_{\bphimu}\ELBO$. Since $Q = \LQ\LQ^T$ is already factorized, this amounts to two sparse triangular solves per iteration, at negligible additional cost. Intuitively, the natural gradient rescales the mean update according to the current posterior uncertainty: directions in which the posterior is already well determined receive smaller updates, while directions that are still uncertain receive larger updates. This update is equivalent to the analytic mean natural gradient derived in \cite{tan2025analytic} from the closed-form inverse Fisher information; we implement it through sparse triangular solves with $\LQ$ to avoid materializing $Q^{-1}$ at high dimension.

\paragraph{Natural gradient for the precision Cholesky parameters}
\label{sec:natural_gradient_precision}
For the precision Cholesky parameters $\bphiL$, the full FIM has a complex block structure that is expensive to invert. We therefore use a diagonal approximation of the FIM in the Cholesky factor space. For a diagonal entry $(\LQ)_{kk}$, the corresponding FIM element is $F_{kk} = 1/(\LQ)_{kk}^2 + \Sigma_{kk}$, while for an off-diagonal entry $(\LQ)_{kl}$ ($k > l$), we approximate $F_{kl} \approx \Sigma_{kk}$, where $\Sigma_{kk} = (Q^{-1})_{kk}$ denotes the $k$-th marginal variance.

Computing these marginal variances naively would require forming $Q^{-1}$, which is dense. Instead, we obtain the exact diagonal of $Q^{-1}$ via the Takahashi selected inversion \cite{takahashi1973formation}, which exploits the sparsity of the Cholesky factor $\LQ$ to compute $\mathrm{diag}(Q^{-1})$ in $\mathcal{O}(\mathrm{nnz})$ time. Since $\LQ$ is already factored for the reparameterization, no additional factorization is needed. To account for the softplus transform on the diagonal entries $(\LQ)_{kk} = \log(1 + \exp(\lambda_k))$, the FIM is mapped to the parameter space via the chain rule: $F_{\lambda,kk} = F_{(\LQ),kk}\cdot\sigmoid(\lambda_k)^2$. The natural gradient for $\bphiL$ is then obtained by element-wise division of the Euclidean gradient by $F_\lambda$.

Together with the mean natural gradient from Equation~\eqref{eqn:natural_grad_mean}, this provides natural gradient preconditioning for all variational parameters at a cost that scales linearly with the number of nonzeros in $\LQ$. In practice, the mean natural gradient alone accounts for the dominant improvement in convergence speed and stability; the precision natural gradient via Takahashi selected inversion offers only a moderate additional benefit (see Section~\ref{sec:convergence_analysis} for quantitative evidence). An analytic full-Fisher natural gradient on the precision Cholesky factor has recently been derived in \cite{tan2025analytic}; our diagonal Takahashi approximation is its high-dimensional simplification, empirically justified by the mean-only natural-gradient row of Table~\ref{tab:ablation_results} ($\varepsilon_{\mathrm{pm}}=34.6\,\%$ versus $35.3\,\%$ for the full variant), which shows the precision-Fisher refinement carries limited marginal value in this regime. This diagonal FIM only preconditions the natural-gradient step and does not enter the exact path-derivative ELBO gradient (Equations~\eqref{eqn:grad_mean}--\eqref{eqn:grad_chol}); an inaccurate Fisher approximation can therefore only slow convergence, not bias the converged variational posterior, since at any stationary point the Euclidean gradient, and hence the natural gradient for any positive-definite FIM, vanishes.

\paragraph{Initialization of the variational parameters}
\label{sec:var_init}
A simple initialization sets the mean to the prior mean and all off-diagonal entries to zero:
\begin{equation}
\label{eqn:params_init_1}
\bphimu \leftarrow \Ex{p(\bx)}{\bx},\qquad
\bphiLoffdiag \leftarrow \boldsymbol{0}.
\end{equation}

\vspace{2pt}
\noindent
The diagonal parameters $\bphiLdiag$ are initialized from the prior marginal variance $\sigma_0^2$, so that the initial precision matches the prior:
\begin{equation}
\label{eqn:params_init_2}
(\LQ)_{ii} = \softplus(\bphiLdiag) = \frac{1}{\sigma_0}\cdot\boldsymbol{1}, \qquad \bphiLdiag = \softplus^{-1}\!\left(\frac{1}{\sigma_0}\right)\cdot\boldsymbol{1}.
\end{equation}

\vspace{2pt}
\noindent
This initialization produces a diagonal precision $Q = \sigma_0^{-2}\cdot I$ with zero off-diagonal entries, so that early samples are concentrated around the prior mean. Keeping $\sigma_0$ small prevents (i) high-frequency noise in the stochastic ELBO gradients and (ii) ill-conditioning in the forward solver.

A more informed initialization exploits the fact that the variational precision Cholesky factor $\LQ$ shares the same sparsity pattern as the SPDE operator $\AL^*$. We compute an incomplete Cholesky factorization of $\AL^*$, restricted to the sparsity pattern of $\LQ$, and use the resulting factor to initialize both the diagonal and the off-diagonal variational parameters. This incomplete factorization populates only a subset of the admissible off-diagonal entries with nonzero values, while the remaining entries are set to zero. During optimization, the SVI procedure is free to populate all entries within the prescribed sparsity pattern, and we observe that it does so, confirming that the optimizer utilizes the full sparsity pattern. 

\subsection{Latent variable models: Probabilistic treatment of hyperparameters}
\label{sec:latent_variables}
Both the prior (Section~\ref{sec: Prior}) and the likelihood (Section~\ref{sec: data_log_lik}) contain hyperparameters whose values are unknown a priori. Rather than fixing them, we introduce latent variables $\bz$ with hyperprior $p(\bz)$ and marginalize them out: $p(\bw)=\int p(\bw|\bz)\cdot p(\bz) \dd \bz$. In the high-dimensional settings we target, this probabilistic treatment is essential for balancing regularization and flexibility.
The SVI routine (Equation~\eqref{eqn: chain_rule_grad}) requires the gradient of the log-marginal:
\begin{align}
\label{eqn: latent_var_model}
\log p(\bw) = \log \int p(\bw|\bs{z})\cdot p(\bs{z}) \dd\bs{z}.
\end{align}
Direct differentiation is intractable due to the outer logarithm. Applying Jensen's inequality\footnote{The inequality becomes an equality if we plug in the posterior distribution $p(\bz|\bw)$.} \cite{dempster1977maximum, Beal_Ghahramani_2003} reformulates the expression in terms of the hyperparameter posterior $p(\bz|\bw)$, moving the logarithm inside the expectation:
\begin{align}
\log p(\bw) =\Ex{p(\bz|\bw)}{\log \left(\frac{p(\bw|\bz)\cdot p(\bz)}{p(\bz|\bw)}\right)} =\int \log \left(\frac{p(\bw|\bz)\cdot p(\bz)}{p(\bz|\bw)}\right)\cdot p(\bz|\bw)\dd\bz.
\end{align}
Fisher's identity\footnote{$\Ex{p(\bz|\bw)}{\nabla_{\bw}\log p(\bz|\bw)}=0$: the score function of a log-density, averaged over its own distribution, vanishes.} \cite{fisher1925theory} then yields the gradient:
\begin{align}
\label{eqn:latent_grad}
\nabla_{\bw}\log p(\bw)=\Ex{p(\bz|\bw)}{\nabla_{\bw}\log p(\bw|\bz)}=\int \nabla_{\bw}\log p(\bw|\bz)\cdot p(\bz|\bw)\dd\bz,
\end{align}
\ie the desired gradient is obtained by averaging the conditional log-gradient $\nabla_{\bw}\log p(\bw|\bz)$ over the hyperparameter posterior. With conjugate hyperpriors, $p(\bz|\bw)$ remains in the same family as $p(\bz)$, so both the posterior and the expectation in Equation~\eqref{eqn:latent_grad} are available in closed form. This procedure is known as Expectation-Maximization (EM) \cite{dempster1977maximum}; within a variational inference context, it is referred to as Variational Bayes EM (VB-EM) \cite{Beal_Ghahramani_2003}.

In our setting, both the prior and the likelihood are Gaussians whose precision matrices are scaled by a single hyperparameter ($\delta$ and $\tau$, respectively), each equipped with a conjugate Gamma hyperprior. Conjugacy ensures that both the E-step (hyperparameter posterior) and the M-step (marginal gradient) admit closed-form expressions. The key result is that the gradient of the log-marginal $\nabla_{\bw}\log p(\bw)$, which would otherwise require intractable integration over the hyperparameter, reduces to a standard Gaussian log-density gradient evaluated at a data-adaptive effective precision $\tilde{Q}(\bw)$ that is updated in each SVI iteration.

\paragraph{Application to the GMRF prior.}
The GMRF prior $p(\bx|\delta)=\mathcal{N}\left(\bx\,|\,\bmu_{\bx},\,Q(\delta)^{-1}\right)$ from Equation~\eqref{eqn:gmrf_full} has the SPDE-based precision $Q(\delta)=\delta\cdot\AL^*\,M^{-1}\,\AL^*$ with a scalar scaling $\delta>0$ and conjugate Gamma hyperprior $p(\delta)=\Gamma[\delta|a_0,b_0]$, $a_0=b_0=10^{-9}$. Since $\delta$ enters only as a scalar multiplier of the precision, conjugacy is preserved, and the VB-EM framework yields the marginalized log-prior and its gradient in closed form:
\begin{subequations}
\begin{align}
    \label{eqn:latent_var_model_markov}
    \log p(\bx) &= \log \Ex{p(\delta)}{p(\bx|\delta)}=\Ex{p(\delta|\bx)}{\log\left(\frac{p(\bx|\delta)\cdot p(\delta)}{p(\delta|\bx)}\right)}=\log\mathcal{N}\bigg[\bx\bigg|\bmu_{\bx},\bigg(\underbrace{\frac{a(\bx)}{b(\bx)}\cdot \AL^*\,M^{-1}\,\AL^*}_{\tilde{Q}}\bigg)^{-1}\bigg]\\
    \nabla_{\bx} \log p(\bx) &= \Ex{p(\delta|\bx)}{\nabla_{\bx}\log p(\bx|\delta)}=\nabla_{\bx}\log\mathcal{N}\bigg[\bx\bigg|\bmu_{\bx},\bigg(\underbrace{\frac{a(\bx)}{b(\bx)}\cdot \AL^*\,M^{-1}\,\AL^*}_{\tilde{Q}}\bigg)^{-1}\bigg]\label{eqn:markov_gradient}\\
    \text{with}\;a(\bx)&=a_0 + \frac{\dim(\bx)}{2},\; b(\bx)=b_0 + \frac{1}{2}(\bx-\bmu_{\bx})^T\cdot \AL^*\,M^{-1}\,\AL^* \cdot(\bx-\bmu_{\bx})\label{eqn:markov_delta_params}
\end{align}
\end{subequations}
The effective precision $\tilde{Q} = \frac{a(\bx)}{b(\bx)}\cdot \AL^*\,M^{-1}\,\AL^*$ adapts to the current state of the field $\bx$ through the ratio of the posterior Gamma parameters $a(\bx)$ and $b(\bx)$. This VB-EM update is embedded in the SVI iteration (Algorithm~\ref{alg:svi}) and removes the need for manual tuning of $\delta$. Here $\delta$ is a genuine Gamma hyperprior marginalized analytically (EM serving only as an efficient solver, with $\delta$ never sampled), so the resulting $\bx$-dependence of $\tilde{Q}$ adds neither bias nor Monte Carlo variance: by Fisher's identity (Equation~\eqref{eqn:latent_grad}), $-\tilde{Q}(\bx)(\bx-\bmu_{\bx})$ is exactly the score of the marginal log-prior, a heavier-tailed prior that conservatively allows larger field variances. The gradient~\eqref{eqn:markov_gradient} requires two sparse matrix-vector products with $\AL^*$ and one well-conditioned mass-matrix solve (see Section~\ref{sec: Prior}), and integrates directly into the adjoint-based SVI routine of Section~\ref{sec:adjoint}.

\paragraph{Application to the data log-likelihood.}
Applying the same framework to the conditional Gaussian likelihood $\logl(\bx,\tau)=\log \mathcal{N}\left(\byobs\big|\Mod(\bx),\;\tau^{-1}\cdot I\right)$ from Equation~\eqref{eqn:gauss_lik}, with the conjugate Gamma hyperprior on the noise precision $\tau$, yields the marginalized log-likelihood gradient in closed form. For a simulation input $\bx$ with output $\by=\Mod(\bx)$:
\begin{subequations}
\begin{align}
\label{eqn:lik_marg_1}
\nabla_{\by} \logl(\by)&=\nabla_{\by}\log \mathcal{N}\left(\by\bigg|\byobs,\frac{b(\byobs,\by)}{a(\byobs,\by)}\cdot I\right)\\
\text{with }\; a(\byobs,\by)& =a_0+\frac{1}{2}\cdot\dim\left(\by\right),\ b(\byobs,\by)=b_0+\frac{1}{2}\cdot\|\by-\byobs\|^2 \label{eqn:lik_marg_2}
\end{align}
\end{subequations}
The partial derivative of the log-likelihood \wrt $\by$ can then be directly communicated to the adjoint model according to Equation \eqref{eqn: adjoint_solve}.
In an iterative SVI procedure, the latent-variable model acts as an \emph{automatic continuation scheme}.
Early in the SVI optimization, the residual $\|\by-\byobs\|$ is large, so the data-driven rate $b(\cdot)$
is large and the effective precision $\tilde{\tau}=\frac{a(\byobs,\by)}{b(\byobs,\by)}$ is \emph{small}. Therefore, the log-likelihood is broad and its gradient \emph{gentle} so that the SVI algorithm can traverse the parameter space smoothly and efficiently, avoiding early spurious local maxima. As the SVI samples $\{\bx_i\}$ approach the true posterior, the data residual shrinks and the effective precision converges to the \emph{actual data precision} (often unknown), thereby focusing on finer-scale SVI adjustments. This automatic coarse-to-fine continuation, previewed in Section~\ref{sec: data_log_lik}, makes the gradient-based SVI noticeably more robust and faster, especially in the very high-dimensional setting we target. This VB-EM treatment of the noise precision is integrated into the main SVI loop as detailed in Algorithm~\ref{alg:svi}.

\FloatBarrier
\subsection{Gradient computation via the adjoint method}
\label{sec:adjoint}

The ELBO gradient (Equation~\eqref{eqn: chain_rule_grad}) requires the log-posterior gradient $\nabla_{\bx}\log p(\bx,\byobs) = \nabla_{\bx}\logl(\bx) + \nabla_{\bx}\log p(\bx)$. The log-likelihood contribution $\nabla_{\bx}\logl(\bx)$ involves differentiation through the forward model $\Mod(\bx)$, which we resolve via the adjoint method \cite{bui2023adjoint, givoli2021tutorial} at the cost of a single additional linear solve. For the steady-state forward problem with discrete residual $\bd(\bx,\by)=\mathbf{0}$, where $\by=\Mod(\bx)$, the adjoint system reads:

\begin{equation}
\label{eqn: adjoint_solve}
    \overbrace{\nabla_{\by}\bd^T(\bx, \by)}^{\substack{\text{Partial derivative PDE}\\ \text{\wrt solution variable}\\ \rightarrow \text{usually simple to implement,}\\ \text{or available from linearization}}}\cdot \blambda = -\overbrace{\nabla_{\by}\logl(\by(\bx))^T}^{\substack{\text{Partial derivative of likelihood}\\ \rightarrow\text{simple, but must be communicated}\\ \text{to adjoint solver}}}.
\end{equation}

Solving Equation \eqref{eqn: adjoint_solve} yields the adjoint field $\blambda$, from which the log-likelihood gradient follows to:
\begin{equation}
\label{eqn: final_inner_product}
    \nabla_{\bx}\logl(\bx)=\blambda^T\cdot\underbrace{\nabla_{\bx}\bd(\bx,\by)}_{\substack{\text{Partial derivative of PDE}\\ \text{\wrt input parameters}\\ \rightarrow \text{mostly more cumbersome}}}
\end{equation}

The system matrix $\nabla_{\by}\bd^T$ is typically available from a linearization of the forward problem and, in many cases (\eg porous media flow, diffusion, linear elasticity), coincides with the forward system matrix, allowing the reuse of existing factorizations and preconditioners.

\paragraph{Prior gradient within the adjoint executable.}
The log-prior gradient $\nabla_{\bx}\log p(\bx) = -\tilde{Q}(\bx - \bmu_\bx)$ (Equation~\eqref{eqn:markov_gradient}) requires two sparse matrix-vector products with $\AL^*=\AL + \kappa^2 M$ and one mass-matrix solve, as described in Section~\ref{sec: Prior}. Since both the Laplace matrix $\AL$ and the mass matrix $M$ are already assembled within the solver, we compute the prior gradient within the same adjoint executable and return the full log-posterior gradient $\nabla_{\bx}\log p(\bx,\byobs)$ in a single call, without additional inter-process communication. The adjoint model is thus defined as $\Adj\left(\bx,\nabla_{\by}\logl(\by)\right) = \nabla_{\bx}\log p(\bx,\byobs)$, encapsulating both contributions of Equation~\eqref{eqn: chain_rule_grad}.

For the porous-media example (Section~\ref{sec:demonstration}), the forward and adjoint solvers are available at \url{https://github.com/jnitzler/porous_media_flow_3d}; the FE level derivations are provided in Appendix~\ref{sec:adjoint_porous_media}. Figure~\ref{fig:adjoint_flow_horizontal} visualizes the computational workflow. The framework is implemented in the probabilistic modeling software QUEENS \cite{queens}.

\begin{figure}[ht]
\centering
\begin{tikzpicture}[%
  blockwidth/.store in = \blockwidth,
  blockwidth            = 3.0cm,       
  labelwidth/.store in  = \labelwidth,
  labelwidth            = 3.3cm,       
  arrowraise/.store in  = \arrowraise,
  arrowraise            = 30mm,        
  node distance         = 8mm and 6mm,
  proc/.style           = {draw, rounded corners, align=center,
                           text width=\blockwidth, inner sep=4pt},
  arrow/.style          = {-{Latex[length=2mm]}, thick},
  arwlabel/.style       = {align=center, text width=\labelwidth},
  boxlabel/.style       = {align=center, text width=\labelwidth}
]

\node[proc] (fwd)
  {\textbf{1.\,Forward solve}\\[2pt]
   $\displaystyle \by_i = \Mod(\bx_i)$\\
   FE solve with\\parameter sample $\bx_i$};

\node[proc, right=of fwd] (sens)
  {\textbf{\shortstack[c]{2.\,Compute partial\\gradient of\\ log-likelihood}}\\[2pt]
   $\displaystyle \nabla_{\by}\logl(\by_i)$\\
   Analytic gradient;\\no extra PDE solve};

\node[proc, right=of sens] (adj)
  {\textbf{3.\,Solve adjoint model\\for $\blambda$}\\[2pt]
   $\nabla_{\by}\bd^{T}(\bx_i,\by_i)\blambda_i =
   -\nabla_{\by}\logl(\by_i)$\\
   One more FE solve;\\system often similar to forward model};

\node[proc, right=of adj] (grad)
  {\textbf{4.\,Log-posterior\\gradient}\\[2pt]
   $\nabla_{\bx}\log p(\bx_i,\byobs) =$\\
   $\nabla_{\bx}\logl(\bx) + \nabla_{\bx}\log p(\bx_i)$\\
   Prior reuses $\AL, M$\\from FE solver};

\coordinate[left=6mm of fwd] (param);


\draw[arrow] (fwd) -- (sens);
\coordinate (midFS) at ($(fwd)!0.5!(sens)$);
\node[arwlabel] (labFS) at ($(midFS)+(0,\arrowraise)$)
      {gather model output $\by_i$\\from forward solve};
\draw[thin] (labFS.south) -- (midFS);

\draw[arrow] (sens) -- (adj);
\coordinate (midSA) at ($(sens)!0.5!(adj)$);
\node[arwlabel] (labSA) at ($(midSA)+(0,\arrowraise)$)
      {communicate partial gradient $\nabla_{\by}\logl(\by_i)$\\to adjoint model};
\draw[thin] (labSA.south) -- (midSA);

\draw[arrow] (adj) -- (grad);

\begin{scope}[on background layer]
  \node[draw, dashed, thick, rounded corners,
        fit=(fwd), inner sep=6pt] (boxFwd)   {};
  \node[draw, dashed, thick, rounded corners,
        fit=(adj)(grad), inner sep=6pt] (boxAdj) {};
\end{scope}

\node[above=1mm of boxFwd, boxlabel]  {\textbf{Forward model $\Mod(\bx)$}};
\node[above=1mm of boxAdj, boxlabel]  {\textbf{Adjoint model $\Adj(\bx)$}};

\end{tikzpicture}

\caption{Computational workflow of the adjoint model $\Adj(\bx)$, which returns the full log-posterior gradient $\nabla_{\bx}\log p(\bx_i,\byobs)$. \textbf{Steps~1--3} solve the forward problem, compute the partial likelihood gradient, and solve the adjoint system for $\blambda_i$. \textbf{Step~4} assembles the log-posterior gradient: the likelihood contribution $\nabla_{\bx}\logl(\bx)$ (Equation \eqref{eqn: final_inner_product}) via element-wise contraction of $\blambda_i$ with $\nabla_{\bx}\bd$, and the prior gradient via the SPDE precision $\AL^*\,M^{-1}\,\AL^*$, computed through two sparse products with $\AL^*$ and one mass-matrix solve, reusing matrices already assembled in the FE solver.}
\label{fig:adjoint_flow_horizontal}
\end{figure}
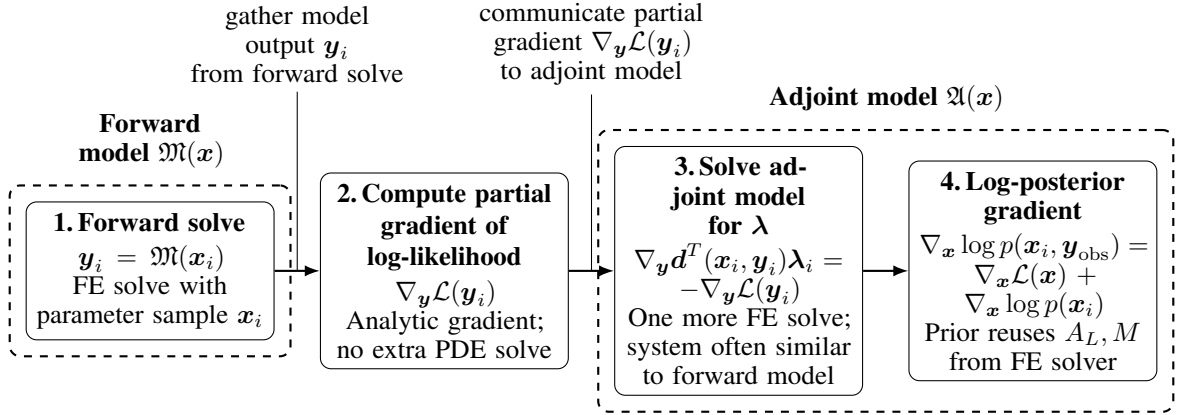

\FloatBarrier

\subsection{Positioning relative to alternative posterior covariance approximations}
\label{sec:positioning}

Every method for scalable Bayesian inversion at $\dim(\bx) > 10^{5}$ trades approximation quality against computational cost, and a principled contribution of any new method is to characterize explicitly where its compromise sits relative to alternatives. Methods proven at this scale predominantly share a Gaussian parametric family and locate similar posterior modes, so the methodological competition between them largely reduces to a comparison of the structural restriction imposed on the covariance. The dominant compromises at extreme scale combine Gaussianity with one of three structural restrictions: exact Hessian curvature at the MAP (Laplace approximation), a rank-$r$ data-informed update to the prior (low-rank Laplace), or an empirical covariance from a finite derivative-free ensemble (ensemble Kalman inversion, EKI). Our approach combines Gaussianity with a sparsity pattern on the precision Cholesky factor, fit globally by stochastic variational inference (stochastic VI or SVI), yielding an implicit dense covariance through the inverse of a sparse factor. Table~\ref{tab:positioning} summarizes the choices.

The Laplace approximation centers a Gaussian at the MAP estimate and matches its covariance to the inverse Hessian of the negative log-posterior at that point; this is exact for linear-Gaussian problems and a local second-order approximation otherwise. The Hessian decomposes into the prior precision plus a data-misfit term, and the latter typically has small effective rank, since sparse observations inform only a limited number of directions in parameter space. Scalability is therefore achieved by retaining the leading $r$ eigenpairs of the prior-preconditioned data-misfit Hessian, computed by randomized matrix-free algorithms \cite{flath2011fast, bui2013computational, Bui_Thanh_2012, henneking2025gordonbell}. The resulting Laplace covariance equals the prior covariance minus a symmetric rank-$r$ correction confined to the top-$r$ data-informed directions: the posterior coincides with the prior where the data are uninformative and is tightened along those informed directions. Note, however, that this covariance approximation is inherently local around the MAP and does not capture the global covariance structure of the posterior, which can lead to severe underestimation of posterior variance and, more broadly, miscalibrated credible intervals when the log-posterior departs markedly from its quadratic approximation away from the mode.

Ensemble Kalman inversion represents the posterior as an empirical ensemble of $J$ members, propagated via Kalman-like updates, avoiding derivatives of the forward model altogether. The ensemble covariance has rank at most $J$ and tends to collapse in very high stochastic dimensions \cite{iglesias2013ensemble, schillings2017analysis}, which limits EKI at the scale targeted here.

Two further families lie adjacent to this Gaussian-covariance comparison. Likelihood-informed and dimension-reduced samplers, preconditioned Crank-Nicolson \cite{cotter2013mcmc}, the likelihood-informed subspace (LIS) \cite{cui2014likelihood} and dimension-independent likelihood-informed (DILI) MCMC \cite{cui2016dimension}, including certified data-free variants \cite{cui2020data}, confine the expensive sampling to a low-dimensional, data-informed subspace and fall back on the prior elsewhere, thereby sharing the small-effective-rank assumption of low-rank Laplace rather than representing the full covariance. Richer variational families, such as gradient-free black-box VI \cite{rei2025blackbox} and normalizing flows \cite{rezende2015variational, papamakarios2021normalizing}, instead capture non-Gaussian structure but have so far been demonstrated only at moderate stochastic dimensions, below the regime targeted here.

Within SVI, the choice of variational family controls the covariance structure. Mean-field Gaussian VI uses a diagonal covariance, which is inexpensive but ignores all posterior correlations. Sparse-covariance parameterizations \cite{franck2016sparse, koutsourelakis2016variational, bruder2018beyond} use a banded Cholesky factor of the covariance but cannot represent correlations beyond the bandwidth, producing rough samples at high stochastic dimensions. Our sparse-precision variational Gaussian instead imposes a sparsity pattern inherited from the FE Laplacian $\AL^{*}$ of the forward mesh on the precision Cholesky factor $\LQ$. The implied marginal covariance $Q^{-1} = (\LQ\LQ^{T})^{-1}$ is dense and encodes correlations between non-neighboring mesh nodes through the inverse of this sparse factor, retaining long-range prior-mediated structure at $\mathcal{O}(\ndofs)$ memory cost. The precision parameterization additionally admits a sparse diagonal approximation of the Fisher information on $\LQ$ via Takahashi selected inversion, enabling natural-gradient preconditioning of the Cholesky parameters at $\mathcal{O}(\ndofs)$ cost; the covariance parameterization has no analogous sparse construction, since $Q = \Sigma^{-1}$ is dense even when $\Sigma$ is sparse.

Across these families, our approach is closest to low-rank Laplace in producing a Gaussian on the FE DoFs with comparable MAP estimates, but differs in truncating the covariance by a sparsity pattern on the precision rather than by a low rank in the Hessian eigenbasis, and in fitting the ELBO globally rather than locally. Here \emph{globally} refers to fitting the covariance over the whole posterior, not to attaining a global optimizer of the non-convex, stochastically optimized ELBO. The global fit preserves calibration when the log-posterior is non-Gaussian away from the mode; the sparsity pattern captures long-range correlations that a low-rank or diagonal covariance cannot resolve. Low-rank Laplace excels for linear or weakly nonlinear forwards with small data-informed effective rank, particularly in offline-online settings; the sparse-precision approach excels for genuinely nonlinear forwards with sparse observations and prior-mediated long-range posterior structure, the regime targeted in Section~\ref{sec:demonstration}. A controlled-variable empirical comparison against mean-field VI, sparse banded-covariance VI, and the Laplace approximation, run on the same forward problem and prior, is reported in Section~\ref{sec:convergence_analysis}.

\begin{table}[htbp]
\centering
\small
\caption{Overview of posterior covariance approximations. Methods are characterized by their treatment of the forward model, the basis used for covariance truncation, the resulting covariance structure, and the memory cost in terms of the stochastic field dimension $\ndofs$, with $r$ the retained rank (low-rank Laplace) and $J$ the ensemble size (EKI).}
\label{tab:positioning}
\begin{tabularx}{\textwidth}{l X X X l}
\toprule
\textbf{Method} & \textbf{Forward treatment} & \textbf{Truncation basis} & \textbf{Covariance form} & \textbf{Cost in $\ndofs$} \\
\midrule
Full Laplace & Local quadratic at MAP & None (infeasible at scale) & Dense inverse Hessian & $\mathcal{O}(\ndofs^{2})$ storage \\[2pt]
Low-rank Laplace \cite{flath2011fast,bui2013computational,Bui_Thanh_2012,henneking2025gordonbell} & Local quadratic at MAP & Rank $r$ in prior-preconditioned Hessian eigenbasis & $\Gamma_{\mathrm{prior}}$ + rank-$r$ data-informed correction & $\mathcal{O}(\ndofs\,r)$ \\[2pt]
EKI \cite{iglesias2013ensemble,schillings2017analysis} & Derivative-free & Ensemble rank $J$ & Empirical $\Sigma$ from $J$ ensemble members & $\mathcal{O}(\ndofs\,J)$ \\[2pt]
Mean-field VI & Global ELBO fit & Diagonal & Diagonal $\Sigma$, no correlations & $\mathcal{O}(\ndofs)$ \\[2pt]
Sparse-covariance VI \cite{franck2016sparse,koutsourelakis2016variational,bruder2018beyond} & Global ELBO fit & Banded covariance Cholesky & Banded $\Sigma$, short-range correlations & $\mathcal{O}(\ndofs)$ \\[2pt]
\textbf{Sparse-precision VI (this work)} & Global ELBO fit & Sparsity pattern of $\AL^{*}$ in mesh basis & Dense $Q^{-1}$ via sparse $\LQ$, long-range correlations & $\mathcal{O}(\ndofs)$ \\
\bottomrule
\end{tabularx}
\end{table}

\subsection{Final algorithmic overview}
\label{sec:overview}
The concepts that were introduced in the previous sections are now summarized and put into context in Algorithm \ref{alg:svi}. We reference the respective sections for more context on the individual steps. For the convenient implementation and orchestration of the involved models and subroutines, we utilize our freely available probabilistic modeling framework, QUEENS \cite{queens}. The specific implementation reproducing this paper's results is available at \url{https://github.com/jnitzler/high_dim_bia_queens}.

\begin{algorithm}[htbp]
\DontPrintSemicolon
\SetAlgoLined
\KwIn{Observed data $\byobs$, forward model $\Mod(\cdot)$, adjoint model $\Adj(\cdot)$, GMRF prior $p(\bx|\delta)$ with $\AL^*$, $M$}
\KwOut{Optimized posterior approximation $q(\bx|\bphi)=\mathcal{N}\!\left(\bx\,|\,\bmu,\,(\LQ\LQ^T)^{-1}\right)$}
\BlankLine
\makebox[0.26\linewidth][r]{Initialize $\bphi$}\makebox[0.42\linewidth][l]{${}=[\bphimu,\bphiL]^T$}\makebox[0.32\linewidth][l]{\emph{Eqs.~\eqref{eqn:params_init_1}--\eqref{eqn:params_init_2}}}\;
\BlankLine
\While{\textnormal{not converged}}{
\tcc{Step 1 -- Draw samples and run forward model, for $i=1,\ldots,\nsamples$}
\makebox[0.26\linewidth][r]{$\br_i$}\makebox[0.42\linewidth][l]{${}\sim\mathcal{N}(\boldsymbol{0},I)$}\makebox[0.32\linewidth][l]{\emph{Eq.~\eqref{eqn:rep_1}}}\;
\makebox[0.26\linewidth][r]{$\bx_i$}\makebox[0.42\linewidth][l]{${}\gets\bmu + \LQ^{-T}\br_i$}\makebox[0.32\linewidth][l]{\emph{Eq.~\eqref{eqn:rep_2}}}\;
\makebox[0.26\linewidth][r]{$\by_i$}\makebox[0.42\linewidth][l]{${}\gets\Mod(\bx_i)$}\makebox[0.32\linewidth][l]{\emph{Sec.~\ref{sec:porous_media_flow}}}\;
\BlankLine
\tcc{Step 2 -- Update effective hyperparameters (VB-EM)}
\makebox[0.26\linewidth][r]{prior precision~~$\tilde{Q}_i$}\makebox[0.42\linewidth][l]{${}\gets\tfrac{a(\bx_i)}{b(\bx_i)}\,\AL^*\,M^{-1}\,\AL^*$}\makebox[0.32\linewidth][l]{\emph{Eq.~\eqref{eqn:markov_delta_params}}}\;
\makebox[0.26\linewidth][r]{noise precision~~$\tilde{\tau}_i$}\makebox[0.42\linewidth][l]{${}\gets\tfrac{a(\byobs,\by_i)}{b(\byobs,\by_i)}$}\makebox[0.32\linewidth][l]{\emph{Eq.~\eqref{eqn:lik_marg_2}}}\;
\BlankLine
\tcc{Step 3 -- Compute log-posterior gradient via adjoint solve}
\makebox[0.26\linewidth][r]{$\nabla_{\by}\logl(\by_i)$}\makebox[0.42\linewidth][l]{${}\gets\nabla_{\by}\log\mathcal{N}(\by_i\,|\,\byobs,\tilde{\tau}_i^{-1}I)$}\makebox[0.32\linewidth][l]{\emph{Eq.~\eqref{eqn:lik_marg_1}}}\;
\makebox[0.26\linewidth][r]{$\nabla_{\bx}\log p(\bx_i,\byobs)$}\makebox[0.42\linewidth][l]{${}\gets\Adj(\bx_i,\nabla_{\by}\logl(\by_i))$}\makebox[0.32\linewidth][l]{\emph{Sec.~\ref{sec:adjoint}}}\;
\BlankLine
\tcc{Step 4 -- Assemble ELBO gradient via path-derivative estimator}
\makebox[0.26\linewidth][r]{$\boldsymbol{g}_i$}\makebox[0.42\linewidth][l]{${}\gets\nabla_{\bx}\log p(\bx_i,\byobs)-\nabla_{\bx}\log q(\bx_i|\bphi)$}\makebox[0.32\linewidth][l]{\emph{Eq.~\eqref{eqn:stl_control_variate}}}\;
\makebox[0.26\linewidth][r]{$\nabla_{\bphimu}\ELBO,\ \nabla_{\bphiL}\ELBO$}\makebox[0.42\linewidth][l]{${}\gets\mathrm{PathDerivative}(\boldsymbol{g}_i)$}\makebox[0.32\linewidth][l]{\emph{Eqs.~\eqref{eqn:grad_mean}--\eqref{eqn:grad_chol}}}\;
\BlankLine
\tcc{Step 5 -- Apply natural-gradient and Adam update}
\makebox[0.26\linewidth][r]{$\widetilde{\nabla}_{\bphimu}\ELBO$}\makebox[0.42\linewidth][l]{${}\gets Q^{-1}\nabla_{\bphimu}\ELBO$\quad\textnormal{\footnotesize(triangular solves with $\LQ$)}}\makebox[0.32\linewidth][l]{\emph{Eq.~\eqref{eqn:natural_grad_mean}}}\;
\makebox[0.26\linewidth][r]{$\widetilde{\nabla}_{\bphiL}\ELBO$}\makebox[0.42\linewidth][l]{${}\gets F_{\lambda}^{-1}\nabla_{\bphiL}\ELBO$\quad\textnormal{\footnotesize(diag.\ FIM via Takahashi)}}\makebox[0.32\linewidth][l]{\emph{Sec.~\ref{sec:natural_gradient_precision}}}\;
\makebox[0.26\linewidth][r]{$\bphimu$}\makebox[0.42\linewidth][l]{${}\gets\mathrm{Adam}(\bphimu,\widetilde{\nabla}_{\bphimu}\ELBO)$}\makebox[0.32\linewidth][l]{\emph{Eq.~\eqref{eqn: stochastic_ascent}}}\;
\makebox[0.26\linewidth][r]{$\bphiL$}\makebox[0.42\linewidth][l]{${}\gets\mathrm{Adam}(\bphiL,\widetilde{\nabla}_{\bphiL}\ELBO)$}\makebox[0.32\linewidth][l]{\emph{Eq.~\eqref{eqn: stochastic_ascent}}}\;
}
\caption{Sparse variational inference with expectation-maximization.}
\label{alg:svi}
\end{algorithm}

\FloatBarrier
\section{Numerical demonstration}
\label{sec:demonstration}

We demonstrate the proposed framework on a three-dimensional porous-media flow problem that is nonlinear in the parameters $\bx$ and discretized at a resolution that yields $\dim(\bx)> 10^5$ stochastic dimensions. The demonstration is designed with the broader goal of extending our earlier multi-fidelity Bayesian inverse analysis framework \cite{nitzler2024bmfia} toward coupled, nonlinear multi-physics systems, which demands a scalable, fully differentiable single-physics methodology as developed herein. To this end, the forward model operates on a curved, three-dimensional domain that necessitates a FE discretization beyond simple equidistant grids, and the spatial resolution is chosen fine enough to make the resulting inverse problem high-dimensional and computationally challenging.
\FloatBarrier
\subsection{Description of the physics-based forward problem: Three-dimensional porous media flow}
\label{sec:porous_media_flow}
\begin{figure}[htbp]
    \centering
    \begin{tikzpicture}
        \node[anchor=south west, inner sep=0] (img) at (0,0) 
            {\includegraphics[scale=0.1723]{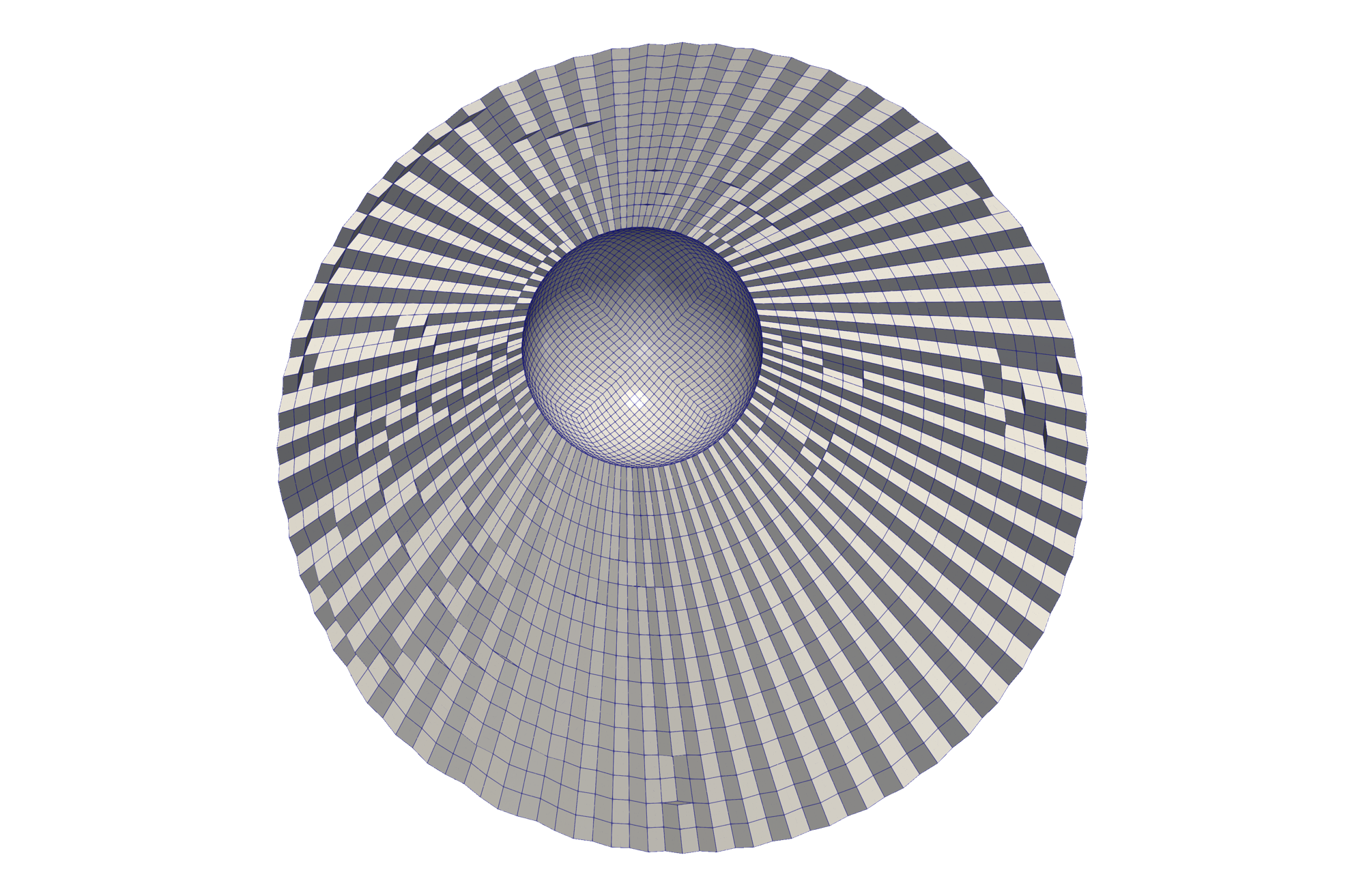}};
        \begin{scope}[x=1cm, y=1cm, shift={(img.south west)}]
            \coordinate (O) at (1, 0); 

            \draw[arrow] (1, 0.0) -- (0.0, 0) node at (-0.2, 0) {\small $c_2$};
            \draw[arrow] (1, 0) -- (1, 1) node at (1, 1.2) {\small $c_3$};

            \node at (O) {$\boldsymbol{\otimes}$};  
            \node at ($(O)+(0.4,0)$) {\small $c_1$};

            \node[red] at (5.7, 4.85) {\Large $\times$}
             node at (6.0, 4.85) {\color{red} \small{ $P_1$}};
             \draw[arrow,red] (5.7, 4.85) -- (4.65, 4.85)
             node[red] at (5.2, 5.0) {\small{$r_1$}};
            \node[red] at (6.1, 4.0) {\Large $\times$}
             node at (6.4, 4.0) {\color{red} \small{ $P_2$}};
            \draw[arrow,red] (6.1, 4.0) -- (2.5, 4.0)
             node[red] at (4.0, 4.2) {\small{$r_2$}};

             \node at (6, 2) {$\Omega$};
             \node at (2.6, 2) {$\Gamma_{p}$};
             \node at (6.1, 5.3) {$\Gamma_{\bu}$};
             \draw[thin] (2.7, 2.1) -- (3.0, 2.2);
             \draw[thin] (6.2, 5.4) -- (6.4, 5.7);

            \draw[arrow] (13.2, 1) -- (12.8, 1.3) node at (12.5, 1.3) {\small $c_2$};
            \draw[arrow] (13.2, 1) -- (13.6, 1.4) node at (13.8, 1.6) {\small $c_1$};
            \draw[arrow] (13.2, 1) -- (13.0, 1.7) node at (13.1, 1.9) {\small $c_3$};
            
        \end{scope}
        \node[anchor=south east, inner sep=0] at ($ (img.south east) + (+2.5cm, 2.5cm) $) {
         \includegraphics[width=4cm]{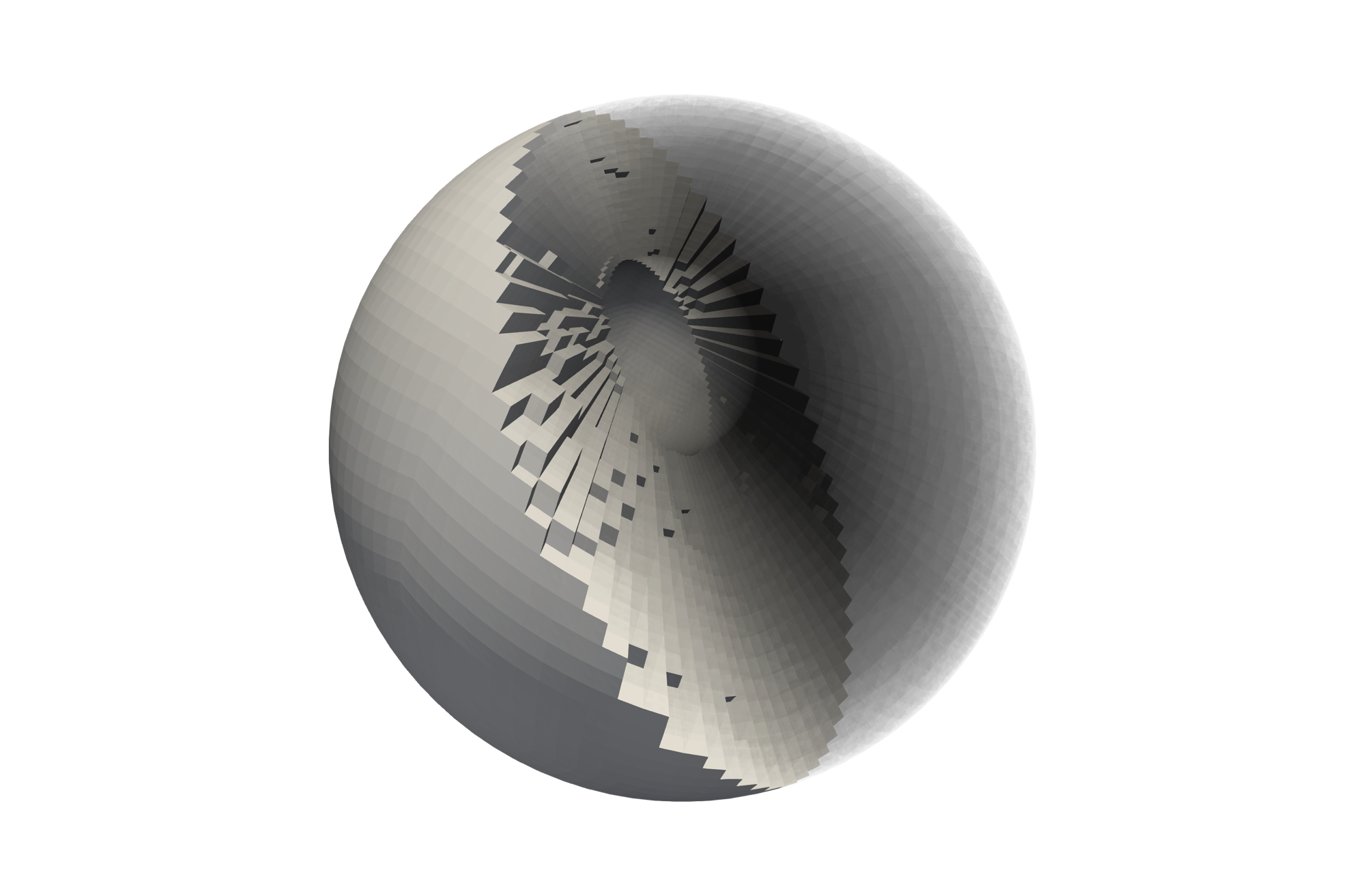}
        };
    \end{tikzpicture}
    \caption{Spherical computational domain $\Omega$ (center $P_2=[0,0,0]$ and radius $r_2=1.0$) with an eccentric spherical void (center $P_1=[0,0.1,0.25]$ and radius $r_1=0.3$), meshed with hexahedral elements. \textbf{Left:} Slice through the $c_2-c_3$ plane, which contains the centers $P_1$ and $P_2$. The outer surface of the domain $\Omega$ is denoted by $\Gamma_{p}$, on which we prescribe the pressure boundary $p(\bc)=b(\bc)=0$. The inner surface of $\Omega$ (surface of the void) is denoted by $\Gamma_{\bu}$ on which we prescribe a no-penetration velocity boundary condition $\bu(\bc)\cdot \bn_{\Gamma_{\bu}}(\bc)=w(\bc)=0$. \textbf{Right:} 3D view of the domain with the void.}
    \label{fig:sphere_domain}
\end{figure}
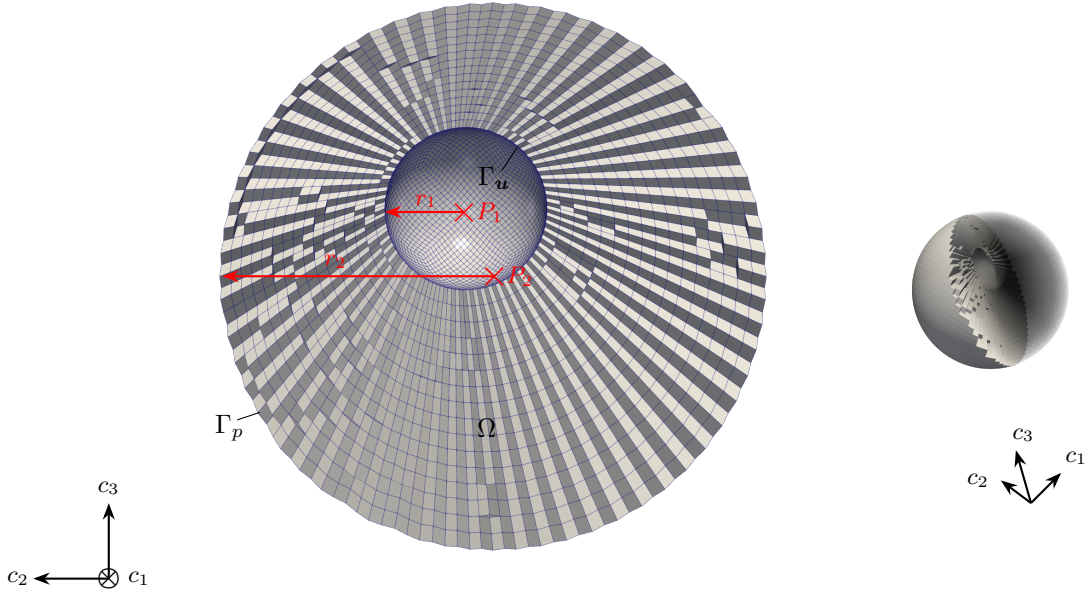

We consider the stationary single-phase flow of an incompressible fluid through a heterogeneous porous medium. Porous media flow problems offer a compelling and widely used benchmark for Bayesian inference \cite{stuart2010inverse, bilionis2013solution} due to their broad relevance and mathematical properties. They arise in many fields of engineering and physics such as biomechanics, particularly tissue modeling, tumor modeling \cite{kremheller2018monolithic, sciume2013multiphase}, modeling of the human lung \cite{K_glmeier_2026} and are also encountered in civil engineering applications such as oil reservoir modeling or groundwater flow applications \cite{zienkiewicz1999computational, rasmussen2021open, chavent1986mathematical, banaei2021numerical, schrefler1993fully, lewis1998finite}.
For the computational domain $\Omega$, we define a three-dimensional sphere with an eccentric spherical void, as depicted in Figure \ref{fig:sphere_domain}.
We implement the subsequent model with the open-source FE library deal.II \cite{africa2024deal}. The forward and adjoint solver implementation is available at \url{https://github.com/jnitzler/porous_media_flow_3d}. 

The governing equations of the steady-state porous media problem are given as follows:
\begin{subequations}
\begin{align}
\label{eqn: strong_form_darcy}
    K^{-1}(\bx, \bc)\cdot \bu  + \nabla p &= \bnil, \ \qquad \text{ in } \Omega,\\
    \text{div } \bu & = a(\bc), \quad \text{ in } \Omega, \\
    p(\bc) & = b(\bc),\   \quad \text{on } \Gamma_{p},\\
    \bu(\bc)\cdot \bn_{\Gamma_{\bu}}(\bc) & = w(\bc),\quad \text{on } \Gamma_{\bu}.
    \label{eqn: strong_form_darcy_3}
\end{align}
\end{subequations}
Here, $\bu$ describes the velocity of a fluid that flows through a porous domain $\Omega$ with spatially variable permeability matrix $K(\bx,\bc)=k(\bx,\bc)\cdot I$. We assume the permeability field $k(\bx,\bc)$ is spatially variable and unknown, and model it as a random field. The gradient $\nabla p$ reflects the spatial pressure gradient in the domain. We assume an incompressible fluid, such that its velocity divergence equals the source term according to $\text{div } \bu  = a(\bc)$. In particular, we choose a no-penetration velocity boundary condition on $\Gamma_{\bu}$ according to $\bu(\bc)\cdot \bn_{\Gamma_{\bu}}(\bc)=w(\bc)=0$; and zero-pressure boundary condition $p(\bc)=b(\bc)=0$ on $\Gamma_{p}$. We furthermore set a positive, constant source term $a(\bc)=1.0$.

We solve the governing Equations \eqref{eqn: strong_form_darcy} - \eqref{eqn: strong_form_darcy_3} in mixed formulation \cite{douglas1985global}. In particular, we use the Bubnov-Galerkin scheme with Lagrangian polynomials for shape and weighting functions, with the settings specified in Table \ref{tab: model_fem_details}. A more detailed derivation of the weak form and discretization is provided in Appendix \ref{sec:darcy_weak}.
\begin{table}[htbp]
\centering
\caption{Overview of employed FEM discretization and settings.}
\label{tab: model_fem_details}
\begin{tabular}{lc} 
\toprule
\textbf{Settings} & \textbf{Value}\\
\midrule
\emph{FEM discretization} & \\
Element type & hexahedral\\
Polynomial degree & $\bu$: 2, $p$: 1, $k$: 2\\
Number of elements & $13\,572$\\
Number of DoFs (forward problem) & $1\,268\,968$ ($1\,216\,710$ velocity, $52\,258$ pressure)\\
$\dim(\bx)$ (stochastic dimension) & $405\,570$\\
\midrule
\emph{Numbers of variational parameters} & \\
Mean parameters $\bphimu$ & $405\,570$\\
Precision Cholesky diagonal $\bphiLdiag$ & $405\,570$\\
Precision Cholesky off-diagonal $\bphiLoffdiag$ & $12\,478\,560$\\
Total variational parameters $\dim(\bphi)$ & $13\,289\,700$\\
\midrule
\emph{Computational resources} & \\
Approx.\ wall time (on 16 cores) & 10\,s (forward) / 11\,s (adjoint)\\
Processor type & AMD Epyc 9354 Zen 4 @ 3.25\,GHz, 32 Core\\
\bottomrule
\end{tabular}
\end{table}
While the stochastic dimension of the random field is $\dim(\bx)=405\,570$, the total number of variational parameters is substantially larger: the lower-triangular precision Cholesky factor $\LQ$ contributes approximately $12.5$ million off-diagonal entries that encode the correlation structure inherited from the FE mesh connectivity. In addition to the mean and diagonal parameters, SVI optimizes approximately $13.3$ million in total variational parameters. The sparse precision parameterization introduced in Section~\ref{sec: appendix_svi} is essential to keep this number tractable. Each SVI iteration draws $\nbatch=4$ Monte Carlo samples and is dominated by the resulting $4$ forward and $4$ adjoint PDE solves (Table~\ref{tab:bia_setup}); the prior, Takahashi-inversion, and triangular-solve operations all scale linearly in the field dimension $\ndofs$ and are negligible by comparison, while memory is likewise $\mathcal{O}(\ndofs)$ since the dense $Q^{-1}$ is never formed. Convergence within $\sim\!200$ iterations (Section~\ref{sec:convergence_analysis}) therefore costs roughly $800$ forward and $800$ adjoint solves in total, well within the $2000$-call budget; this solve count is independent of mesh refinement, and the $\mathcal{O}(\ndofs)$ inference operations above grow only linearly with it, never dominating the FE forward and adjoint solves themselves.

We initialize the variational parameters using the prior-based incomplete Cholesky strategy described in Section~\ref{sec:var_init}. The incomplete factorization of the prior precision matrix populates $6\,848\,180$ of the $12\,478\,560$ admissible off-diagonal entries with nonzero values, while the remaining entries start at zero. After convergence of the SVI optimization, all off-diagonal entries have moved away from their initial values, confirming that the optimizer utilizes the full sparsity pattern dictated by the mesh connectivity.

\FloatBarrier

\subsection{Ground-truth and observables of the Bayesian inverse problem}
\label{sec:observables}
\begin{figure}[htbp]
    \centering
    \begin{tikzpicture}
    \node[anchor=south west,inner sep=0] (background) at (0,0) {\includegraphics[scale=0.07]{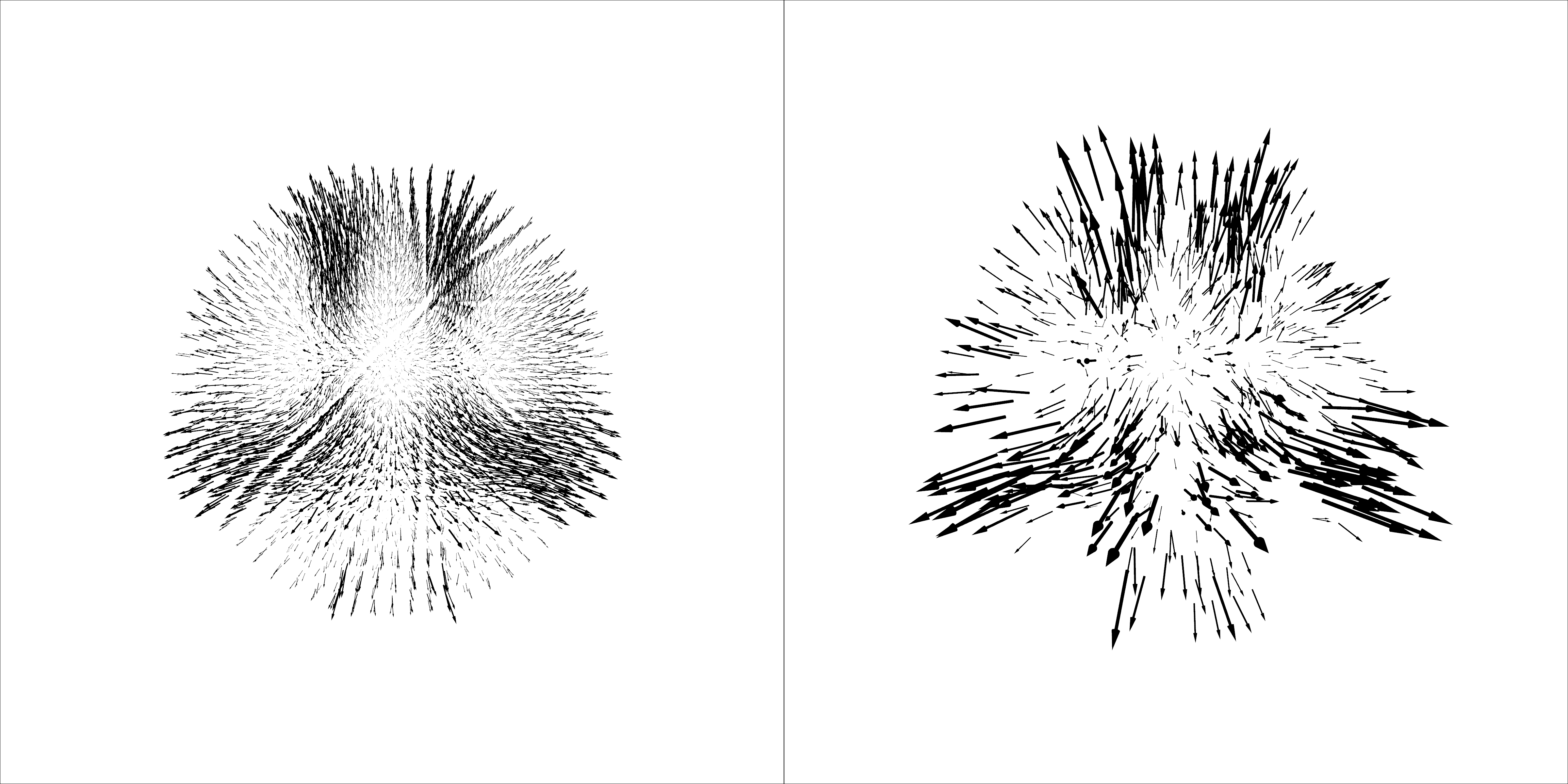}};    
    \begin{scope}[x=1cm, y=1cm, shift={(img.south west)}]
    \coordinate (O) at (1, 0); 

    \draw[arrow] (1, 0.0) -- (0.0, 0) node at (-0.2, 0) {\small $c_2$};
    \draw[arrow] (1, 0) -- (1, 1) node at (1, 1.2) {\small $c_3$};

    \node at (O) {$\boldsymbol{\otimes}$};  
    \node at ($(O)+(0.4,0)$) {\small $c_1$};

    \node at ($(3.0, 6.1)$) {All velocity observations};
    \node at ($(8.8, 6.1)$) {Every $10^{\text{th}}$ velocity observation (scaled)};
    \end{scope}

    \end{tikzpicture}
    \caption{Noise-polluted velocity observations, depicted as magnitude-scaled vector field. \textbf{Left:} Image of all noisy observation vectors. \textbf{Right:} Rescaled and thinned out representation for a better impression of the noise level.}
    \label{fig:observations}
\end{figure}
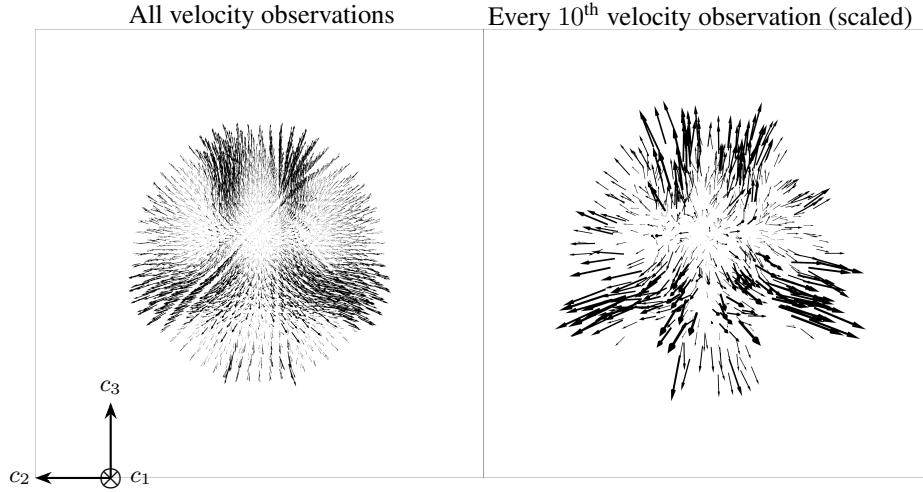

Our objective is to infer the spatially variable isotropic permeability field $k(\bc)$, in $K = k(\bc)\cdot I$ from Equation~\eqref{eqn: strong_form_darcy}, from potentially noisy flow (velocity) observations $\byobs$ as shown in Figure \ref{fig:observations}. As the permeability field is strictly positive, we furthermore introduce a positive parameterization of the following form:
\begin{equation}
    \label{eqn:pos_permeability}
    k(\bx, \bc) = \exp\left(\tx(\bx,\bc)\right),
\end{equation}
with $\tx(\bx,\bc)$ defined according to the FEM discretization specified in Equations \eqref{eqn: rf_fe}.
Although the governing equations are linear in the state variables, the parameterization of the permeability field $\bx$ enters the system nonlinearly due to the parameterization and the inverse operator $K^{-1}$.

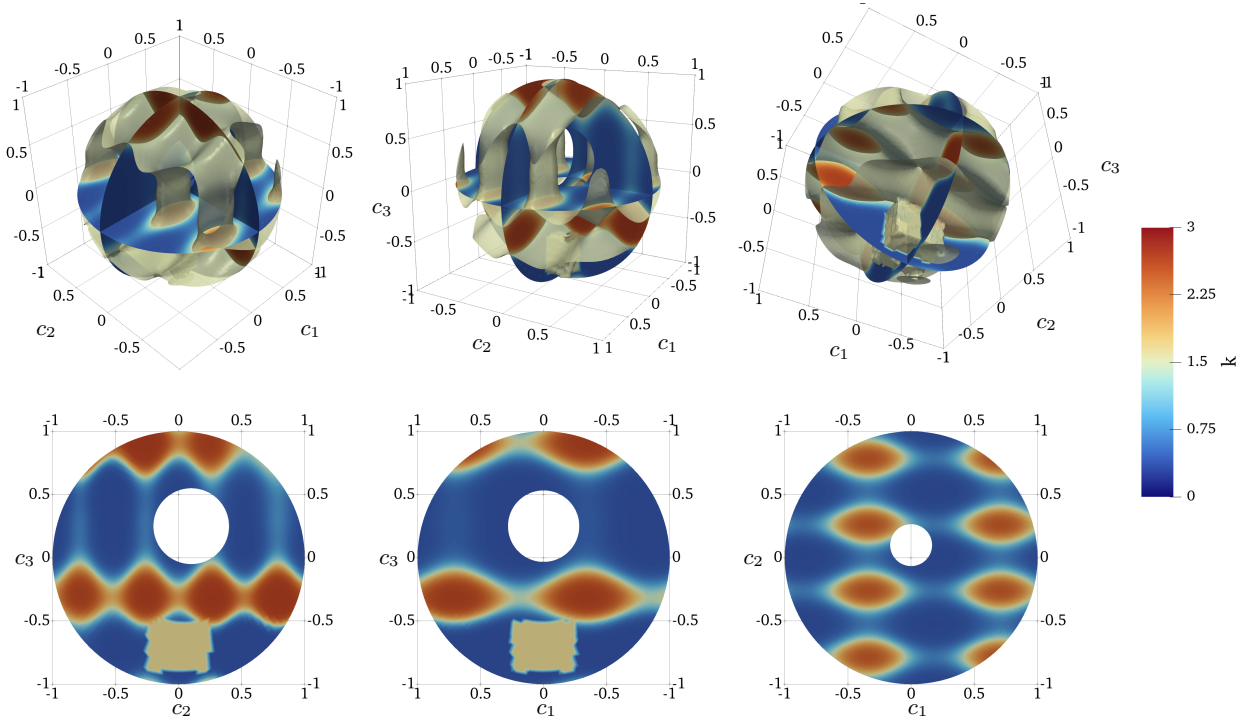
\begin{figure}[htbp]
    \centering
    \resizebox{\textwidth}{!}{%
    \begin{tikzpicture}
    \def\imgwidth{0.4\textwidth}
    \def\dx{\imgwidth - 1.3cm}   
    \def\dy{-0.32\textwidth}    

    \node[inner sep=0pt, anchor=north west] (img1) at (0,0)
      {\includegraphics[scale=0.1]{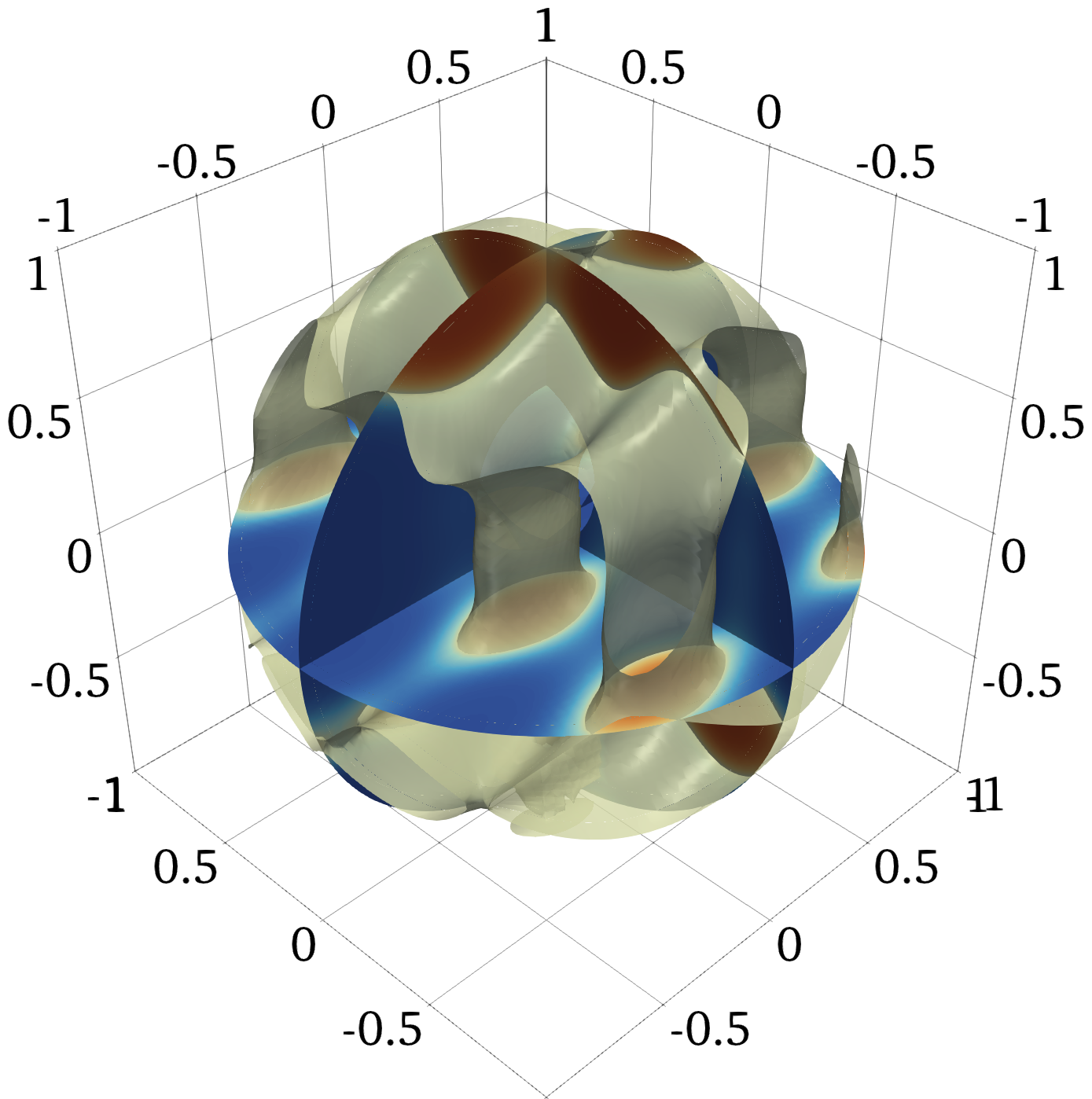}};
    \node at (4.3, -4.4) {\small{$c_1$}};
    \node at (0.6, -4.4) {\small{$c_2$}};
    
    \node[inner sep=0pt, anchor=north west] (img2) at (\dx,-0.4)
      {\includegraphics[scale=0.10]{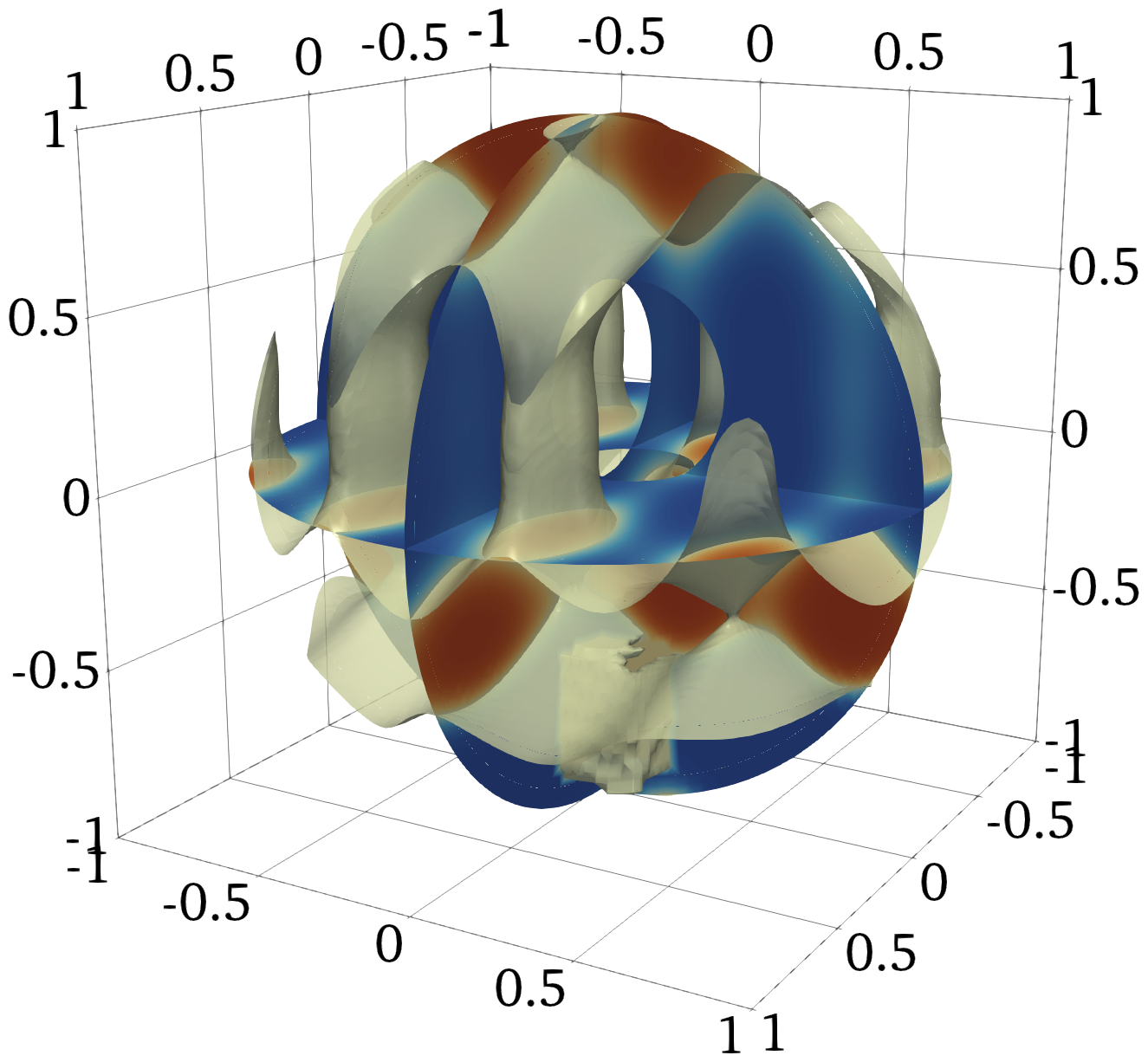}};
    \node at (5.3, -2.7) {\small{$c_3$}};          
    \node at (6.7, -4.6) {\small{$c_2$}};     
    \node at (9.3, -4.6) {\small{$c_1$}};
    
    \node[inner sep=0pt, anchor=north west] (img3) at ({1.75*\dx},0.2)
      {\includegraphics[scale=0.10]{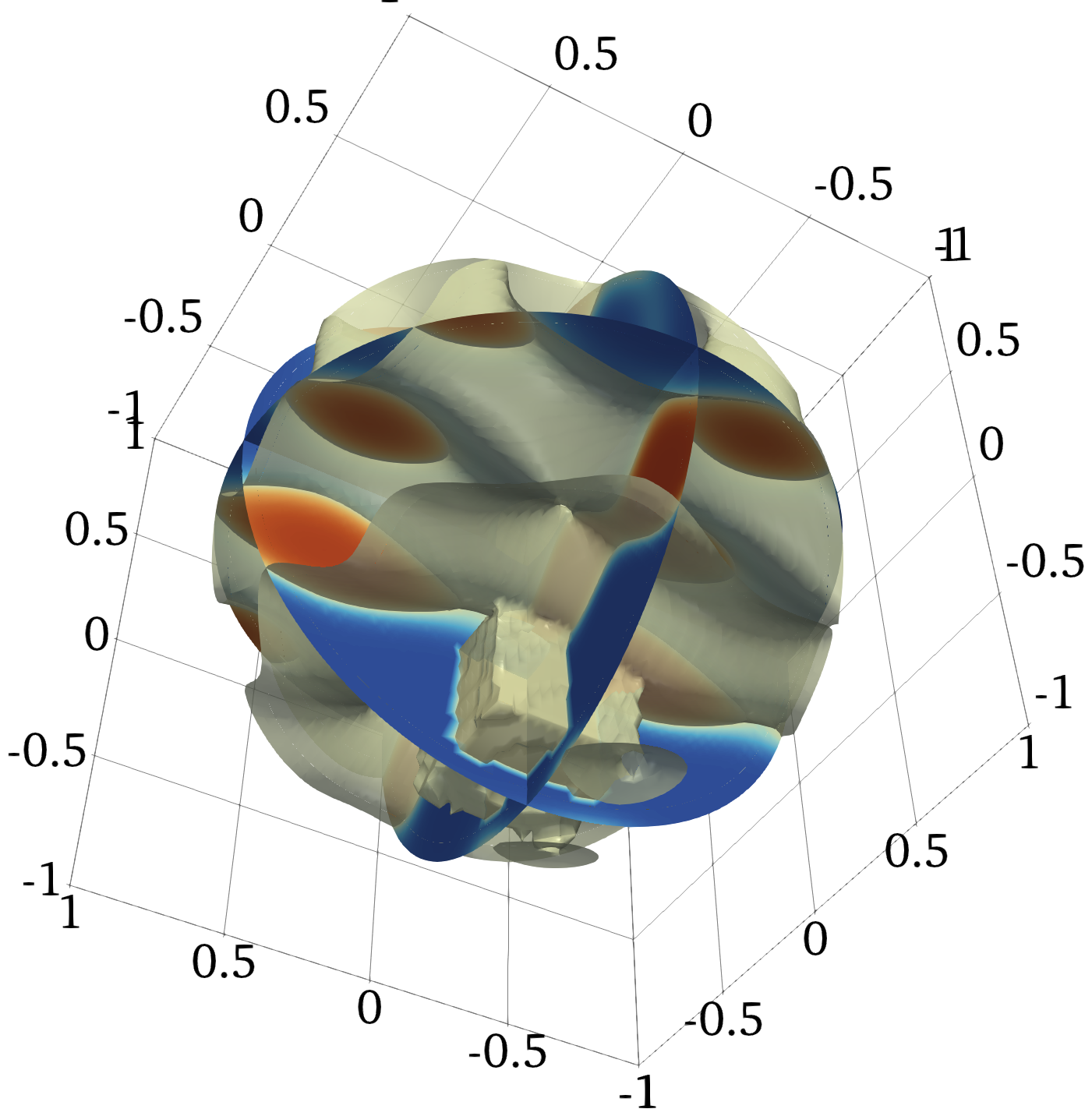}};
    \node at (15.5, -2.1) {\small{$c_3$}};          
    \node at (14.6, -4.3) {\small{$c_2$}};     
    \node at (11.7, -4.7) {\small{$c_1$}};     

    \node[inner sep=0pt, anchor=north west] (img4) at (0,\dy)
      {\includegraphics[scale=0.1]{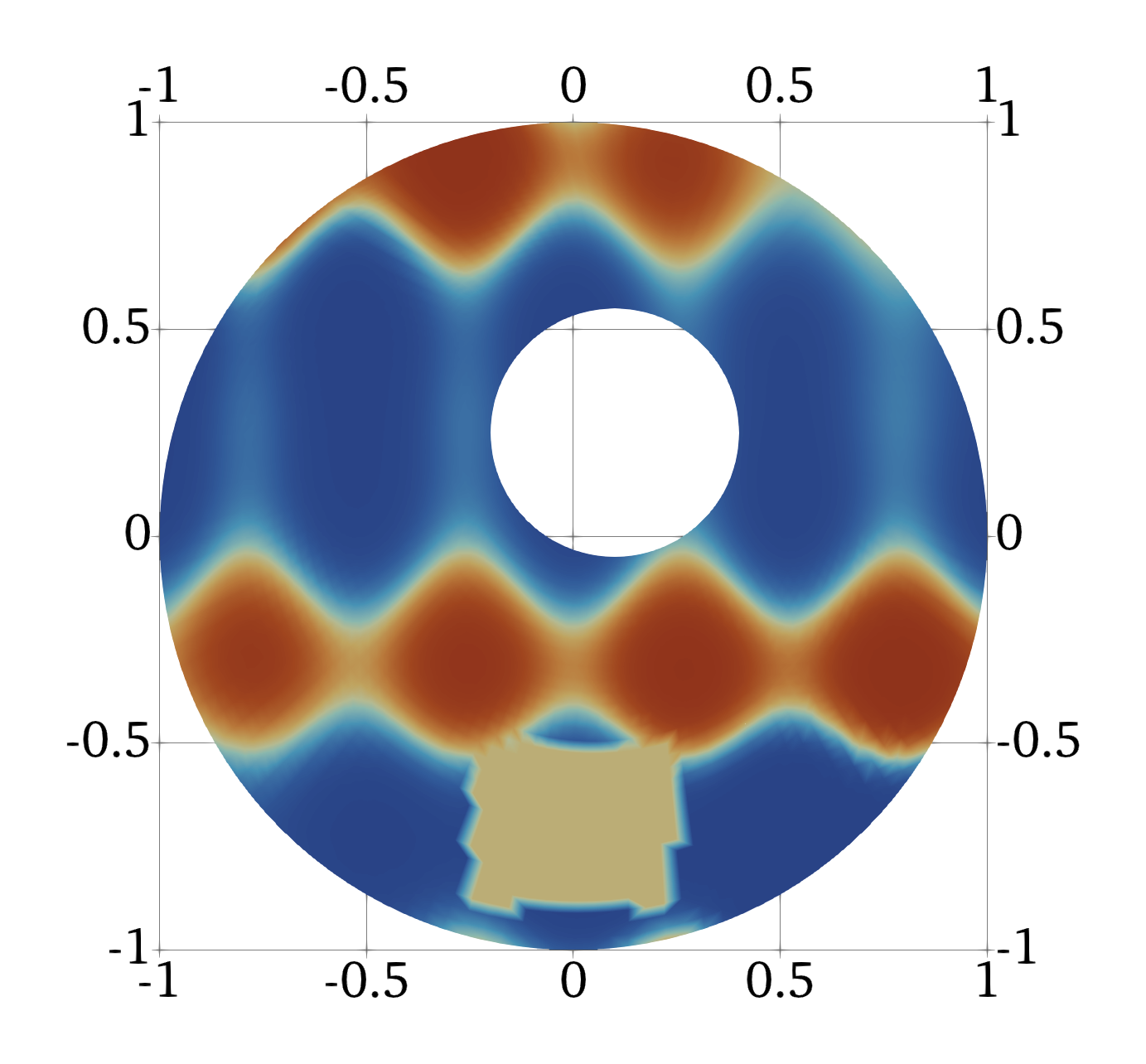}};
    \node at (0.3, -7.6) {\small{$c_3$}};     
    \node at (2.5, -9.7) {\small{$c_2$}};          
    
    \node[inner sep=0pt, anchor=north west] (img5) at (\dx-5.5,\dy)
      {\includegraphics[scale=0.1]{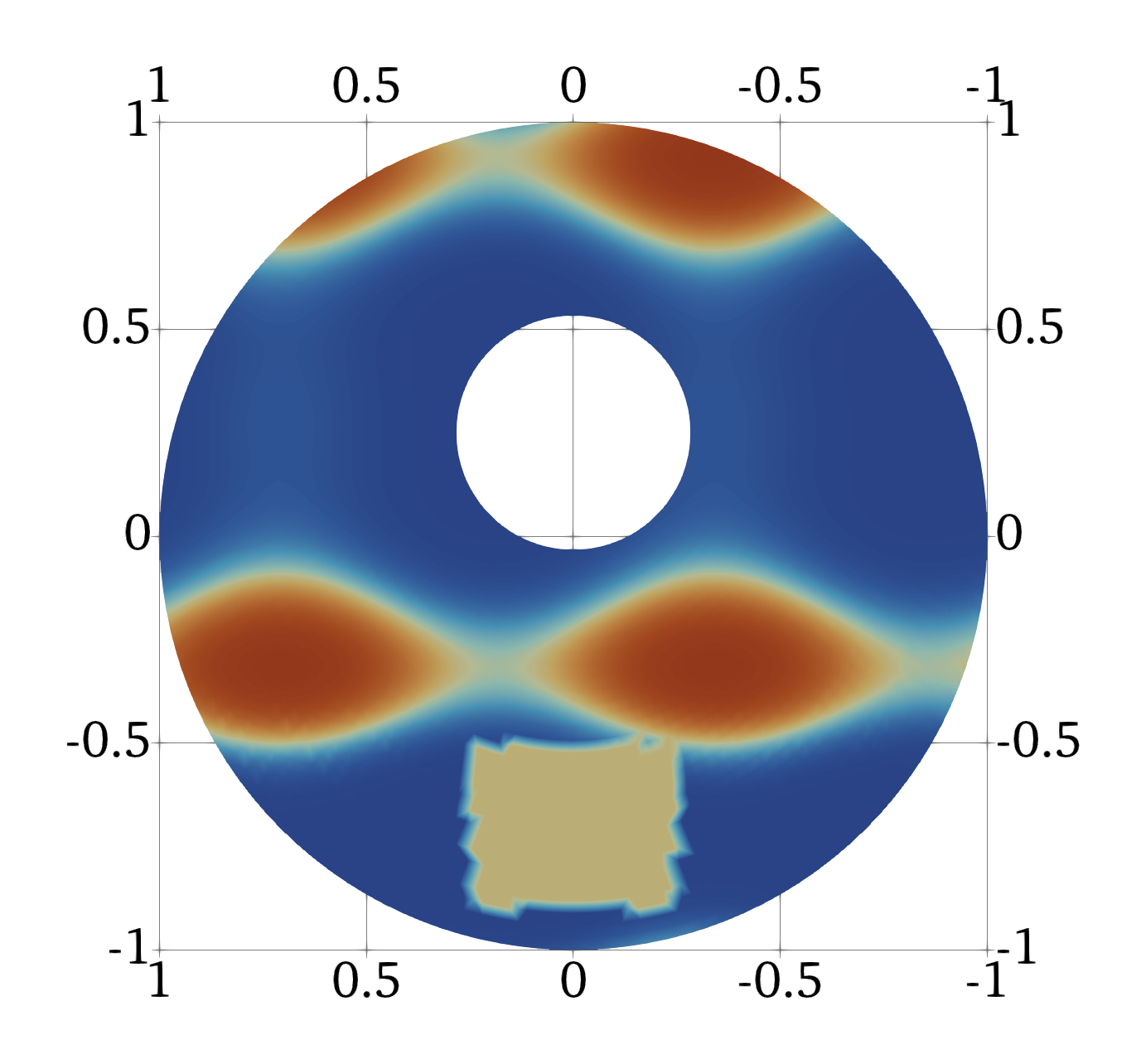}};
    \node at (5.4, -7.6) {\small{$c_3$}};     
    \node at (7.6, -9.7) {\small{$c_1$}};          
    
    \node[inner sep=0pt, anchor=north west] (img6) at ({1.75*\dx},\dy)
      {\includegraphics[scale=0.1]{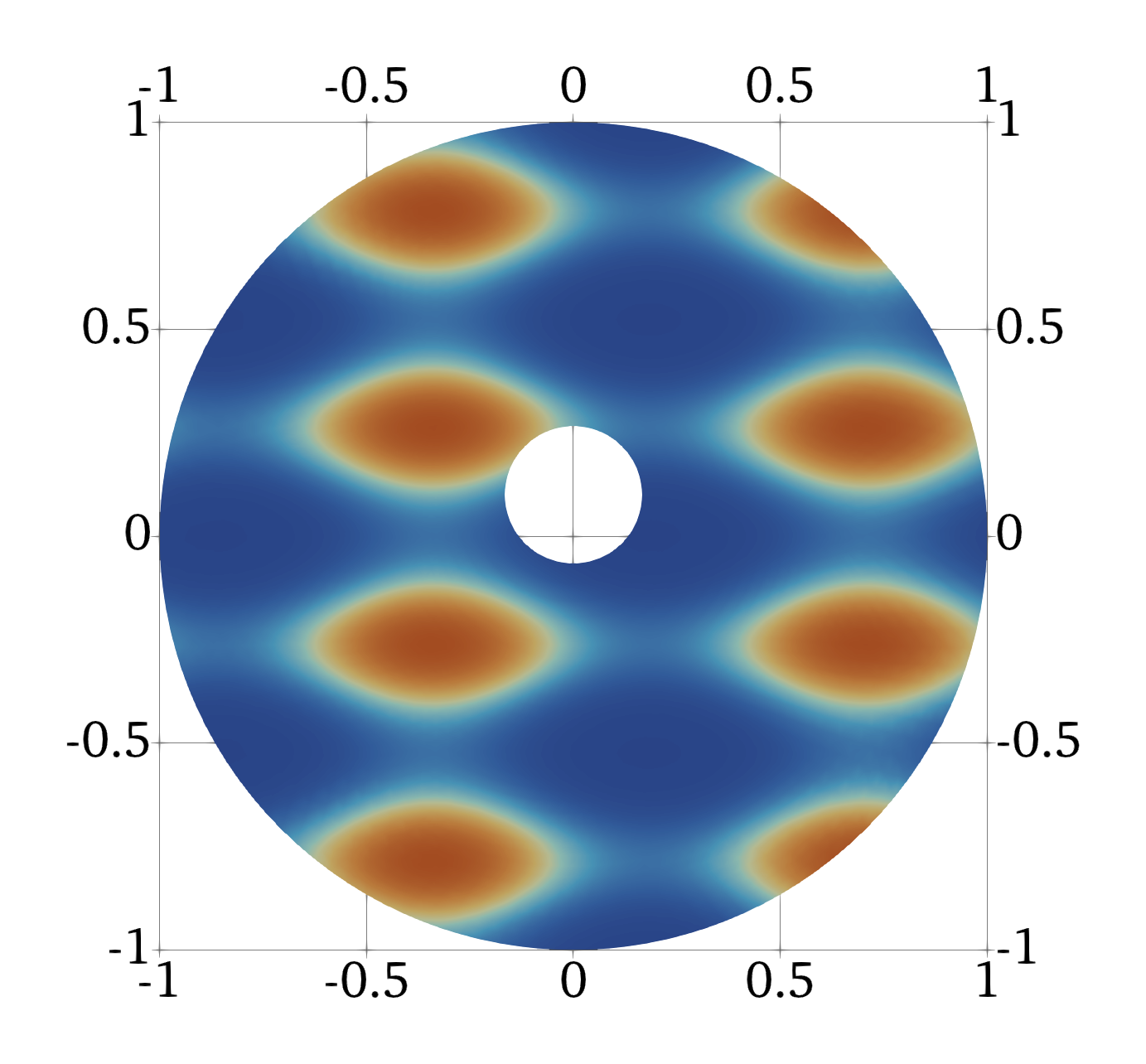}};
    \node at (10.5, -7.6) {\small{$c_2$}};     
    \node at (12.8, -9.7) {\small{$c_1$}};          

    \node[anchor=west, xshift=0.5cm] (colorbar) at ($(img3.north east)!0.5!(img6.south east)$) {
        \includegraphics[height=0.25\textwidth]{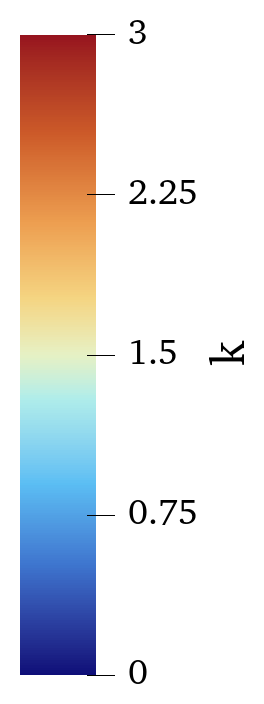}
    };
    \end{tikzpicture}}
\caption{Three-dimensional ground-truth permeability field $\kgt(\bc)$ (scalar field). \textbf{Top row}: Different 3D visualizations with slices through the origin at $(0,0,0)$ along with a slightly transparent iso-surface (beige color) for $\kgt=2.5$. \textbf{Bottom row}: Slices through the domain and origin $(0,0,0)$. From left to right: $c_2-c_3$-plane, $c_1-c_3$-plane, and $c_1-c_2$-plane. The inclusion of $\Omega^*$ is clearly visible as a beige cube.}
\label{fig:gt_permeability}
\end{figure}

\begin{figure}[htbp]
    \centering
    \begin{tikzpicture}
    \def\imgwidth{0.4\textwidth}
    \def\dx{\imgwidth - 1.3cm}   

    \node[inner sep=0pt, anchor=north west] (img1) at (0,0)
      {\includegraphics[scale=0.1]{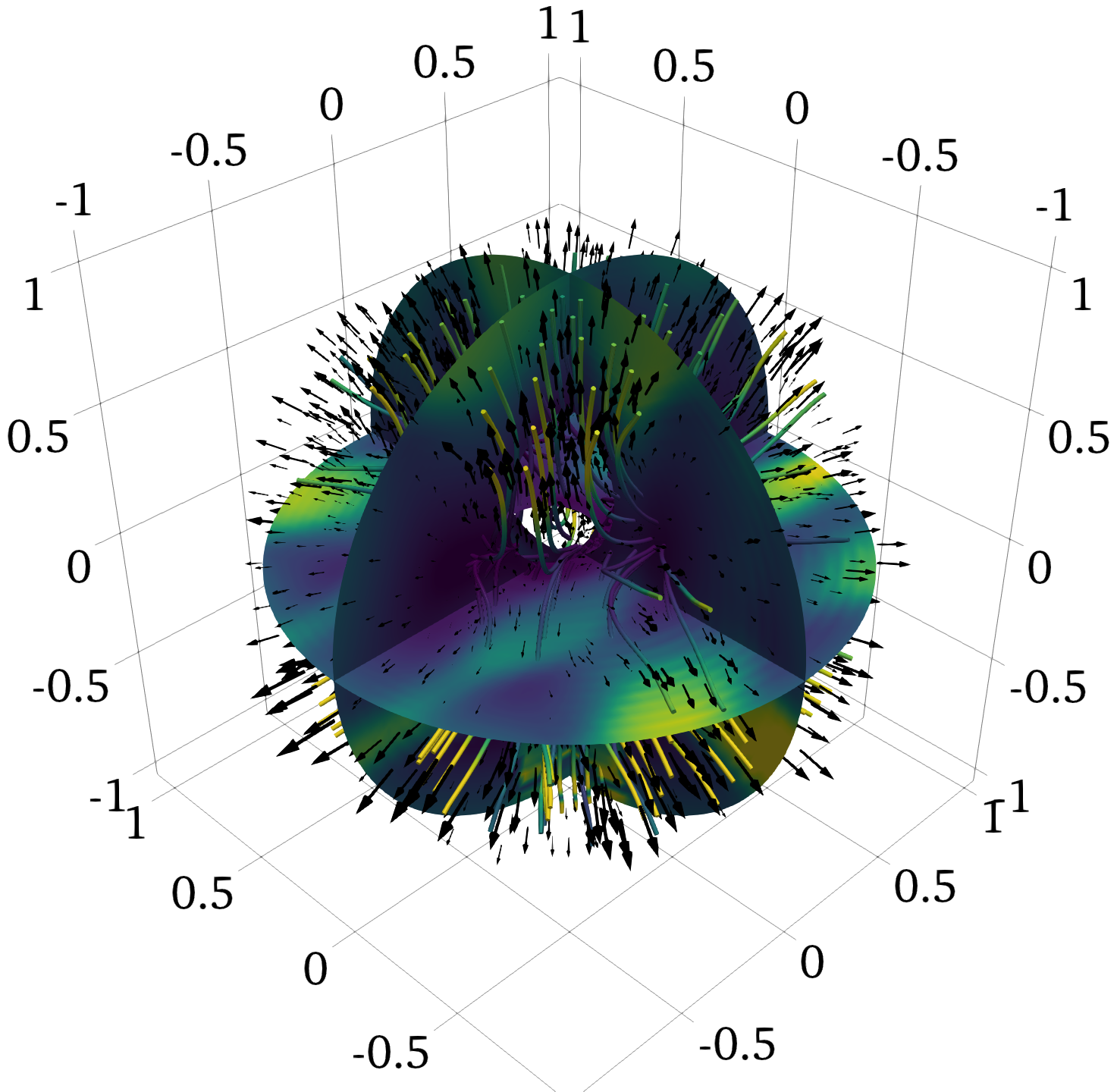}};
    \node at (4.3, -4.4) {\small{$c_1$}};
    \node at (0.6, -4.4) {\small{$c_2$}};
      
    \node[inner sep=0pt, anchor=north west] (img2) at (\dx,-0.4)
      {\includegraphics[scale=0.1]{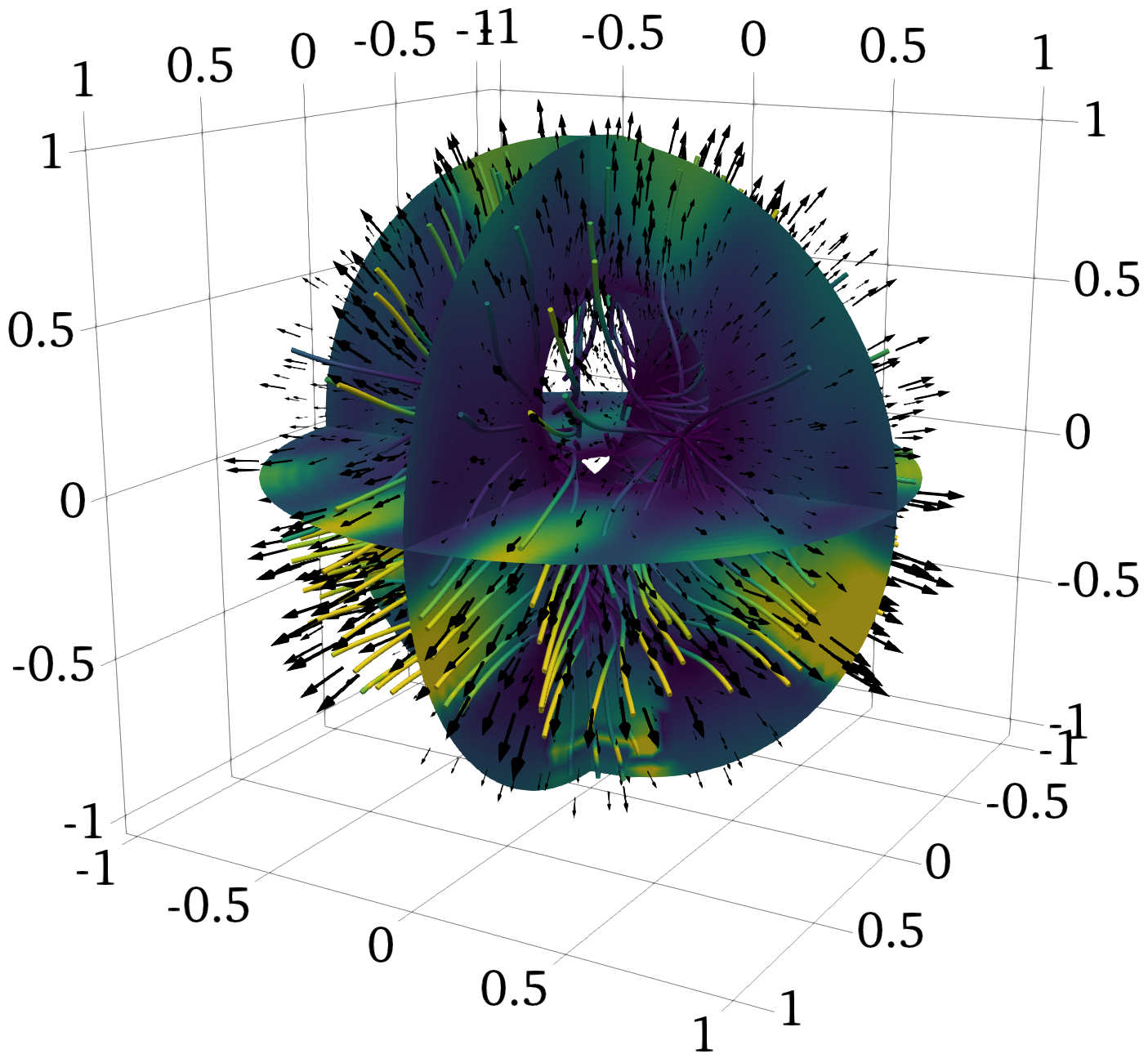}};
    \node at (5.3, -2.7) {\small{$c_3$}};          
    \node at (6.7, -4.7) {\small{$c_2$}};     
    \node at (9.4, -4.7) {\small{$c_1$}};     
      
    \node[inner sep=0pt, anchor=north west] (img3) at ({1.75*\dx},0.2)
      {\includegraphics[scale=0.1]{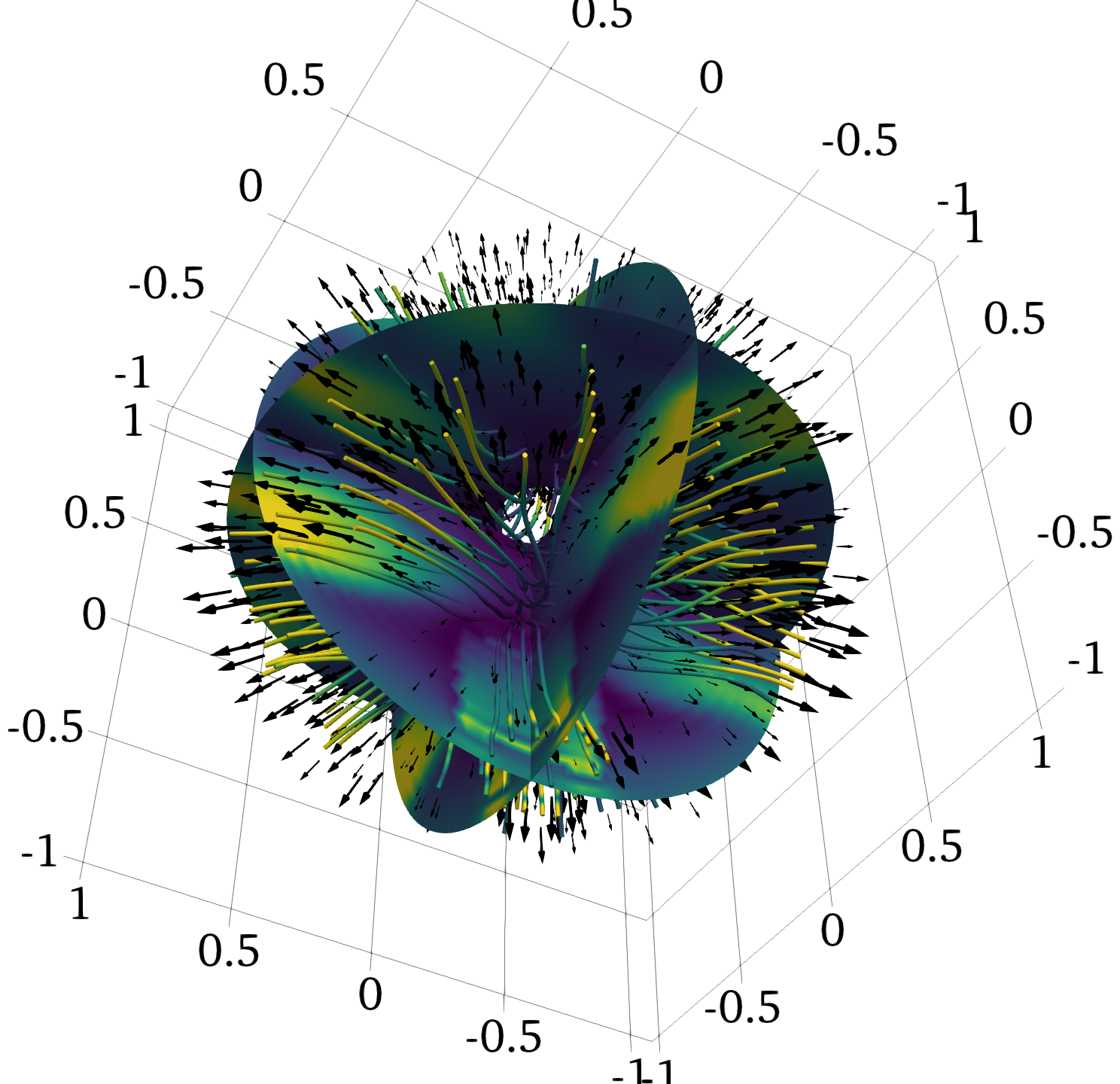}};
    \node at (15.6, -2.1) {\small{$c_3$}};          
    \node at (14.6, -4.3) {\small{$c_2$}};     
    \node at (11.7, -4.7) {\small{$c_1$}};          

    \node[inner sep=0pt, anchor=north west] (img4) at ({0.9*\dx},-5.2)
      {\includegraphics[scale=0.13]{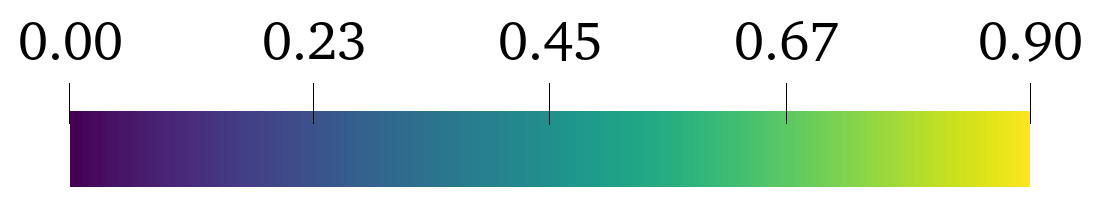}};
      \node at (10, -5.8) {\small{$||u||$}};

    \end{tikzpicture}
    \caption{3D representation of the resulting ground-truth velocity field $\boldsymbol{u}_{\mathrm{gt}}(\bc)$: We afterwards generate the observables $\byobs$ by evaluating the velocity at points defined by a coarser triangulation of the same geometry and corrupting the resulting values with \iid Gaussian noise with an SNR of 20 (not shown here).}
    \label{fig:vel_gt}
\end{figure}
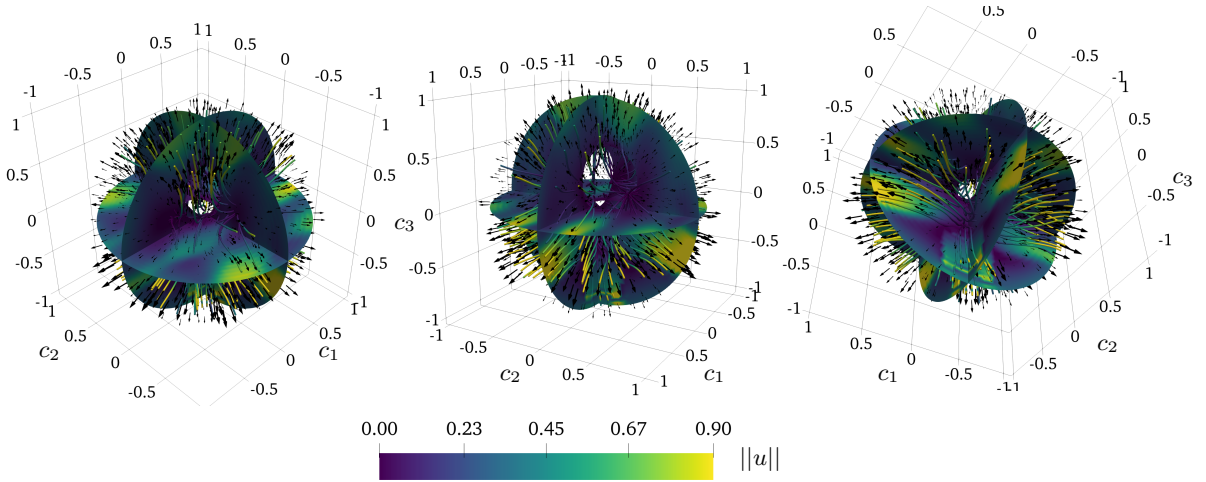

To assess the quality of the posterior, in a controlled environment, we generate the observables $\byobs$ artificially, by:
\begin{enumerate}
\item conducting a forward simulation with a ground-truth parameterization $\bxgt$ for the permeability field (depicted in Figure \ref{fig:gt_permeability}), which results in ground-truth velocity and pressure fields (the output fields or solution variables of the porous media flow problem); only the velocity field is depicted in Figure \ref{fig:vel_gt}.
\item evaluating the velocity field at point locations (based on a coarser triangulation of the same geometry), leading in total to 6930 observation points, as depicted in Figure \ref{fig:observations}.
\item corrupting the resulting ground-truth velocity field with additive, independent and identically distributed (\iid) Gaussian noise with a signal-to-noise ratio (SNR) of 20, which can be considered a \emph{medium} noise pollution. 

\end{enumerate}
The ground-truth permeability field $\kgt(\bxgt,\bc)=\exp\left(\txgt(\bc)\right)$ is analytically given by:
\begin{align}
\begin{split}
    \txgt(\bc)&=
    \begin{cases}
        \tanh\big[\big(\sin(6 c_1 +\phi)+\cos(12c_2)+\sin(6c_3)+\cos(9c_3+\phi) & \\
        +\sin(6c_3-\phi)\big)\cdot(c_2\cdot c_3 -1)\big], & \mathrm{for }\ \bc\notin \Omega^*\\
        0.5, &\mathrm{for }\ \bc \in \Omega^*
    \end{cases}\\
    \mathrm{with }\ \Omega^*&:=\{-0.25\le c_1 \le 0.25, -0.25 \le c_2 \le 0.25 ,-0.9\le c_3\le -0.5\}
\end{split}
\end{align}
The parameterization $\bxgt$ can then be easily computed by evaluating $\txgt(\bc)$ at the locations of the FEM DoFs. We chose the representation for $\txgt$ to include both smooth and non-smooth regions, and subsequently investigated the quality of their Bayesian reconstructions. We use quadratic shape functions to represent the permeability field with the same triangulation as for the primary variable.

\FloatBarrier
\subsection{Posterior permeability field reconstruction}
\label{sec:investigations_bia}
\begin{figure}[htbp]
    \centering
    \resizebox{\textwidth}{!}{%
    \begin{tikzpicture}
    \def\imgwidth{0.4\textwidth}
    \def\dx{\imgwidth - 1.3cm}   
    \def\dy{-0.32\textwidth}    

    \node[inner sep=0pt, anchor=north west] (img1) at (0,0)
      {\includegraphics[scale=0.1]{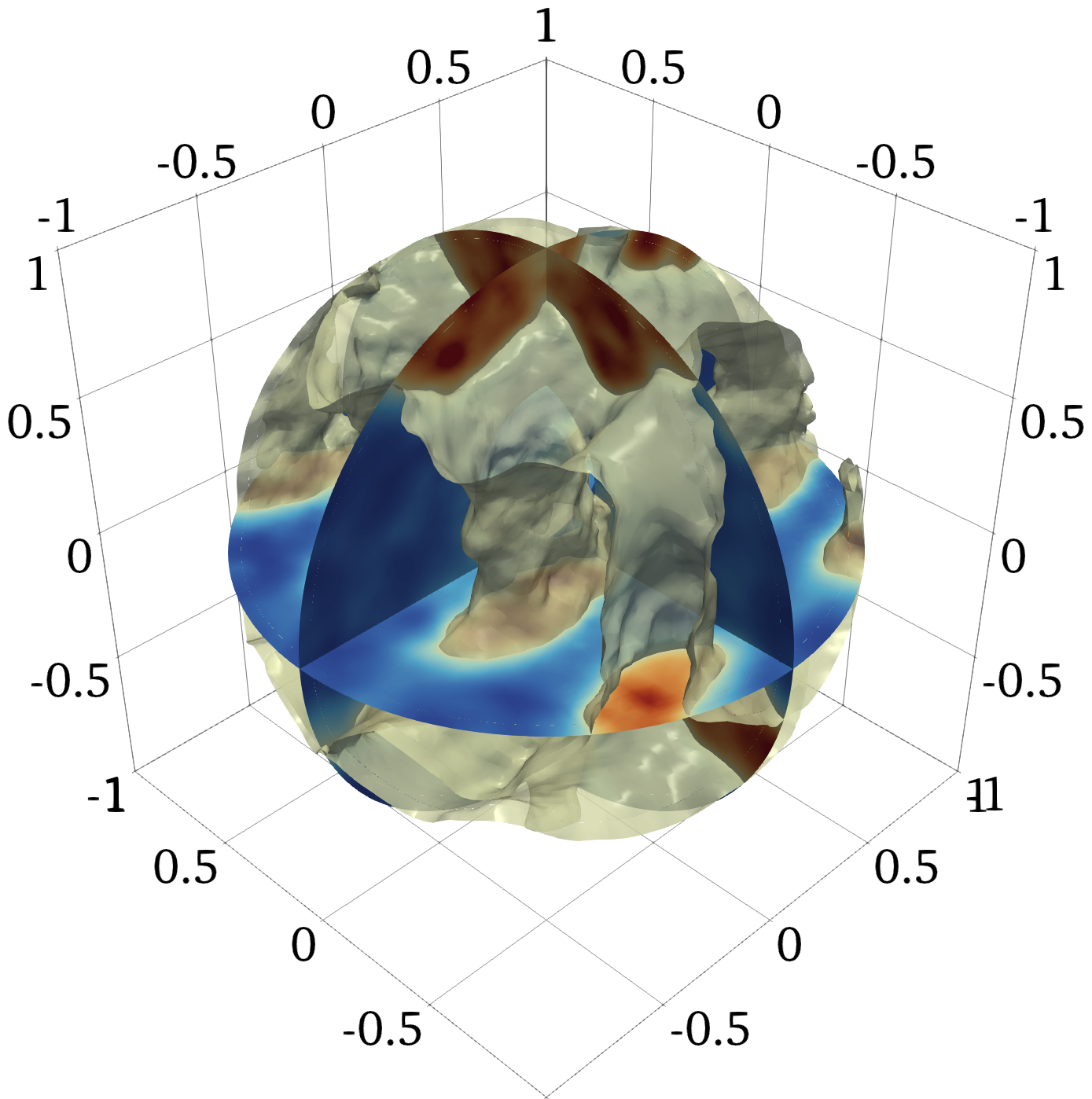}};
    \node at (4.3, -4.4) {\small{$c_1$}};
    \node at (0.6, -4.4) {\small{$c_2$}};

    \node[inner sep=0pt, anchor=north west] (img2) at (\dx,-0.4)
      {\includegraphics[scale=0.10]{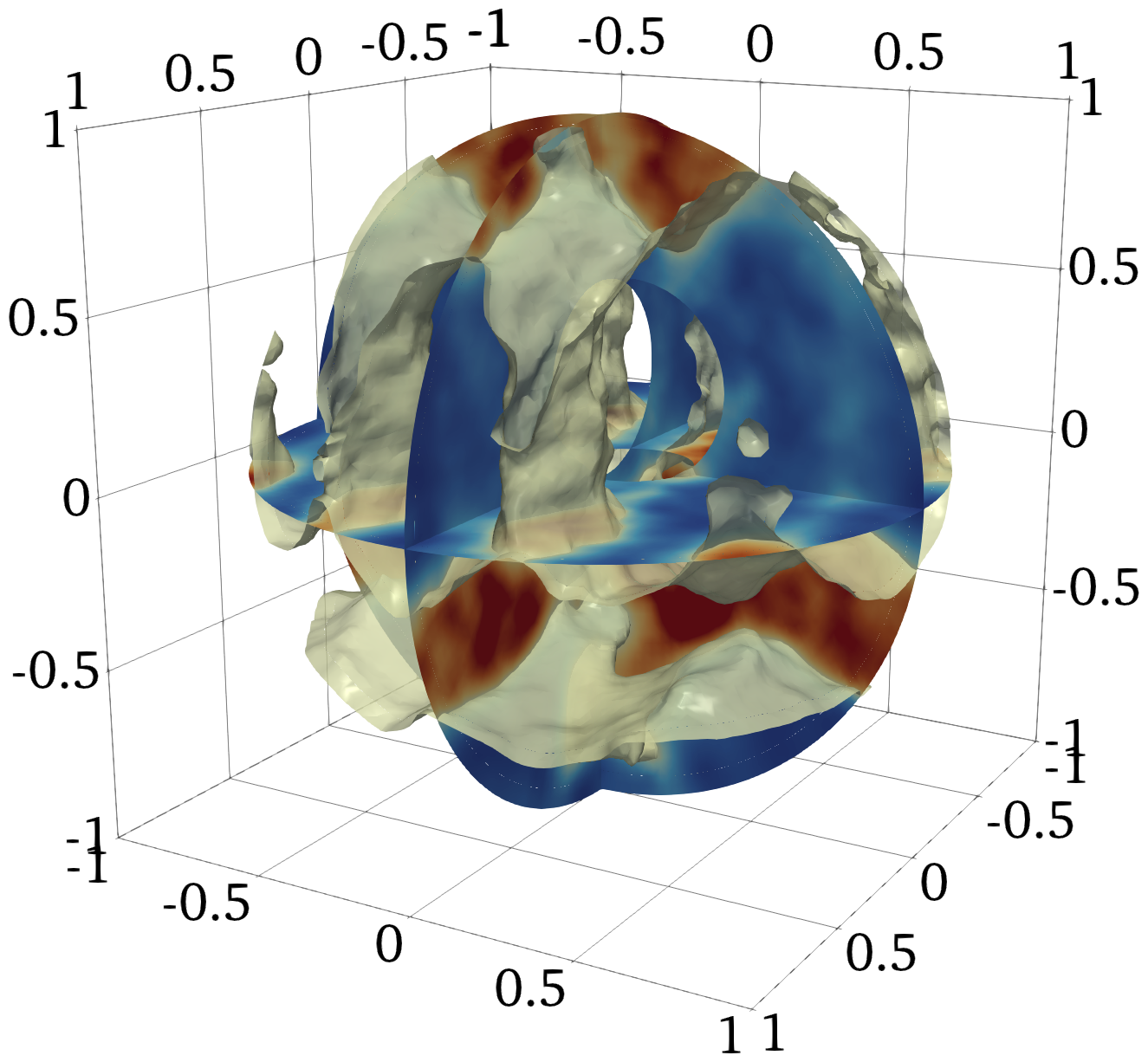}};
    \node at (5.3, -2.7) {\small{$c_3$}};
    \node at (6.7, -4.6) {\small{$c_2$}};
    \node at (9.3, -4.6) {\small{$c_1$}};

    \node[inner sep=0pt, anchor=north west] (img3) at ({1.75*\dx},0.2)
      {\includegraphics[scale=0.10]{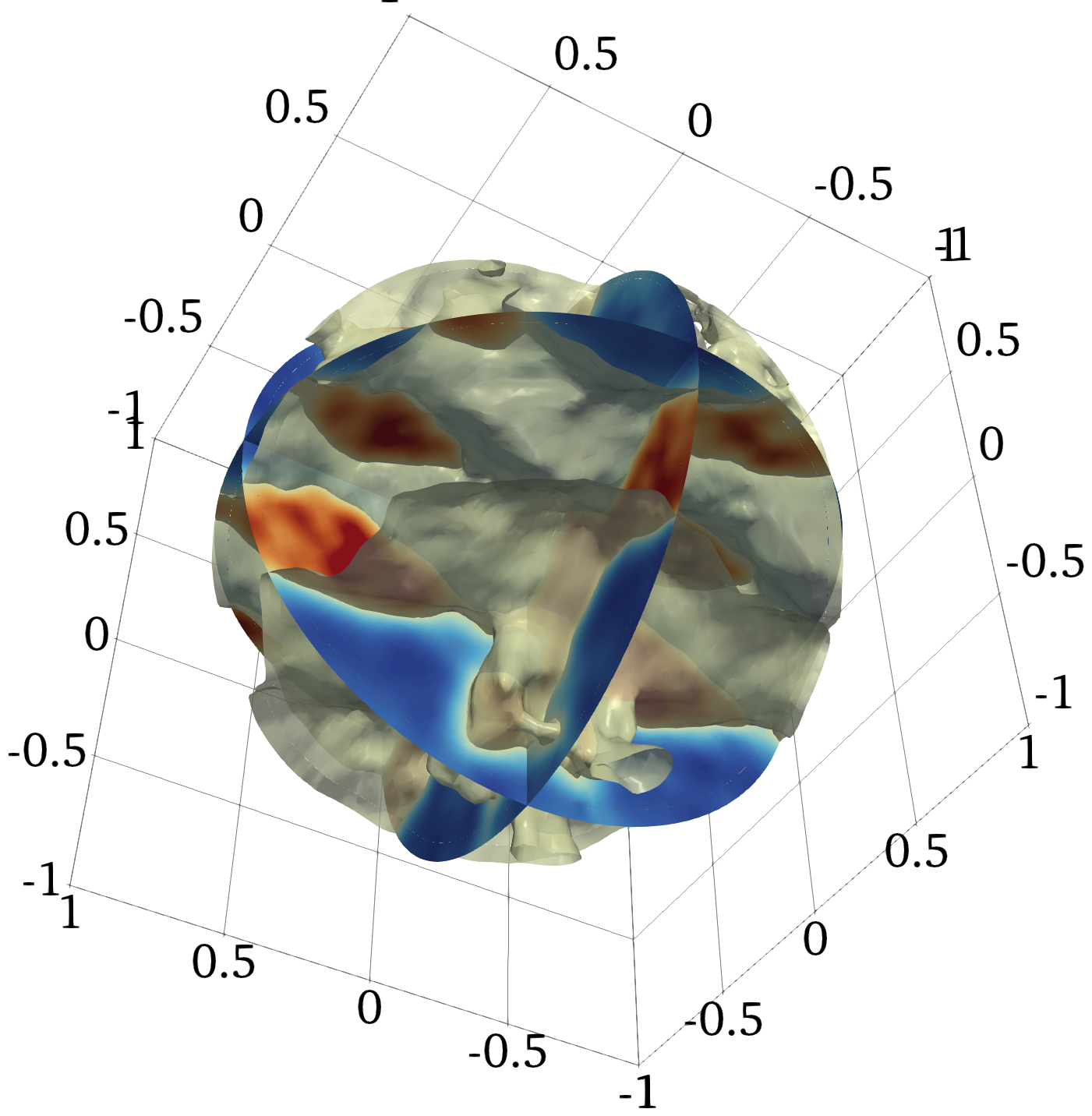}};
    \node at (15.5, -2.1) {\small{$c_3$}};
    \node at (14.6, -4.3) {\small{$c_2$}};
    \node at (11.7, -4.7) {\small{$c_1$}};

    \node[inner sep=0pt, anchor=north west] (img4) at (0,\dy)
      {\includegraphics[scale=0.1]{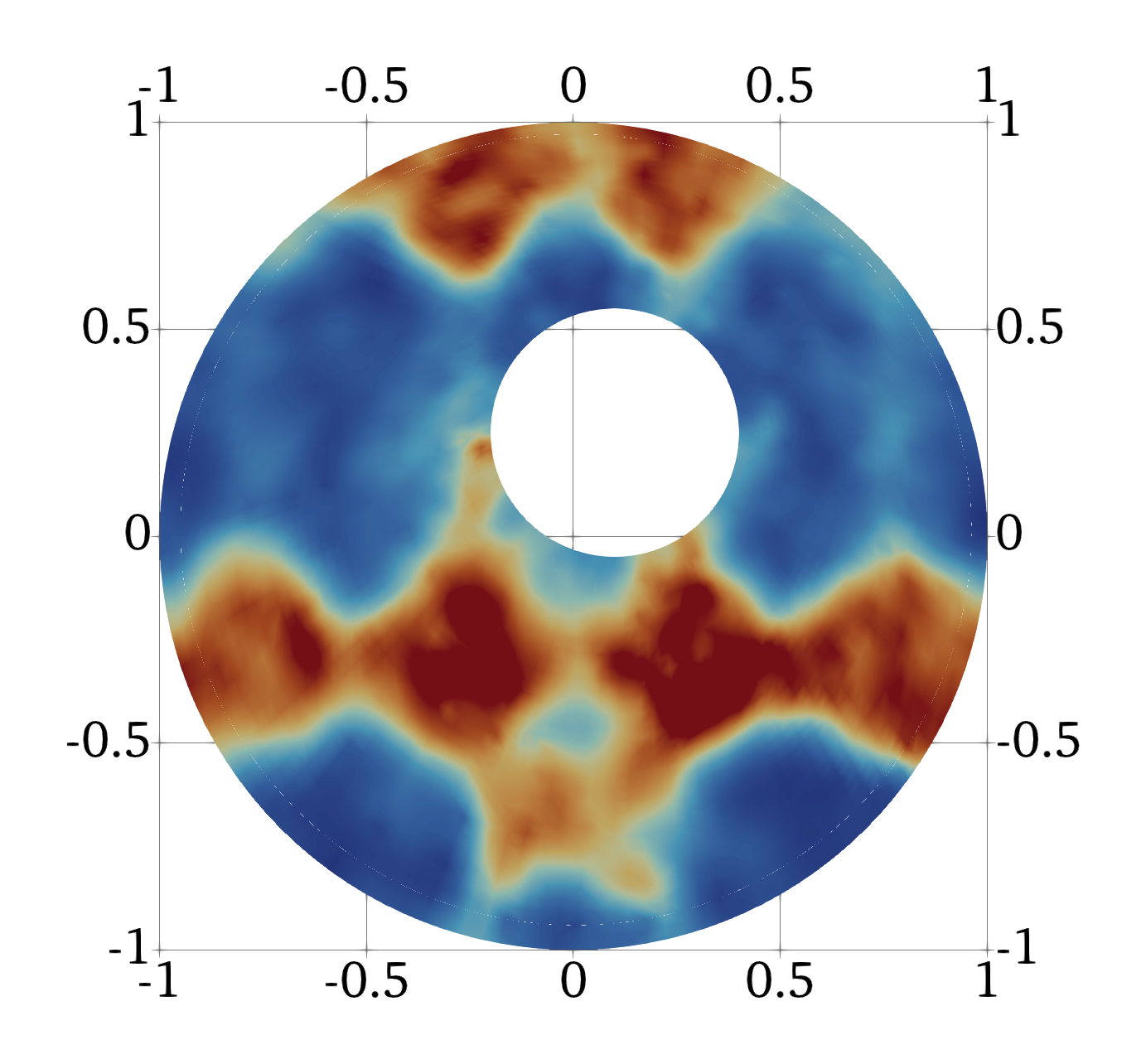}};
    \node at (0.3, -7.6) {\small{$c_3$}};
    \node at (2.5, -9.7) {\small{$c_2$}};

    \node[inner sep=0pt, anchor=north west] (img5) at (\dx-5.5,\dy)
      {\includegraphics[scale=0.1]{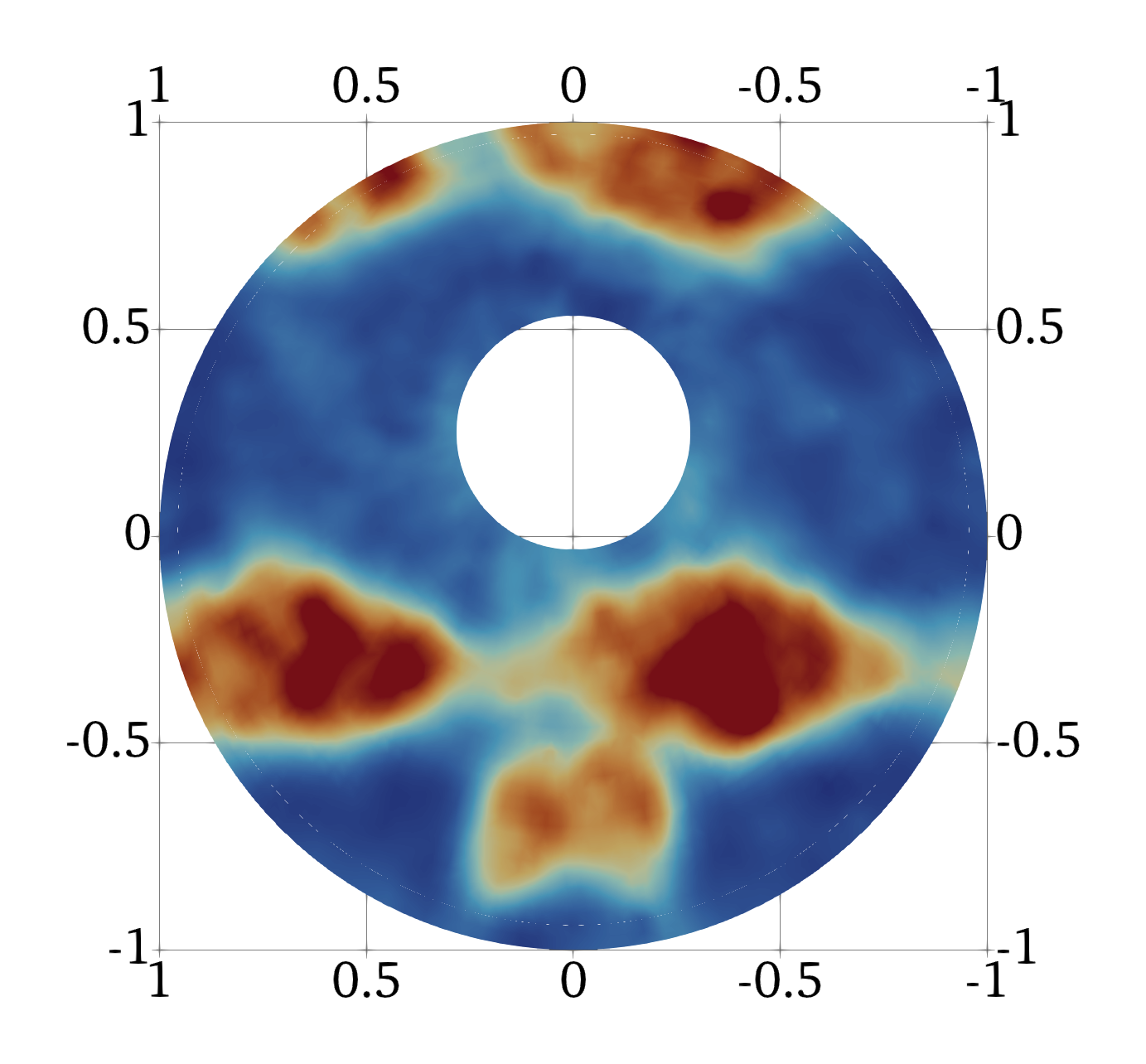}};
    \node at (5.4, -7.6) {\small{$c_3$}};
    \node at (7.6, -9.7) {\small{$c_1$}};

    \node[inner sep=0pt, anchor=north west] (img6) at ({1.75*\dx},\dy)
      {\includegraphics[scale=0.1]{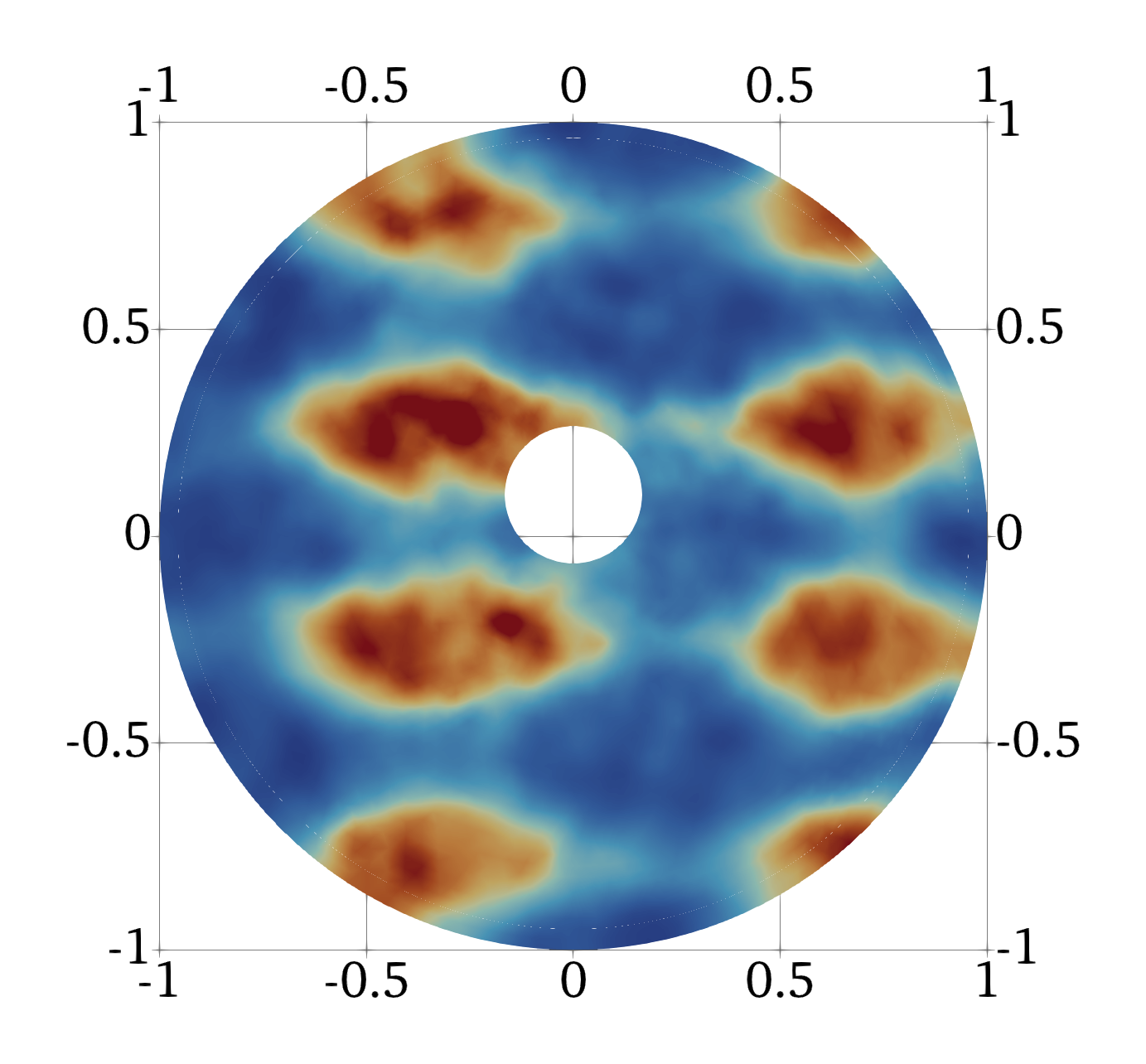}};
    \node at (10.5, -7.6) {\small{$c_2$}};
    \node at (12.8, -9.7) {\small{$c_1$}};

    \node[anchor=west, xshift=0.5cm] (colorbar) at ($(img3.north east)!0.5!(img6.south east)$) {
        \includegraphics[height=0.25\textwidth]{colorbar_standalone.png}
    };
    \end{tikzpicture}}
\caption{Posterior mean of the reconstructed permeability field $\Ex{q(\bx|\bphi^*)}{k(\bx,\bc)}$ (scalar field). \textbf{Top row}: 3D visualizations with slices through the origin at $(0,0,0)$. \textbf{Bottom row}: Slices through the domain and origin $(0,0,0)$. From left to right: $c_2$-$c_3$-plane, $c_1$-$c_3$-plane, and $c_1$-$c_2$-plane. Compare with the ground-truth field in Figure~\ref{fig:gt_permeability}.}
    \label{fig:posterior_mean_baseline}
\end{figure}

Given the observations $\byobs$, we formulate and solve the Bayesian inverse problem for the reconstruction of the permeability field $k(\bx,\bc)$, \ie we approximate the posterior density $p(\bx|\byobs)$ using the sparse stochastic variational inference framework developed in Section~\ref{sec:methodology} and summarized in Algorithm~\ref{alg:svi}. The complete problem setup is summarized in Table~\ref{tab:bia_setup} (and Table \ref{tab: model_fem_details}).
\begin{table}[htbp]
\centering
\caption{Bayesian inverse problem setup for the high-dimensional 3D porous media flow example.}
\label{tab:bia_setup}
\resizebox{\linewidth}{!}{%
\begin{tabular}{lll}
\toprule
\textbf{Settings} & \textbf{Value} & \textbf{Reference}\\
\midrule
\emph{SVI} & & Section \ref{sec: appendix_svi}\\
Number of samples per batch & $\nbatch=4$\\
Max.\ number of forward/adjoint solver calls (each) & 2000 \\
Variational distribution & Precision-param.\ sparse Gaussian & \\
Sparsity pattern of $\LQ$ & FE triangulation sparsity (Laplacian structure) & \\
Initialization of $\LQ$ & Incomplete Cholesky of prior precision & \\
Natural gradient ($\bphimu$ and $\bphiL$) & yes & \\
Stochastic optimizer & Adam \\
Learning rate & $5\cdot 10^{-2}$\\
\midrule
\emph{Prior model $p(\bx)$} & & Section \ref{sec: Prior}\\
Form & $p(\bx)=\int p(\bx|\delta)\cdot p(\delta)\dd \delta$ & \\
Type & GMRF: $p(\bx|\delta)=\mathcal{N}\left(\bx|\bmu_{\bx},Q^{-1}(\delta)\right)$ & \\
Prior mean & $\bmu_{\bx}=0.1$ \\
Hyperprior for precision $\delta$ & Gamma: $p(\delta)=\Gamma(\delta|a_0,b_0)$, $a_0=b_0=10^{-9}$ & \\
SPDE parameter & $\kappa^2=10^{-4}$ & \\
\midrule
\emph{Likelihood model $p(\byobs|\bx)$} & & Section \ref{sec: data_log_lik}\\
Form & $p(\byobs|\bx)=\int p(\byobs|\bx,\tau)\cdot p(\tau)\dd\tau$& \\
Type & Diag.\ Gaussian: $p(\byobs|\bx,\tau)=\mathcal{N}\left(\byobs|\by(\bx),\tau^{-1}\cdot I\right)$& \\
Underlying assumption & Gaussian \iid additive noise & \\
Hyperprior for precision $\tau$ & Gamma: $p(\tau)=\Gamma(\tau|a_0,b_0)$, $a_0=b_0=10^{-9}$ & \\
\midrule
\emph{Forward model $\Mod(\bx)$} && Section \ref{sec:porous_media_flow}\\
Parameterization & $k(\bx,\bc)=\exp\left(\tx(\bx,\bc)\right)$ & \\
FE basis (random field) & quadratic Lagrangian polynomials & \\
Stochastic dimension & $\dim(\bx)=405\,570$ & \\
\bottomrule
\end{tabular}}
\end{table}
The resulting posterior mean field is shown in Figure~\ref{fig:posterior_mean_baseline} and the corresponding two posterior standard deviations in Figure~\ref{fig:posterior_2std_baseline}. Individual posterior samples are provided in Appendix~\ref{sec:posterior_samples_3d}.
Comparing the posterior mean (Figure~\ref{fig:posterior_mean_baseline}) with the ground-truth field (Figure~\ref{fig:gt_permeability}), the reconstruction recovers all major spatial features of the permeability with high fidelity despite the noisy observations and the $405\,570$-dimensional parameter space. The alternating high- and low-permeability bands in the $c_2$-$c_3$ slice, the localized high-permeability regions in the $c_1$-$c_2$ plane, and the large-scale structure around the eccentric enclosure are clearly resolved. The 3D views confirm that the isosurfaces' 3D topology closely matches the ground truth. Minor deviations in fine-scale features are visible but expected, as they fall within the posterior uncertainty. A notable example is the sharp rectangular boundary of the eccentric enclosure, which the ground truth shows as a permeability discontinuity. The GMRF/SPDE prior is inherently smoothing, and the posterior mean accordingly rounds off this discontinuity, yielding a diffused transition zone around the enclosure. This is a direct consequence of the prior's regularity assumptions and illustrates the trade-off between smoothness and the ability to resolve sharp interfaces.

We emphasize that comparing the posterior mean point-wise against the ground truth does not fully characterize the quality of the Bayesian solution. The relevant criterion is whether the posterior distribution encapsulates the ground truth within its credible region. The posterior samples in Appendix~\ref{sec:posterior_samples_3d} confirm that the ground-truth field is consistent with the variability captured by the approximate posterior $q(\bx|\bphi^*)$.

The two posterior standard deviations in Figure~\ref{fig:posterior_2std_baseline} reveal a spatially structured uncertainty field. The 2D slices exhibit a characteristic patchy pattern that reflects the epistemic uncertainty between observation locations: regions between data points show elevated uncertainty, whereas regions near observations are more tightly constrained. The 3D views corroborate this, with elevated uncertainty visible on the outer surface and in regions far from the observation grid. Notably, uncertainty is higher in regions of large absolute permeability values, which is consistent with the exponential parameterization $k = \exp(\tx)$ amplifying posterior variance in high-permeability zones. A quantitative ablation of the algorithmic components and a comparison with alternative inference methods are presented in Section~\ref{sec:convergence_analysis}.

\begin{figure}[htbp]
    \centering
    \resizebox{\textwidth}{!}{%
    \begin{tikzpicture}
    \def\imgwidth{0.4\textwidth}
    \def\dx{\imgwidth - 1.3cm}   
    \def\dy{-0.32\textwidth}    

    \node[inner sep=0pt, anchor=north west] (img1) at (0,0)
      {\includegraphics[scale=0.1]{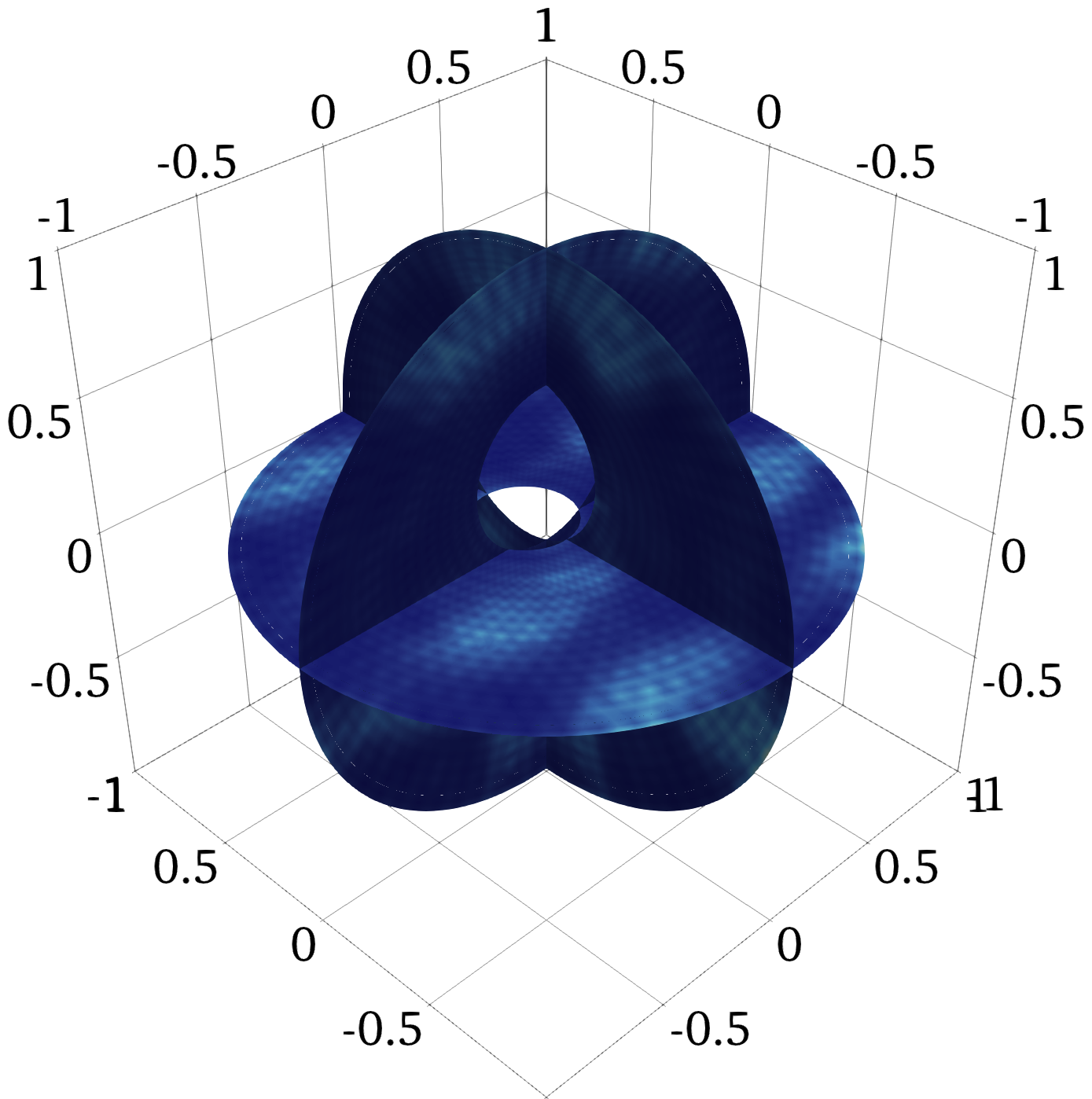}};
    \node at (4.3, -4.4) {\small{$c_1$}};
    \node at (0.6, -4.4) {\small{$c_2$}};

    \node[inner sep=0pt, anchor=north west] (img2) at (\dx,-0.4)
      {\includegraphics[scale=0.10]{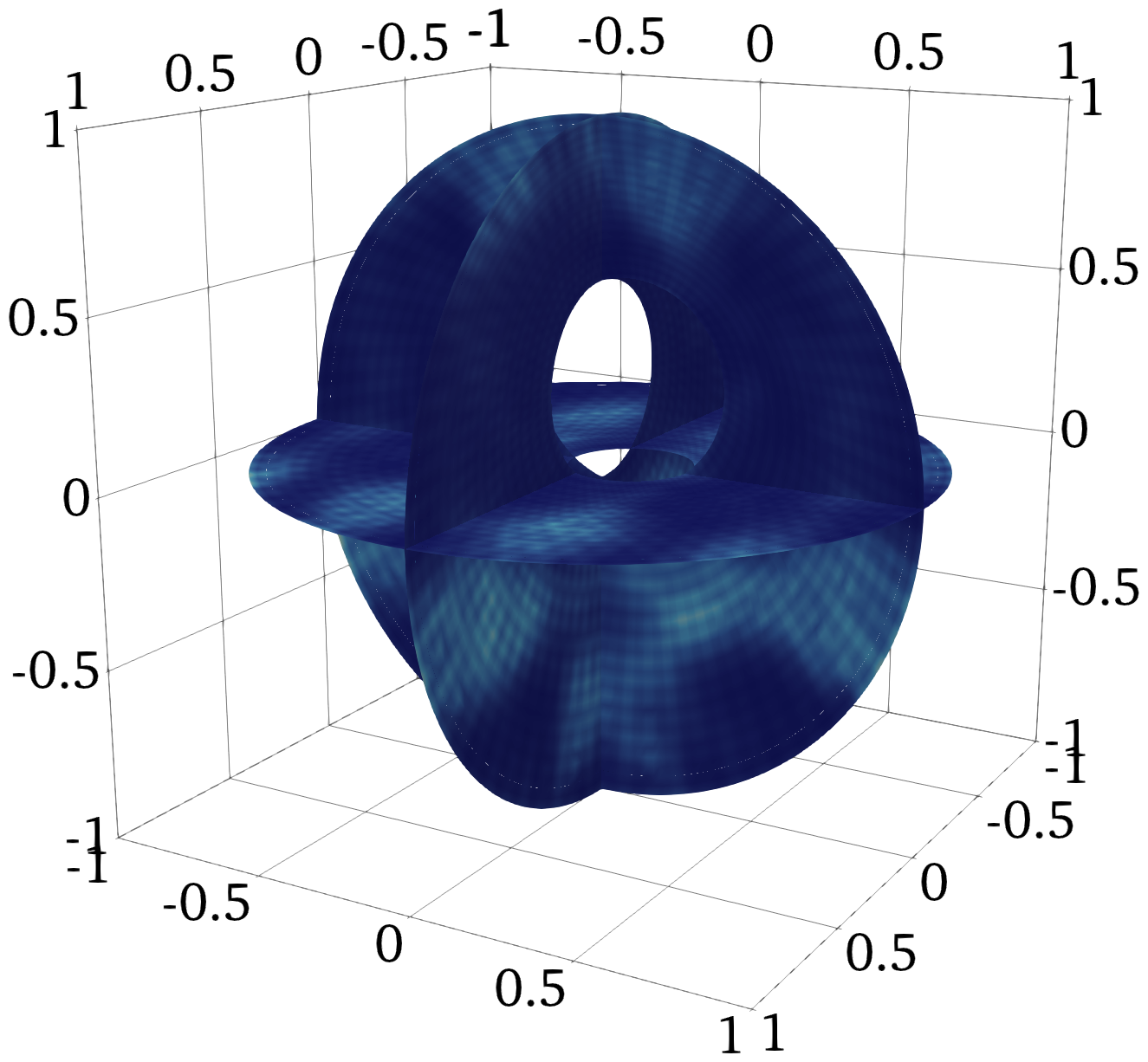}};
    \node at (5.3, -2.7) {\small{$c_3$}};
    \node at (6.7, -4.6) {\small{$c_2$}};
    \node at (9.3, -4.6) {\small{$c_1$}};

    \node[inner sep=0pt, anchor=north west] (img3) at ({1.75*\dx},0.2)
      {\includegraphics[scale=0.10]{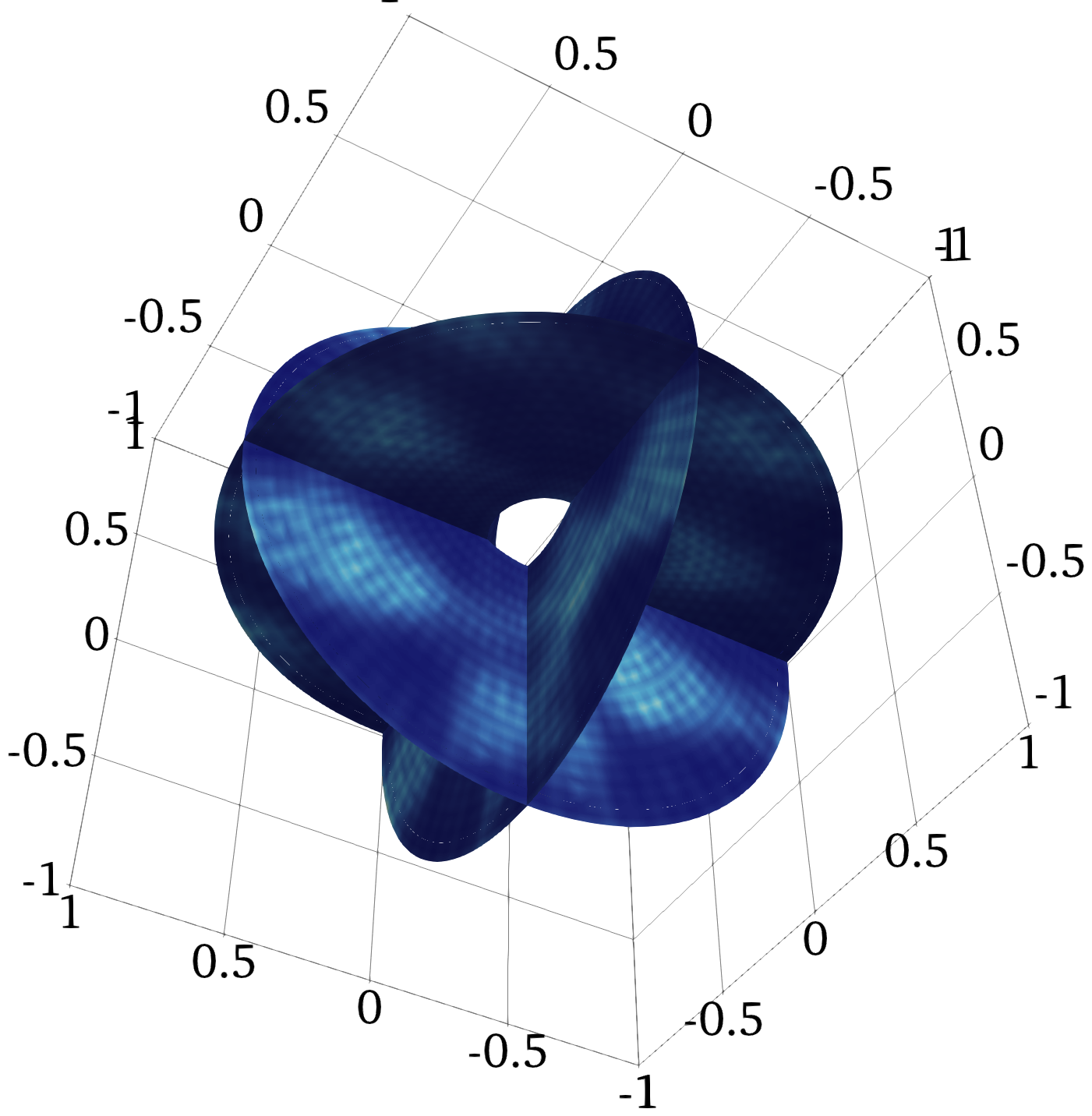}};
    \node at (15.5, -2.1) {\small{$c_3$}};
    \node at (14.6, -4.3) {\small{$c_2$}};
    \node at (11.7, -4.7) {\small{$c_1$}};

    \node[inner sep=0pt, anchor=north west] (img4) at (0,\dy)
      {\includegraphics[scale=0.1]{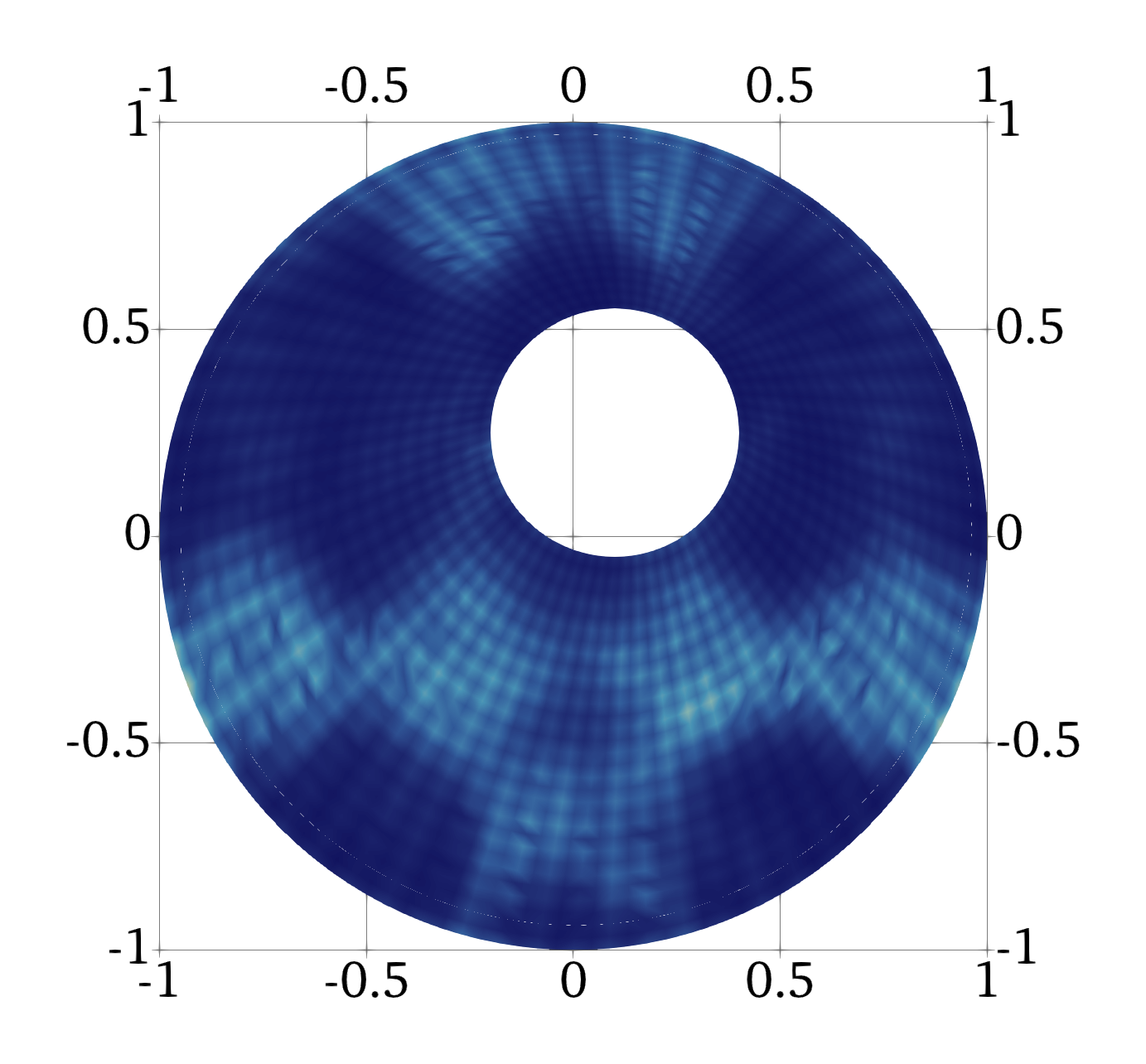}};
    \node at (0.3, -7.6) {\small{$c_3$}};
    \node at (2.5, -9.7) {\small{$c_2$}};

    \node[inner sep=0pt, anchor=north west] (img5) at (\dx-5.5,\dy)
      {\includegraphics[scale=0.1]{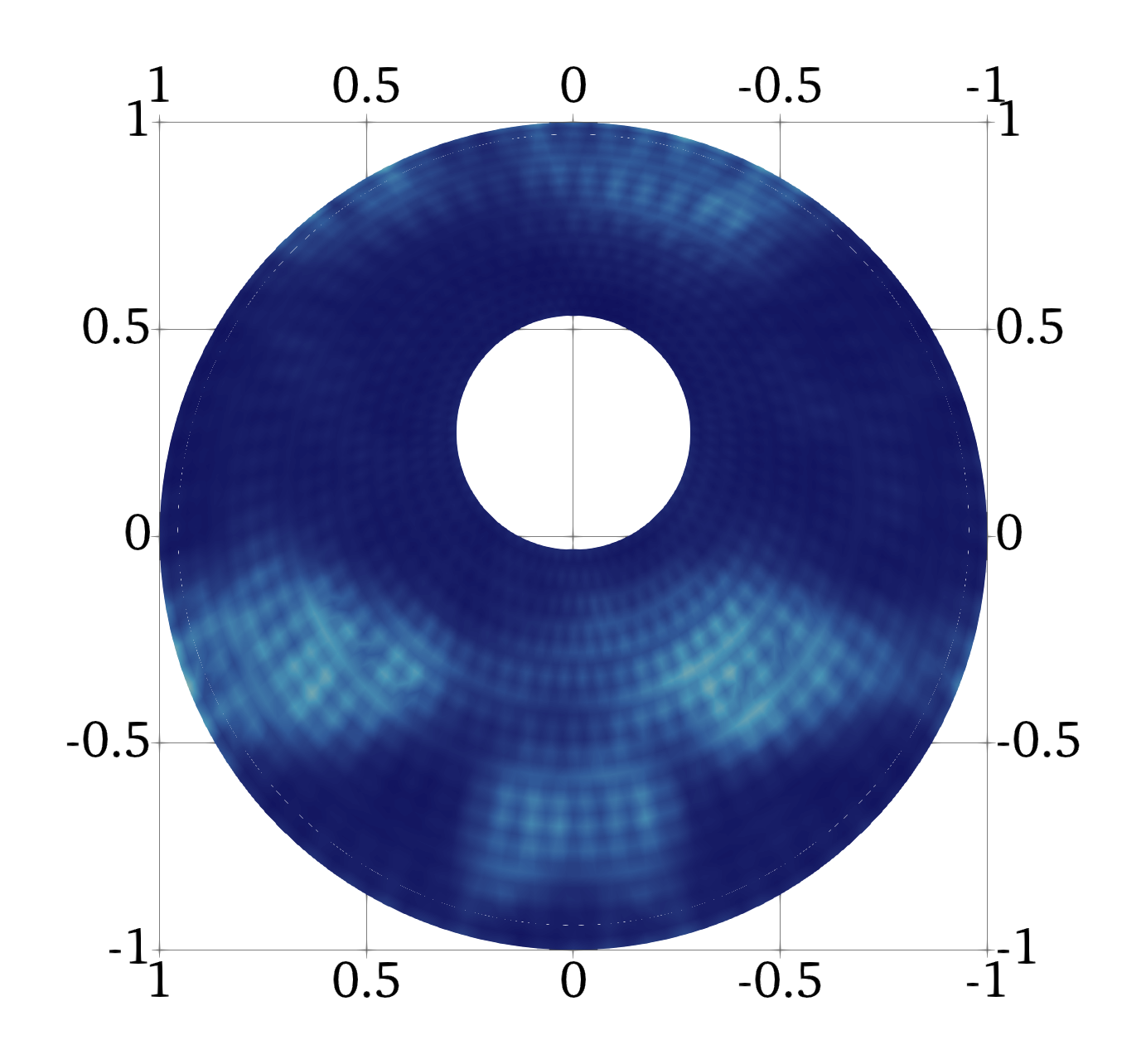}};
    \node at (5.4, -7.6) {\small{$c_3$}};
    \node at (7.6, -9.7) {\small{$c_1$}};

    \node[inner sep=0pt, anchor=north west] (img6) at ({1.75*\dx},\dy)
      {\includegraphics[scale=0.1]{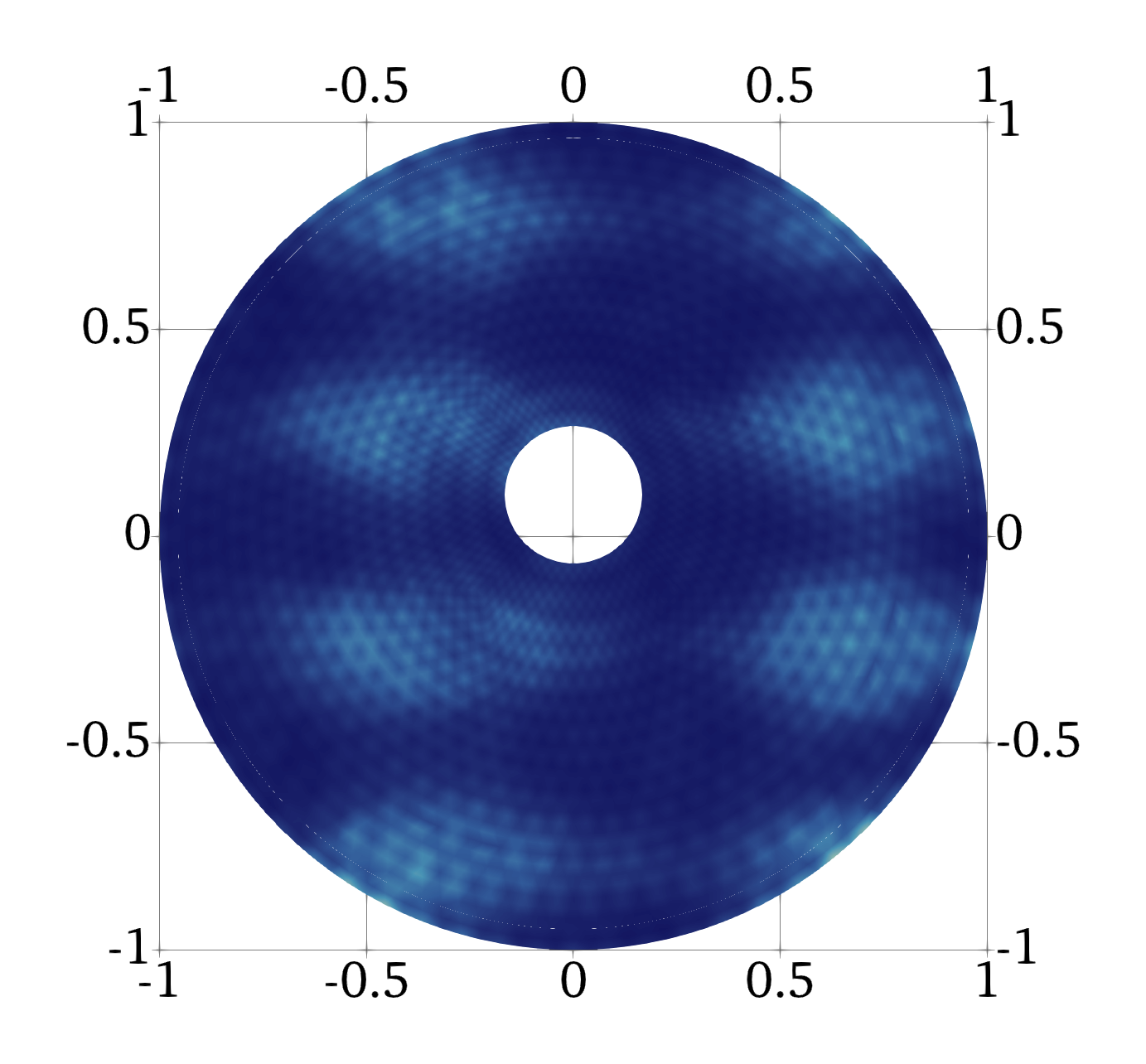}};
    \node at (10.5, -7.6) {\small{$c_2$}};
    \node at (12.8, -9.7) {\small{$c_1$}};

    \node[anchor=west, xshift=0.5cm] (colorbar) at ($(img3.north east)!0.5!(img6.south east)$) {
        \includegraphics[height=0.25\textwidth]{colorbar_standalone.png}
    };
    \end{tikzpicture}}
\caption{Two posterior standard deviations $2\,\mathrm{STD}_{q(\bx|\bphi^*)}[k(\bx,\bc)]$ of the reconstructed permeability field (scalar field). Regions with high uncertainty indicate areas where the observational data provide limited information about the permeability. \textbf{Top row}: 3D visualizations with slices through the origin at $(0,0,0)$. \textbf{Bottom row}: Slices through the domain and origin $(0,0,0)$. From left to right: $c_2$-$c_3$-plane, $c_1$-$c_3$-plane, and $c_1$-$c_2$-plane.}
    \label{fig:posterior_2std_baseline}
\end{figure}

\begin{remark}[From inference to decisions]
\label{sec:posterior_decisions}
The approximate posterior $q(\bx|\bphi^*)$ is the starting point for downstream decision problems. The sparse precision parameterization enables efficient sampling via Equation~\eqref{eqn:rep_1}, which in turn supports (i)~risk assessment through exceedance probabilities $P(k(\bx,\bc) < k_{\mathrm{crit}} \mid \byobs)$, (ii)~uncertainty propagation to derived quantities of interest with calibrated credible intervals, and (iii)~Bayesian optimal experimental design \cite{chaloner1995bayesian}, where the implicit posterior covariance informs sensor placement strategies. These applications inherently require the full uncertainty characterization provided by the posterior distribution, rather than a point estimate alone.
\end{remark}

\FloatBarrier
\subsection{Quantitative ablation and comparison with alternative inference methods}
\label{sec:convergence_analysis}

\begin{figure}[htbp]
\centering
\includegraphics[scale=0.8]{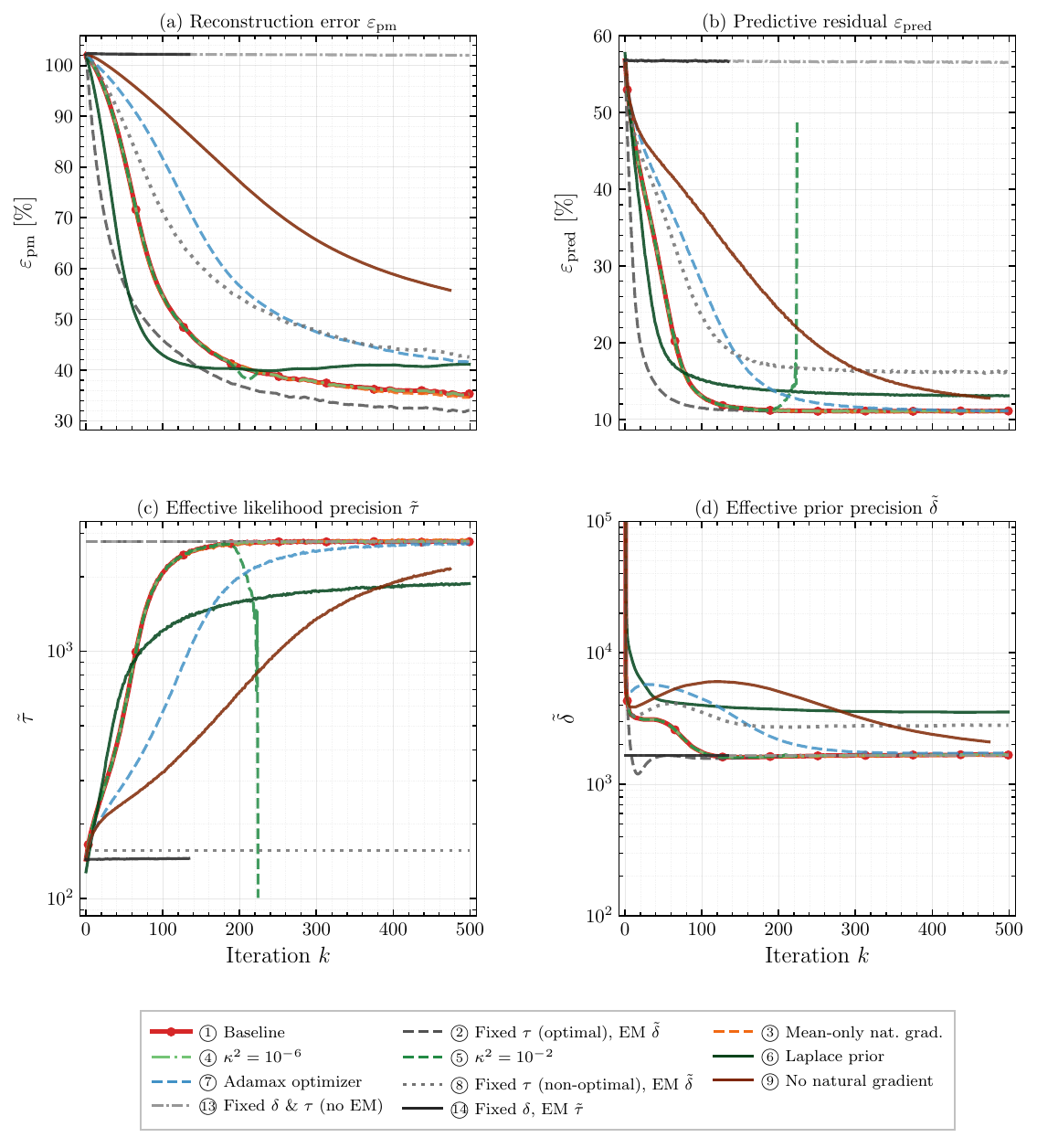}
\caption{Algorithmic ablation: convergence diagnostics for the ten configurations that each modify a single internal axis of Algorithm~\ref{alg:svi} relative to the baseline. \textbf{(a)}~posterior-mean discrepancy $\varepsilon_{\mathrm{pm}}=\|\boldsymbol{\mu}-\bxgt\|_2/\|\bxgt\|_2$; \textbf{(b)}~relative predictive residual $\varepsilon_{\mathrm{pred}}=\|\bar{\by}-\byobs\|_2/\|\byobs\|_2$; \textbf{(c)}~effective likelihood precision $\tilde{\tau}$ (log scale); \textbf{(d)}~effective prior precision $\tilde{\delta}$ (log scale). Corresponding converged numbers are reported in the algorithmic-ablation rows of Table~\ref{tab:ablation_results}.}
\label{fig:convergence_ablation}
\end{figure}

\begin{figure}[htbp]
\centering
\includegraphics[scale=0.8]{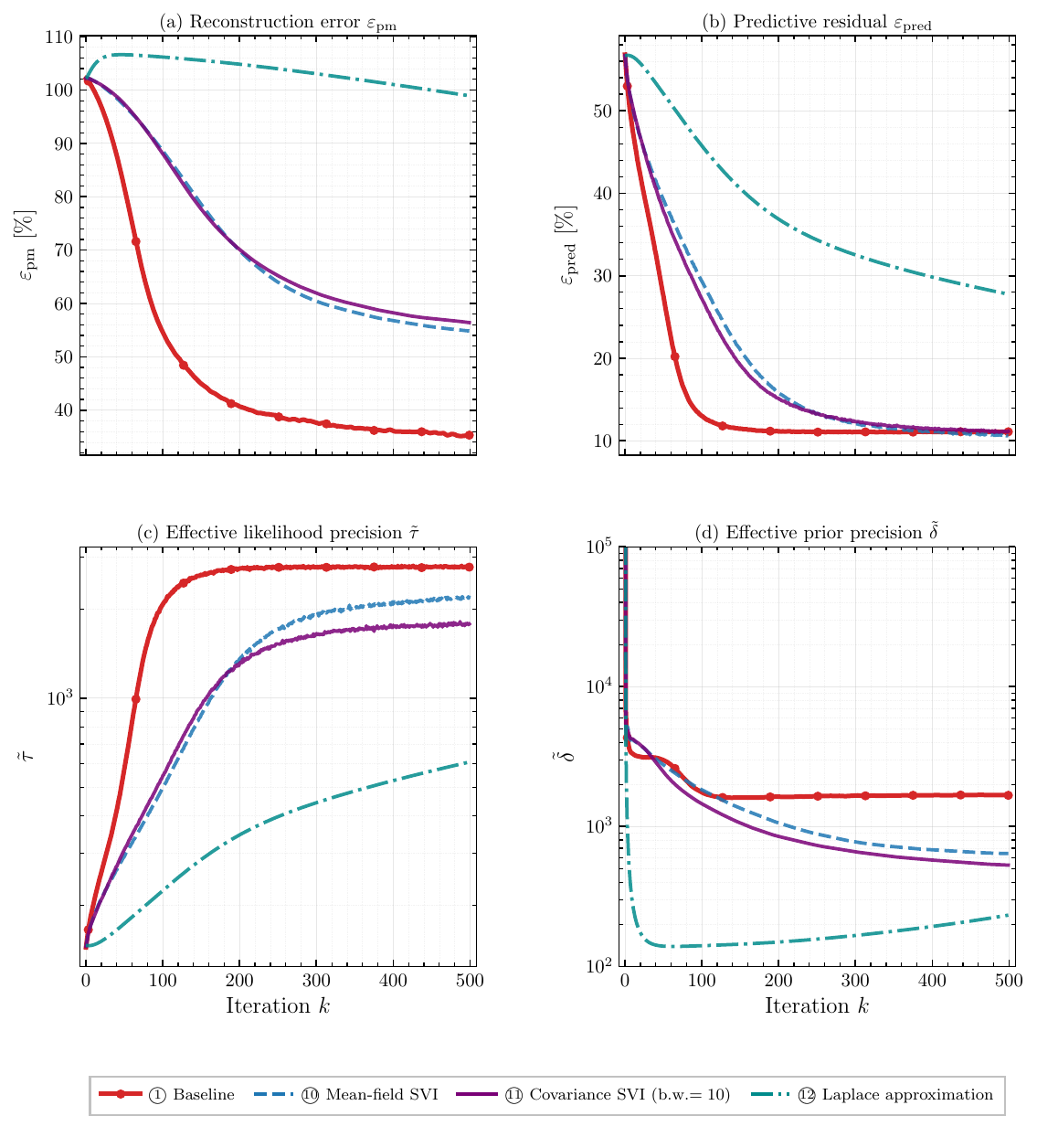}
\caption{Comparison with alternative inference methods: convergence diagnostics for the proposed sparse-precision SVI against three alternative inference methods (mean-field SVI, sparse banded-covariance SVI with off-diagonal bandwidth $10$, and the Laplace approximation), run on the same forward problem, GMRF prior, and adjoint pipeline under otherwise identical settings. The Laplace baseline is realized within the present framework by collapsing the variational covariance to a near-Dirac diagonal, replacing the VB-EM update on $\delta$ with MacKay's evidence approximation \cite{mackay1992bayesian}, and disabling natural-gradient preconditioning (which requires a non-degenerate variational covariance); see the corresponding paragraph in Section~\ref{sec:convergence_analysis} for details. \textbf{(a)}~posterior-mean discrepancy $\varepsilon_{\mathrm{pm}}=\|\boldsymbol{\mu}-\bxgt\|_2/\|\bxgt\|_2$; \textbf{(b)}~relative predictive residual $\varepsilon_{\mathrm{pred}}=\|\bar{\by}-\byobs\|_2/\|\byobs\|_2$; \textbf{(c)}~effective likelihood precision $\tilde{\tau}$ (log scale); \textbf{(d)}~effective prior precision $\tilde{\delta}$ (log scale).}
\label{fig:convergence_comparison}
\end{figure}

We complement the qualitative assessment of Section~\ref{sec:investigations_bia} with a quantitative study that isolates the contribution of each algorithmic component and contrasts the framework against alternative inference methods on the same forward problem. Figure~\ref{fig:convergence_ablation} tracks four diagnostics across SVI iterations for ten algorithmic ablation configurations relative to the baseline of Algorithm~\ref{alg:svi}: the posterior-mean discrepancy $\varepsilon_{\mathrm{pm}}=\|\boldsymbol{\mu}-\bxgt\|_2/\|\bxgt\|_2$, the relative predictive residual $\varepsilon_{\mathrm{pred}}=\|\bar{\by}-\byobs\|_2/\|\byobs\|_2$, and the two effective VB-EM hyperparameters $\tilde{\tau}$ and $\tilde{\delta}$. Figure~\ref{fig:convergence_comparison} reports the same diagnostics for the proposed method against three alternative inference methods (mean-field VI, sparse banded-covariance VI, and the Laplace approximation at the MAP). Here $\boldsymbol{\mu}$ is the variational posterior mean of the latent field and $\bar{\by}\equiv\Ex{q_{\bphi}}{\by}$ the corresponding posterior-predictive mean of the observations. Table~\ref{tab:ablation_results} reports the converged quantities after $500$ iterations, grouped into two blocks (algorithmic ablation and comparison with alternative inference methods).

\begin{remark}[Interpretation of $\tilde{\tau}$ and $\tilde{\delta}$]
In keeping with the fully Bayesian VB-EM formulation of Section~\ref{sec:latent_variables}, the prior over $\bx$ is obtained by marginalization, $p(\bx)=\int p(\bx|\delta)p(\delta)\dd\delta$, and analogously for the $\tau$ precision parameter in the likelihood. The quantities $\tilde{\tau}$ and $\tilde{\delta}$ reported throughout this section are the posterior expectations $a/b$ under the Gamma hyperprior that the variational update plugs into the model, not values to which the hyperparameters have been conditioned.
\end{remark}

\begin{remark}[On the role of $\varepsilon_{\mathrm{pm}}$]
\label{rem:rec_error}
At $\dim(\bx) > 4\cdot 10^{5}$, a high-fidelity reference posterior, \eg via MCMC, is computationally infeasible (Section~\ref{sec:introduction}); the diagnostics reported here are therefore necessarily indirect, comparing the ground-truth field $\bxgt$ against the variational posterior mean (through $\varepsilon_{\mathrm{pm}}$) and qualitatively against the inferred posterior's credible region (Section~\ref{sec:investigations_bia}, Appendix~\ref{sec:posterior_samples_3d}).
The quantity $\varepsilon_{\mathrm{pm}}$ is a \emph{posterior-mean discrepancy}: it is reported here as a relative comparator across the configurations of Table~\ref{tab:ablation_results}, not as an absolute accuracy metric for a Bayesian reconstruction. Two structural features of the test problem inflate it by construction: $\txgt$ has small magnitude with extended near-zero regions, so any local deviation contributes disproportionately to the relative norm; and the rectangular inclusion $\Omega^*$ is structurally unresolvable by a smooth GMRF prior. The criterion that is actually relevant for a Bayesian reconstruction is whether the ground truth lies within the inferred posterior's credible region, which the qualitative comparison of Section~\ref{sec:investigations_bia} and the posterior samples of Appendix~\ref{sec:posterior_samples_3d} confirm across the entire domain. A formally calibrated assessment, via coverage probabilities, posterior-predictive checks (\eg Bayesian $p$-values), or proper scoring rules, would require a reference posterior that is unavailable at this PDE scale; it is therefore best pursued by density matching against constructed targets with a known posterior, a complement to the real-PDE demonstration here that is left to future work.
\end{remark}

\begin{table}[htbp]
\centering
\caption{Ablation and comparison study: converged diagnostics after $500$ SVI iterations ($\dim(\bx)=405\,570$). The table is divided into two blocks: \emph{(i)~Algorithmic ablation}, configurations that each modify a single internal component of Algorithm~\ref{alg:svi} relative to the baseline; and \emph{(ii)~Comparison with alternative inference methods}, all sharing the proposed prior, data, and adjoint pipeline (the proposed sparse-precision SVI is the baseline of block~(i) and is not duplicated here). Within block~(i), rows are sorted by $\varepsilon_{\mathrm{pm}}$ after the pinned baseline; the two trailing rows correspond to prior-basin failures, in which the variational mean does not leave the prior basin. \textbf{Bold} marks the baseline and the smallest value in each error column of that block. The third column reports the effective likelihood precision $\tilde{\tau}$ (posterior mean of $\tau$ under the Gamma hyperprior); the corresponding $\tilde{\delta}$ trajectories are in panel~(d) of Figure~\ref{fig:convergence_ablation} for block~(i) and Figure~\ref{fig:convergence_comparison} for block~(ii).}
\label{tab:ablation_results}
\begin{tabular}{lccc}
\toprule
\textbf{Configuration} & $\varepsilon_{\mathrm{pm}}$ [\%] & $\varepsilon_{\mathrm{pred}}$ [\%] & $\tilde{\tau}$\\
\midrule
\multicolumn{4}{l}{\emph{(i) Algorithmic ablation}}\\
\cnum{1}~\textbf{Baseline (full configuration, Algorithm~\ref{alg:svi})} & $\mathbf{35.3}$ & $\mathbf{11.1}$ & $\mathbf{2.77\!\cdot\!10^{3}}$\\
\cnum{2}~Fixed $\tau$ (optimal), VB-EM $\tilde{\delta}$ & $\mathbf{32.1}$ & $11.1$ & $2.77\!\cdot\!10^{3}$\\
\cnum{3}~Mean-only nat.\ grad.\ (no Takahashi) & $34.6$ & $\mathbf{11.0}$ & $2.83\!\cdot\!10^{3}$\\
\cnum{4}~SPDE, $\kappa^2 = 10^{-6}$ & $35.3$ & $11.1$ & $2.79\!\cdot\!10^{3}$\\
\cnum{5}~SPDE, $\kappa^2 = 10^{-2}$ & $40.1$ & $48.7^{\dagger}$ & $1.01\!\cdot\!10^{2}$\\
\cnum{6}~Laplacian prior (no SPDE, Remark~\ref{rem:gradient_penalty}) & $41.1$ & $13.1$ & $1.88\!\cdot\!10^{3}$\\
\cnum{7}~AdaMax (instead of Adam) & $41.5^{*}$ & $11.1^{*}$ & $2.72\!\cdot\!10^{3}$\\
\cnum{8}~Fixed $\tau$ (non-optimal), VB-EM $\tilde{\delta}$ & $42.6$ & $16.1$ & $1.56\!\cdot\!10^{2}$\\
\cnum{9}~No natural gradient & $55.7$ & $12.8$ & $2.15\!\cdot\!10^{3}$\\
\cnum{10}~Fixed $\delta$ \& $\tau$ (no VB-EM) & $102.0$ & $56.6$ & $2.77\!\cdot\!10^{3}$\\
\cnum{11}~Fixed $\delta$, VB-EM $\tilde{\tau}$ & $102.2$ & $56.7$ & $1.45\!\cdot\!10^{2}$\\
\midrule
\multicolumn{4}{l}{\emph{(ii) Comparison with alternative inference methods}}\\
\cnum{12}~Mean-field VI & $55.0$ & $11.0$ & $2.50\!\cdot\!10^{3}$\\
\cnum{13}~Sparse banded-covariance VI (bandwidth $=10$) & $56.4$ & $11.2$ & $1.78\!\cdot\!10^{3}$\\
\cnum{14}~Laplace approximation (MAP) & $98.9^{*}$ & $27.8^{*}$ & $6.08\!\cdot\!10^{2}$\\
\bottomrule
\multicolumn{4}{l}{\footnotesize ${}^{*}$Not yet converged at $500$ iterations; trajectory still evolving in the corresponding panel of Figure~\ref{fig:convergence_ablation} or~\ref{fig:convergence_comparison}.}\\
\multicolumn{4}{l}{\footnotesize ${}^{\dagger}$Predictive residual destabilized by an excessive nugget; see panels (b)+(c) of Figure~\ref{fig:convergence_ablation}.}
\end{tabular}
\end{table}

\paragraph{\cnum{1}~Baseline convergence.}
The baseline configuration converges within roughly $200$ iterations: $\varepsilon_{\mathrm{pm}}$ drops from its initial value near $100\,\%$ to $35.3\,\%$, $\varepsilon_{\mathrm{pred}}$ from approximately $57\,\%$ to $11.1\,\%$, and $\tilde{\tau}$ settles at $2.77\!\cdot\!10^{3}$. This corresponds to a noise standard deviation $\sigma_{\varepsilon}=1/\sqrt{\tilde{\tau}}\approx 1.9\!\cdot\!10^{-2}$ that is consistent with the synthetic SNR. The hyperparameter trajectories realize the coarse-to-fine continuation predicted in Section~\ref{sec:latent_variables}: $\tilde{\tau}$ rises monotonically as the variational mean concentrates around the data, while $\tilde{\delta}$ relaxes from its initial overshoot toward $\tilde{\delta}\approx 1.7\!\cdot\!10^{3}$ as the prior precision adapts to the residual scale. Notably, the two hyperparameters move in opposite directions: $\tilde{\delta}$ starts high and decreases, so that the prior enforces strong smoothness early on when the inferred signal is still weak, and progressively relaxes its regularization as the variational mean acquires structure; $\tilde{\tau}$, by contrast, rises monotonically throughout, so that the effective likelihood variance $\tilde{\tau}^{-1}$ starts broad and contracts, providing implicit smoothing in early SVI iterations. The converged $\varepsilon_{\mathrm{pred}}\approx 11\,\%$ is only roughly twice the irreducible noise floor of $1/\mathrm{SNR}=5\,\%$ on the observations and is therefore essentially data-noise-limited; together with Remark~\ref{rem:rec_error} and the qualitative assessment of Section~\ref{sec:investigations_bia}, this confirms the baseline accuracy. Across the ten ablation configurations, no row dominates the baseline jointly on both $\varepsilon_{\mathrm{pm}}$ and $\varepsilon_{\mathrm{pred}}$; the two rows that attain a marginally lower $\varepsilon_{\mathrm{pm}}$ --- \cnum{2} via an oracle pin on $\tau$ and \cnum{3} by dropping the precision-Fisher term --- are discussed below. Since the ELBO is non-convex and is optimized stochastically and jointly with the VB-EM hyperparameter updates, convergence is to a local optimum rather than to a provably global Gaussian fit. This non-convexity is intrinsic to nonlinear Bayesian inversion and is equally faced by the optimization-based alternatives; in particular, the MAP estimate underlying the Laplace approximation is itself the outcome of a non-convex optimization, after which its covariance is fitted only locally at that mode. From the common prior-based initialization of Section~\ref{sec:var_init}, every run with adaptive VB-EM and a non-degenerate variational covariance reaches the same data-noise-limited basin (Figures~\ref{fig:convergence_ablation} and~\ref{fig:convergence_comparison}); only the fixed-hyperparameter ablations and the covariance-collapsed Laplace baseline stall in the prior basin, by design rather than from initialization sensitivity.

\paragraph{\cnum{2}\,\cnum{8}\,\cnum{10}\,\cnum{11}~VB-EM and hyperparameter adaptation.}
The four VB-EM ablations together establish that a fully Bayesian treatment of $\delta$ \emph{and} $\tau$ is indispensable. Holding both hyperparameters fixed, or holding $\delta$ fixed while adapting only $\tau$, traps the optimizer near $\varepsilon_{\mathrm{pm}}\approx 102\,\%$ and $\varepsilon_{\mathrm{pred}}\approx 57\,\%$ (panels (a) and (b) of Figure~\ref{fig:convergence_ablation}): an over-strong manually chosen prior prevents the variational mean from leaving the prior, and adapting $\tilde{\tau}$ alone cannot compensate. Adapting only $\delta$ with $\tau$ frozen at a non-optimal value is still costly, inflating $\varepsilon_{\mathrm{pm}}$ to $42.6\,\%$ and $\varepsilon_{\mathrm{pred}}$ to $16.1\,\%$; since the true $\tau$ is not known a priori, this is precisely the failure mode that VB-EM eliminates. Row~\cnum{2} attains the lowest $\varepsilon_{\mathrm{pm}}=32.1\,\%$ in the table, but uses an oracle pin: $\tau$ is held at the value that the baseline's VB-EM update itself recovered. It therefore serves as a sanity check on VB-EM rather than as a substitute for it. Preliminary experiments without VB-EM on the Laplacian prior failed to converge altogether, reinforcing the same conclusion.

\paragraph{\cnum{3}\,\cnum{9}~Natural gradient preconditioning.}
Disabling the natural gradient entirely raises $\varepsilon_{\mathrm{pm}}$ from $35.3\,\%$ to $55.7\,\%$ and noticeably slows the rise of $\tilde{\tau}$ in panel~(c), confirming that a scalar learning rate cannot accommodate the spatially varying posterior curvature (Section~\ref{sec:natural_gradient_mean}). The mean natural gradient alone accounts for essentially all of this improvement; the additional Takahashi-based precision natural gradient (Section~\ref{sec:natural_gradient_precision}) confers no measurable benefit ($35.3\,\%$ with Takahashi versus $34.6\,\%$ mean-only, within the stochastic noise of the SVI estimator). The cheaper mean-only variant suffices in practice.

\paragraph{\cnum{7}~Optimizer choice.}
Replacing Adam with AdaMax under otherwise identical settings slows convergence noticeably: at $500$ iterations Adam has settled at $\varepsilon_{\mathrm{pm}}=35.3\,\%$ while AdaMax has only reached $41.5\,\%$ and is still decreasing in panel~(a) of Figure~\ref{fig:convergence_ablation}. Both are first-order adaptive optimizers and are expected to reach a comparable converged basin under sufficient iterations; Adam's per-coordinate second-moment estimate adapts more aggressively to the spatially varying posterior curvature in this regime, however, and therefore reaches that basin within the iteration budget considered.

\paragraph{\cnum{4}\,\cnum{5}\,\cnum{6}~Prior formulation and the role of $\kappa^2$.}
Replacing the SPDE-based prior with the simpler Laplacian prior $Q_{\mathrm{simple}}=\delta\cdot\AL$ (Remark~\ref{rem:gradient_penalty}) raises $\varepsilon_{\mathrm{pm}}$ from $35.3\,\%$ to $41.1\,\%$ and $\varepsilon_{\mathrm{pred}}$ from $11.1\,\%$ to $13.1\,\%$, confirming that the mass-matrix correction provides improved regularization through longer-range spatial correlations (Section~\ref{sec: Prior}). Given how much simpler the Laplacian prior is, requiring only the bare discrete Laplacian with no mass-matrix correction or shift parameter, the $\sim\!6$ percentage point degradation is comparatively small, and the Laplacian prior remains a viable lightweight alternative. The reconstruction is, by contrast, insensitive to the SPDE shift parameter $\kappa^2$ as long as it is kept in the small-nugget regime: setting $\kappa^2=10^{-6}$ instead of the baseline $10^{-4}$ produces results that are numerically indistinguishable from the baseline. Once $\kappa^2$ is no longer negligible the picture changes: at $\kappa^2=10^{-2}$, panel~(c) of Figure~\ref{fig:convergence_ablation} shows $\tilde{\tau}$ collapsing to the order $10^{2}$ and the predictive residual rebounding to $48.7\,\%$ as the variational update destabilizes. We therefore recommend treating $\kappa^2$ as a small nugget ($\kappa^2\lesssim 10^{-4}$) that ensures positive-definiteness of the SPDE operator; the VB-EM update of $\tilde{\delta}$ already absorbs the relevant prior-strength variability.

\paragraph{\cnum{12}~Comparison against mean-field VI.}
Mean-field VI \cite{franck2016sparse, koutsourelakis2016variational} restricts the variational family to a diagonal covariance, eliminating all posterior correlations. Under conditions otherwise identical to the baseline, it converges to $\varepsilon_{\mathrm{pm}}\approx 55.0\,\%$ and $\varepsilon_{\mathrm{pred}}\approx 11.0\,\%$ at $500$ iterations, with $\tilde{\tau}$ rising to $\sim\!2.5\!\cdot\!10^{3}$ (Figure~\ref{fig:convergence_comparison}). The $\sim\!20$ percentage-point gap in $\varepsilon_{\mathrm{pm}}$ relative to the baseline is structural: the GMRF posterior carries dense long-range correlations whose representation at this stochastic dimension requires inverting a sparse precision factor, and a diagonal covariance encodes none of that structure. The variational mean therefore cannot concentrate around the data as efficiently as in the full sparse-precision case.

\paragraph{\cnum{13}~Comparison against sparse banded-covariance VI.}
The closest existing variational competitor is the sparse banded-covariance parameterization of \cite{bruder2018beyond}, transferred into the present framework with an off-diagonal bandwidth of $10$ and run with settings otherwise identical to the baseline of Algorithm~\ref{alg:svi}. The predictive residual converges to $\varepsilon_{\mathrm{pred}}\approx 11.2\,\%$, indistinguishable from the baseline and essentially data-noise-limited; the posterior-mean discrepancy, however, reaches only $\varepsilon_{\mathrm{pm}}\approx 56.4\,\%$ at $500$ iterations and is still slowly decreasing, with $\tilde{\tau}$ reaching only $\sim\!1.8\!\cdot\!10^{3}$ rather than settling at the baseline's $2.77\!\cdot\!10^{3}$ (panels~(a) and~(c) of Figure~\ref{fig:convergence_comparison}). The banded-covariance row is numerically indistinguishable from the mean-field row above on the error panels and is, on the $\tilde{\tau}$ panel, even slightly farther from the baseline value than mean-field: adding short-range correlations through a bandwidth-$10$ covariance Cholesky factor on top of a strictly diagonal mean-field confers no measurable advantage on $\varepsilon_{\mathrm{pm}}$ at this stochastic dimension. The structural gap therefore lies between any covariance-truncation parameterization and the proposed sparse-precision parameterization, not within the covariance family across bandwidths; encoding the GMRF posterior's dense long-range correlations through a band of width $\lesssim 10$ in the covariance Cholesky truncates the bulk of that correlation structure away. Since all other axes were held identical, the gap in $\varepsilon_{\mathrm{pm}}$ is attributable to the covariance-truncation basis itself. An additional consequence of the parameterization choice, beyond the convergence gap measured here, is that cheap natural-gradient preconditioning of the Cholesky parameters via Takahashi selected inversion (Section~\ref{sec:natural_gradient_precision}) has no analogue at this scale in the banded-covariance family (Section~\ref{sec:positioning}).

\paragraph{\cnum{14}~Comparison against Laplace approximation.}
The Laplace approximation centers a Gaussian at the MAP estimate $\bx^*$ with covariance equal to the inverse Hessian of the negative log-posterior at $\bx^*$ (Section~\ref{sec:positioning}). We realize this baseline within the present framework by (i)~collapsing the variational covariance to a fixed near-Dirac diagonal with small constant $\sigma_0$, so that the variational mean $\bphimu$ converges to the MAP estimate, (ii)~replacing the conjugate-Gamma VB-EM update on $\delta$ with MacKay's evidence approximation \cite{mackay1992bayesian}, the standard MAP-regime substitute in which the parameter dimension $\dim(\bx)$ in the numerator is replaced by the data-informed effective degrees of freedom $\gamma\approx N_{\mathrm{obs}}$, and (iii)~disabling natural-gradient preconditioning, since the Fisher information of the variational mean, $Q=\LQ\LQ^T$, degenerates as $\sigma_0\!\to\!0$. The conjugate-Gamma update on $\tau$ is retained throughout: the numerator $\dim(\by)/2$ already references the observation count and is therefore well posed in the MAP regime. All remaining settings (prior, forward and adjoint models, learning rate, batch size, iteration budget) match the baseline. Figure~\ref{fig:convergence_comparison} reports the resulting trajectories: $\tilde\tau$ rises monotonically as VB-EM on $\tau$ remains active, $\tilde\delta$ drops toward $\sim\!2.3\!\cdot\!10^{2}$ in line with the smaller MacKay numerator, $\varepsilon_{\mathrm{pred}}$ settles near $27.8\,\%$, and $\varepsilon_{\mathrm{pm}}$ remains near $98.9\,\%$ at $500$ iterations, having barely left the prior basin. We additionally verified that fixing $\tau$ at the known noise level rather than adapting it via VB-EM yields essentially identical converged $\varepsilon_{\mathrm{pm}}$ and $\varepsilon_{\mathrm{pred}}$, confirming that the choice between adaptive and oracle $\tau$ is not the determining factor here. Adapting $\delta$ via MacKay's update, by contrast, remains essential: fixing $\delta$ would trap the optimizer in the prior basin, as the algorithmic-ablation rows~\cnum{10} and~\cnum{11} ($\varepsilon_{\mathrm{pm}}\approx 102\,\%$) show.

Two factors explain this slow convergence. First, the natural gradient employed in the baseline (Section~\ref{sec:natural_gradient_mean}) is the probabilistic analog of a second-order update, with $Q$ playing the role of posterior curvature; it degenerates as the variational variance collapses, so MAP optimization within this framework forfeits the curvature preconditioning the baseline relies on. State-of-the-art MAP-Laplace inversions at extreme scale \cite{Bui_Thanh_2012, henneking2025gordonbell, villa2021hippylib} recover this preconditioning via Newton-Krylov optimization with low-rank Hessian operators, machinery that we do not implement here; the convergence rate observed in Figure~\ref{fig:convergence_comparison} therefore reflects first-order MAP optimization at this stochastic dimension and is slower than such second-order solvers would deliver on the same problem. Second, the maintained variational variance in the baseline confers an optimization-side benefit beyond uncertainty quantification by providing the broad-likelihood annealing through which $\tilde\tau$ rises from below (Section~\ref{sec: data_log_lik}); a milder version of this stabilizing role is visible in the mean-field row (block~(ii)), which retains a diagonal variational variance and converges noticeably faster than the Laplace configuration despite carrying no posterior correlations. Beyond convergence speed, the proposed sparse-precision Gaussian fits the ELBO globally, whereas the Laplace approximation matches the negative-log-posterior curvature locally at $\bx^*$: for an exactly Gaussian posterior the two coincide, but for posteriors that depart from local quadratic behavior the global Gaussian variational fit attains a smaller Kullback-Leibler divergence to the true posterior, with the gap widening for skewed or non-Gaussian posteriors where a local Hessian fit at the MAP under- or over-estimates the dominant moments.

\paragraph{Summary of ablation findings.}
Five conclusions emerge from the algorithmic ablation. (i)~Across the ten ablation runs, no row dominates the baseline jointly on both $\varepsilon_{\mathrm{pm}}$ and $\varepsilon_{\mathrm{pred}}$. (ii)~Fully Bayesian VB-EM for both $\tau$ and $\delta$ is indispensable, since any fixed-hyperparameter configuration either traps the optimizer at the prior or incurs a measurable accuracy penalty. (iii)~Natural gradient preconditioning of the variational mean is essential, while the additional precision-parameter variant is dispensable in practice. (iv)~Adam~\cite{kingma2014adam} outperforms AdaMax at the iteration budget considered. (v)~The SPDE shift $\kappa^2$ should be kept as a small nugget rather than promoted to a tunable parameter.

\paragraph{Summary of comparison findings.}
The sparse-precision parameterization outperforms both standard variational alternatives by approximately $20$ percentage points on $\varepsilon_{\mathrm{pm}}$: the sparse banded-covariance VI of \cite{bruder2018beyond} (bandwidth $10$) stagnates near $56.4\,\%$ and the diagonal mean-field VI of \cite{franck2016sparse, koutsourelakis2016variational} near $55.0\,\%$, both at the baseline's data-noise-limited predictive residual ($\varepsilon_{\mathrm{pred}}\approx 11\,\%$). The two rows are numerically indistinguishable on all panels of Figure~\ref{fig:convergence_comparison}, so the gap is structural to the covariance-truncation basis rather than to the chosen bandwidth. The Laplace baseline trails by a wider margin still within the available iteration budget; the convergence-rate and global-vs-local Kullback-Leibler arguments are developed in the Laplace paragraph above.

\FloatBarrier
\section{Conclusion and outlook}
\label{sec:conclusion}
We presented a unified FE-native inference framework for fully Bayesian reconstruction of spatial fields on three-dimensional domains, including physics-based problems that are nonlinear in the inferred parameters. Its four components, the SPDE/GMRF prior, the sparse-precision Gaussian variational family, the VB-EM hyperparameter updates, and the adjoint-based likelihood gradient, are all expressed through FE operators shared with the forward problem, in particular the discrete Laplacian $\AL$ and the mass matrix $M$, so that no auxiliary data structures are introduced for the inference. The variational family inherits the Markov sparsity of the GMRF prior and represents dense posterior covariances implicitly through its sparse precision Cholesky factor. To our knowledge, this is the first fully Bayesian, full-covariance variational reconstruction of a spatial random field combining a PDE forward model nonlinear in the inferred parameters, a three-dimensional FE discretization on a curved domain, and a stochastic dimension exceeding $4\cdot 10^{5}$. The contribution is twofold: three genuinely new methodological ingredients (the sparsity inheritance from the FE Laplacian, the path-derivative ELBO gradient for the precision parameterization, and the FE-native unification of these four components), combined with the careful deployment of established techniques at the three-dimensional FE scale. We, therefore, complement the low-rank Hessian-Laplace approaches \cite{Bui_Thanh_2012, henneking2025gordonbell} that dominate extreme-scale Bayesian inversion by learning the full covariance globally rather than a local quadratic approximation at the mode, which leads to an increased overall posterior approximation accuracy.

The framework was applied to a three-dimensional Darcy flow problem on a curved domain, reconstructing the log-permeability field from $6\,930$ noisy velocity observations at a stochastic dimension of $4.05\cdot 10^{5}$. Unlike deterministic inversion, the method yields a full posterior distribution, providing uncertainty quantification, a basis for Bayesian optimal experimental design, and statistical decision capabilities (Remark~\ref{sec:posterior_decisions}) that are essential when decisions must account for uncertainty in inferred material properties, \eg in medical or other critical settings. 
The baseline configuration converges within approximately $200$ SVI iterations to $\varepsilon_{\mathrm{pm}}=35.3\,\%$ and $\varepsilon_{\mathrm{pred}}=11.1\,\%$, the latter within a factor of two of the irreducible noise floor $1/\mathrm{SNR}=5\,\%$, so that the reconstruction is essentially data-noise-limited. The algorithmic ablation of Section~\ref{sec:convergence_analysis} (Table~\ref{tab:ablation_results}) identifies two components as indispensable: VB-EM of both $\delta$ and $\tau$, whose absence traps the optimizer in the prior basin at $\varepsilon_{\mathrm{pm}}\approx 102\,\%$; and natural gradient preconditioning of the variational mean. The additional Takahashi-based precision-Cholesky natural gradient retained in the baseline confers no measurable benefit ($35.3\,\%$ with Takahashi versus $34.6\,\%$ mean-only, within stochastic noise), so the cheaper mean-only variant is a viable alternative. The accompanying comparison with alternative inference methods further establishes that the sparse-precision parameterization outperforms both the diagonal mean-field and the sparse banded-covariance variants by approximately $20$ percentage points on $\varepsilon_{\mathrm{pm}}$, since neither can encode the long-range correlations that the GMRF posterior demands. The Laplace baseline lags by a wider margin still within the available iteration budget (Figure~\ref{fig:convergence_comparison}, panel~(a)), reflecting that first-order MAP optimization within this framework cannot exploit the natural-gradient preconditioning that the maintained variational uncertainty enables; a Newton-Krylov solver as in \cite{Bui_Thanh_2012, henneking2025gordonbell, villa2021hippylib} would be required to recover competitive convergence outside the present framework. Beyond convergence rate, the global ELBO minimization here attains a smaller Kullback-Leibler divergence to the true posterior than a local Hessian-Laplace fit at the mode, with the gap widening for skewed or strongly non-Gaussian posteriors.

The method has several limitations. The Gaussian variational family is unimodal and cannot represent multi-modal posteriors arising in severely ill-posed problems. The GMRF prior enforces spatial smoothness and therefore diffuses sharp discontinuities, as observed for the rectangular inclusion $\Omega^*$. The numerical investigation is furthermore restricted to a single forward model, and the behavior of other physics remains to be assessed. A full posterior-sampling reference at $\dim(\bx) > 4\cdot 10^{5}$ is computationally out of budget for this study; the controlled-variable comparison with alternative inference methods in Section~\ref{sec:convergence_analysis} provides the in-scope quantitative reference. Finally, the validation rests on a single synthetic ground truth at one observation density and noise level, and the inference likelihood (\iid Gaussian, of the same forward-model class as the data generator) matches the data-generating process, so model error, correlated noise, and likelihood misspecification are not probed; assessing this robustness across these axes and on a real or experimental dataset is left to future work.

Several directions follow naturally. On the computational side, the current implementation, built on the open-source QUEENS framework \cite{queens} and the deal.II FE library \cite{africa2024deal}, serves as a feasibility demonstration; caching system matrices across iterations, relaxing solver tolerances early in the optimization, and adaptively controlling solver precision based on the state of the variational parameters would reduce wall-clock time without affecting the converged solution. Methodologically, the fact that the baseline predictive residual is already close to the noise floor motivates richer likelihoods or denser sensor placements via Bayesian optimal experimental design rather than further inference-side improvements. 
Promising algorithmic extensions include spatially adaptive priors via edge-specific scaling of the Laplacian entries, with a VB-EM update on $\delta$ to enable spatially varying smoothness; richer variational families, such as normalizing flows on sparse graph structures, for capturing posterior asymmetry; an analytic full-Fisher natural gradient on the precision Cholesky factor in the form recently derived by \cite{tan2025analytic}, as a refinement of our diagonal Takahashi approximation under the FE-Laplacian sparsity pattern; and generalization to transient and coupled multi-physics inverse problems. The most immediate application target is patient-specific reconstruction of tissue properties in the human lung, which requires a coupled, nonlinear poroelastic forward model for which full-fidelity Bayesian inference remains prohibitively expensive. In \cite{nitzler2024bmfia}, a Bayesian multi-fidelity inverse analysis was demonstrated in two spatial dimensions; extending this approach to three dimensions using the single-physics SVI method developed here as the low-fidelity component would be an exciting next direct step towards complex patient-specific modeling. Combining these ideas with multi-physics observations, as suggested in \cite{haeusel2026multi}, would further enhance reconstruction accuracy for future work.
\section*{Acknowledgements}
The authors gratefully acknowledge financial support from the Deutsche Forschungsgemeinschaft (DFG, German Research Foundation) in the project WA1521/26-1, and by BREATHE, a Horizon—ERC–2020–ADG project (grant agreement No. 101021526-BREATHE). We furthermore want to thank Lea J. Haeusel for helpful discussions on probabilistic modeling and didactic suggestions on the presentation of the theory and results.

\printbibliography
\appendix
\section{Weak form of PDE and FEM discretization}
\label{sec:darcy_weak}
We derive the weak form of Equation \eqref{eqn: strong_form_darcy} to subsequently solve it with FEM in mixed form \cite{douglas1985global}. 
For convenience, we first repeat the strong form (first without boundary conditions):
\begin{align*}
    K^{-1}(\bx, \bc)\cdot \bu  + \nabla p &= \bnil, \qquad \text{ in } \Omega,\\
    \text{div } \bu & = a, \qquad \text{ in } \Omega.
\end{align*}
Multiplication with test functions $q$ and $\bv$ for pressure $p$ and velocity $\bu$ along with integration over the domain $\Omega$, leads to the following residual form:
\begin{subequations}
\begin{align}
    \label{eqn: residual_form}
    \left(\bv, K^{-1}\bu\right)_{\Omega} + \left(\bv,\nabla p\right)_{\Omega}  &=\boldsymbol{0}\\
 \left(q,\text{div}\bu\right)_{\Omega}  &= \left(q,a\right)_{\Omega}.
\end{align}
\end{subequations}

As we want to prescribe weak boundary conditions for pressure and velocity, we furthermore conduct integration by parts for the terms $\left(\bv,\nabla p\right)_{\Omega}$ and $\left(q,\text{div}\bu\right)_{\Omega}$ to expose boundary contributions of the primary variables:
\begin{subequations}
\begin{align}
 \label{eqn: weak_form}   
 \left(\bv,K^{-1}\bu\right)_{\Omega} \underbrace{- \left(\text{div}\bv,p\right)_{\Omega} + \left(\bv\cdot\bn,p\right)_{\Gamma}}_{=\left(\bv,\nabla p\right)_{\Omega}}  &= \boldsymbol{0}\\
 \underbrace{- \left(\nabla q,\bu\right)_{\Omega} + \left(q,\bu\cdot \bn\right)_{\Gamma}}_{=\left(q,\text{div}\bu\right)_{\Omega}} &= \left(q,a\right)_{\Omega}.
\end{align}
\end{subequations}
We can now use the boundary terms to impose weak boundary conditions for pressure $p=b$ on $\Gamma_p$ and for the normal flux $\bu\cdot\bn=w$ on $\Gamma_{\bu}$.
After the FE discretization, this leads to the following block system, with system matrix $A$, solution vector $\by$, and \rhs:
\begin{equation}
 \label{eqn:dary_matrix_vector}
 \underbrace{
 \begin{bmatrix}
 M & B\\
 B^T & 0
 \end{bmatrix}}_{\substack{\text{system}\\ \text{matrix } A}}\cdot \underbrace{
 \begin{bmatrix}
 \bu^d\\
 \bp^d
 \end{bmatrix}}_{\by}
 - \underbrace{\begin{bmatrix}
 \boldsymbol{f}\\
 \bh
 \end{bmatrix}}_{\text{\rhs}} = \bd(\bx,\by)=\boldsymbol{0},
\end{equation}
The element-level contributions to the system matrix blocks are:
\begin{subequations}
\begin{align}
\label{eqn:darcy_ele_matrices}
M_{\mathrm{ele},ij} &= \left(\bphi^{\bu}_i,\,K^{-1}\bphi^{\bu}_j\right)_{\Omega_{\mathrm{ele}}}, \\
B_{\mathrm{ele},ij} &= -\left(\mathrm{div}\,\bphi^{\bu}_i,\,\phi^{p}_j\right)_{\Omega_{\mathrm{ele}}} + \left(\bphi^{\bu}_i\cdot\bn,\,\phi^{p}_j\right)_{\Gamma_{\bu,\mathrm{ele}}},
\end{align}
\end{subequations}
as confirmed by integration by parts on $B_{\mathrm{ele},ij}$. 
The element-level \rhs contributions are
\begin{align}
\label{eqn:darcy_ele_rhs}
\boldsymbol{f}_{\mathrm{ele},i} &= \underbrace{-\left(\bphi^{\bu}_i\cdot\bn,\,b\right)_{\Gamma_{p,\mathrm{ele}}}}_{\substack{\text{pressure}\\ \text{B.C.}}},\qquad
\bh_{\mathrm{ele},i} = \underbrace{-\left(\phi^{p}_i,\,a\right)_{\Omega_{\mathrm{ele}}}}_{\substack{\text{source}\\ \text{term}}} + \underbrace{\left(\phi^{p}_i,\,w\right)_{\Gamma_{\bu,\mathrm{ele}}}}_{\substack{\text{normal flux}\\ \text{B.C.}}}.
\end{align}
All global matrices and vectors are assembled from these element contributions:
\begin{equation}
\label{eqn:darcy_assembly}
M = \Assembly\!\left[M_{\mathrm{ele},ij}\right],\quad B = \Assembly\!\left[B_{\mathrm{ele},ij}\right],\quad  \boldsymbol{f} = \AssemblyVec\!\left[\boldsymbol{f}_{\mathrm{ele},i}\right],\quad \bh = \AssemblyVec\!\left[\bh_{\mathrm{ele},i}\right].
\end{equation}
We implemented the problem efficiently in C++ using the deal.II library \cite{africa2024deal}.

\section{On preconditioner and solution strategy}
\label{sec:preconditioner}
The saddle-point matrix $A$ is symmetric (its $(2,1)$ block equals $B^T$ by the sign convention in the weak form) but indefinite (the $(2,2)$ block vanishes). The block lower-triangular preconditioner~\eqref{eqn:block_precond} is non-symmetric, making the preconditioned system $P^{-1}A$ non-symmetric; we therefore solve with preconditioned GMRES with a Krylov subspace dimension of $200$ and a relative residual tolerance of $10^{-6}$.

We employ a block lower-triangular preconditioner based on the Schur complement factorization:
\begin{equation}
\label{eqn:block_precond}
P = \begin{bmatrix} \tilde{M} & 0 \\ B^T & -\tilde{S} \end{bmatrix},
\end{equation}
where $\tilde{M}$ and $\tilde{S}$ denote preconditioners for the velocity block $M$ and the Schur complement $S = B^T M^{-1} B$, respectively. Computing $S$ is prohibitively expensive. Instead, we approximate it by a permeability-weighted pressure Laplacian with symmetric Nitsche boundary conditions on $\Gamma_p$:
\begin{equation}
\label{eqn:approx_schur}
\tilde{S}_{\mathrm{ele},ij} = \left(K\,\nabla\phi^{p}_i,\, \nabla\phi^{p}_j\right)_{\Omega_{\mathrm{ele}}} - \left(\nabla\phi^{p}_i\cdot\bn,\, \phi^{p}_j\right)_{\Gamma_{p,\mathrm{ele}}} - \left(\phi^{p}_i,\, \nabla\phi^{p}_j\cdot\bn - \tau\,\phi^{p}_j\right)_{\Gamma_{p,\mathrm{ele}}},
\end{equation}
with penalty parameter $\tau = 5(p{+}1)^2/h$ \cite{shahbazi2005explicit}, where $p$ denotes the pressure polynomial degree and $h$ the local cell diameter. The global matrix is assembled via $\tilde{S} = \Assembly[\tilde{S}_{\mathrm{ele},ij}]$, which yields a symmetric positive definite approximation that is spectrally equivalent to $S$ for Darcy-type saddle-point problems. Both $\tilde{M}^{-1}$ and $\tilde{S}^{-1}$ are approximated by a single algebraic multigrid (AMG) V-cycle provided by Trilinos ML, configured for elliptic operators with higher-order FE and two smoother sweeps. Since $A$ is symmetric, the adjoint system reduces to $A\blambda = \br$; it is solved with GMRES using the transposed (upper block-triangular) preconditioner $P^T$.

\section{On the implementation of the adjoint formulation}
\label{sec:adjoint_porous_media}
We specialize the abstract adjoint framework of Section~\ref{sec:adjoint} (Equations~\eqref{eqn: adjoint_solve}--\eqref{eqn: final_inner_product}) to the Darcy problem. The discrete PDE residual from Equation~\eqref{eqn:dary_matrix_vector} reads:
\begin{equation}
\label{eqn:darcy_residuum}
\bd(\bx,\by) = A\,\by - \begin{bmatrix} \boldsymbol{f} \\ \bh \end{bmatrix} = \begin{bmatrix} M & B \\ B^T & 0 \end{bmatrix}\begin{bmatrix} \bu^d \\ \bp^d \end{bmatrix} - \begin{bmatrix} \boldsymbol{f} \\ \bh \end{bmatrix} = \boldsymbol{0}.
\end{equation}
Since the system is linear in the solution variable $\by=[\bu^d,\, \bp^d]^T$, the partial derivative of the residual \wrt $\by$ returns the system matrix itself:
\begin{equation}
\label{eqn: grad_pde_grad_u}
\frac{\partial \bd}{\partial \by} = A = \begin{bmatrix} M & B \\ B^T & 0 \end{bmatrix}.
\end{equation}
The adjoint system~\eqref{eqn: adjoint_solve} therefore becomes $A^T\blambda = -\nabla_{\by}\logl^T$; since $A$ is symmetric ($A = A^T$), we reuse the forward system matrix together with the transposed block preconditioner from Appendix~\ref{sec:preconditioner}. The adjoint \rhs is assembled from the upstream gradient $\nabla_{\by}\logl$, evaluated at the observation points $\bc_{\mathrm{obs}}$ and projected onto the velocity FE basis. Technically, we differentiate \wrt the full solution vector $\by$ and subsequently discard the pressure component of the adjoint solution.

The more involved part is the partial derivative of the residual \wrt the input parameters $\bx$. Only the velocity block $M$ depends on $\bx$ through $K^{-1}(\bx)$; the off-diagonal blocks $B$ and $B^T$ and the right-hand side are independent of $\bx$. The partial derivative is therefore:
\begin{equation}
\label{eqn: grad_pde_x}
\frac{\partial \bd}{\partial \bx} = \frac{\partial A}{\partial \bx}\,\by = \begin{bmatrix} \frac{\partial M}{\partial \bx}\,\bu^d \\[4pt] \boldsymbol{0} \end{bmatrix}.
\end{equation}
Forming the full tensor $\frac{\partial M}{\partial \bx}$ of size $\ndofs\times\ndofs\times\dimx$ is prohibitively expensive. Instead, we compute the gradient $\nabla_{\bx}\logl = \blambda^T\frac{\partial \bd}{\partial \bx}$ (Equation~\eqref{eqn: final_inner_product}) directly on the element level, assembling only the resulting vector of length $\dimx$:
\begin{subequations}
\begin{align}
\label{eqn: final_assembly}
\left(\nabla_{\bx}\logl\right)_k &= \sum_{\text{elements}} \left(\blambda^u_{\mathrm{ele}}\right)^T \frac{\partial M_{\mathrm{ele}}}{\partial x_k}\, \bu^d_{\mathrm{ele}} = \sum_{\text{elements}} \int_{\Omega_{\mathrm{ele}}} \frac{\partial K^{-1}}{\partial x_k}\left(\blambda^u_{\mathrm{ele}} \cdot \bu^d_{\mathrm{ele}}\right) \dd \Omega_{\mathrm{ele}}, \\
\label{eqn: k_derivative}
\text{with}\quad \frac{\partial K^{-1}}{\partial x_k} &= -\exp\!\left(-x(\bc)\right)\cdot \phi_k^{\mathrm{rf}}(\bc),
\end{align}
\end{subequations}
where $\phi_k^{\mathrm{rf}}$ denotes the $k$-th shape function of the random field FE space and $\blambda^u_{\mathrm{ele}}$, $\bu^d_{\mathrm{ele}}$ are the element-local velocity components of the adjoint and forward solutions, respectively. For the isotropic case $K(\bx,\bc) = \exp(x(\bc))\cdot I$, we used $K^{-1}=\exp(-x)\cdot I$ and the chain rule $\frac{\partial}{\partial x_k}\exp(-x(\bc)) = -\exp(-x(\bc))\cdot\phi_k^{\mathrm{rf}}(\bc)$, where $x(\bc) = \sum_k \phi_k^{\mathrm{rf}}(\bc)\, x_k$ is the FE representation of the log-permeability field.

\section{Posterior samples: isosurface comparison}
\label{sec:posterior_samples_3d}

\begin{figure}[htbp]
    \centering
    \resizebox{\textwidth}{!}{%
    \begin{tikzpicture}
    \def\imgwidth{0.4\textwidth}
    \def\dx{\imgwidth - 1.3cm}   
    \def\dy{-0.32\textwidth}    

    \node[inner sep=0pt, anchor=north west] (img1) at (0,0)
      {\includegraphics[scale=0.1]{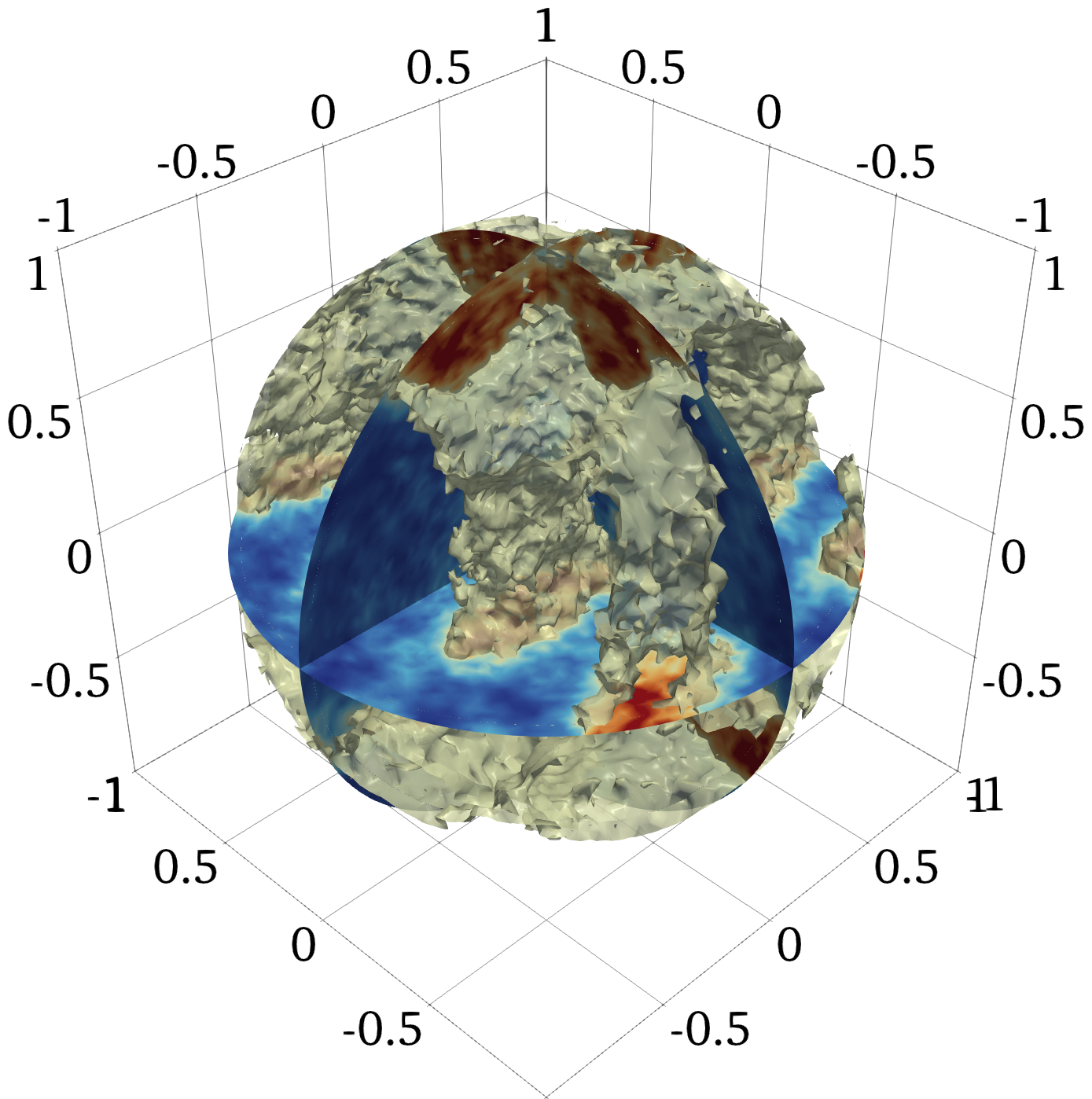}};
    \node at (4.3, -4.4) {\small{$c_1$}};
    \node at (0.6, -4.4) {\small{$c_2$}};

    \node[inner sep=0pt, anchor=north west] (img2) at (\dx,-0.4)
      {\includegraphics[scale=0.10]{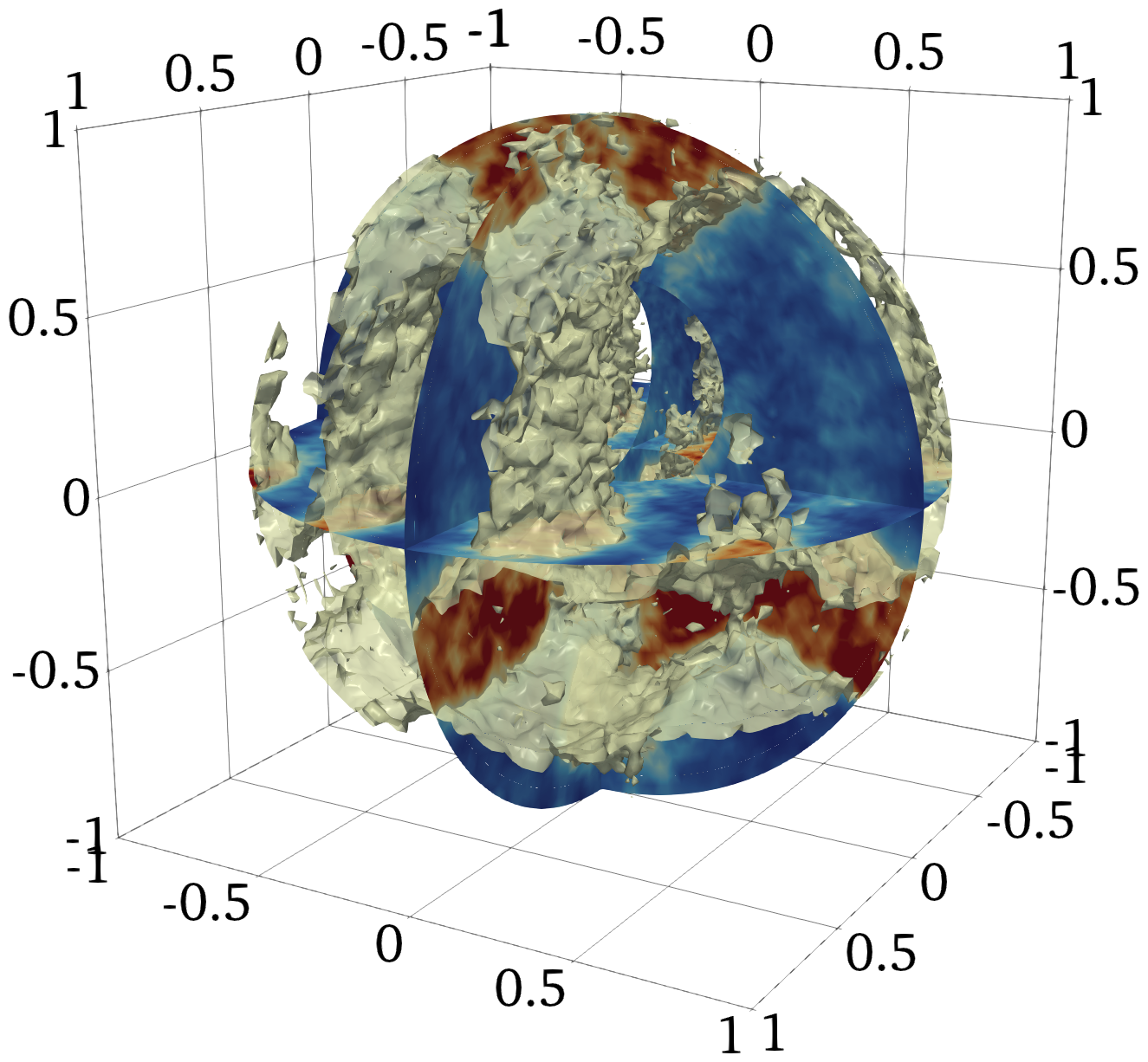}};
    \node at (5.3, -2.7) {\small{$c_3$}};
    \node at (6.7, -4.6) {\small{$c_2$}};
    \node at (9.3, -4.6) {\small{$c_1$}};

    \node[inner sep=0pt, anchor=north west] (img3) at ({1.75*\dx},0.2)
      {\includegraphics[scale=0.10]{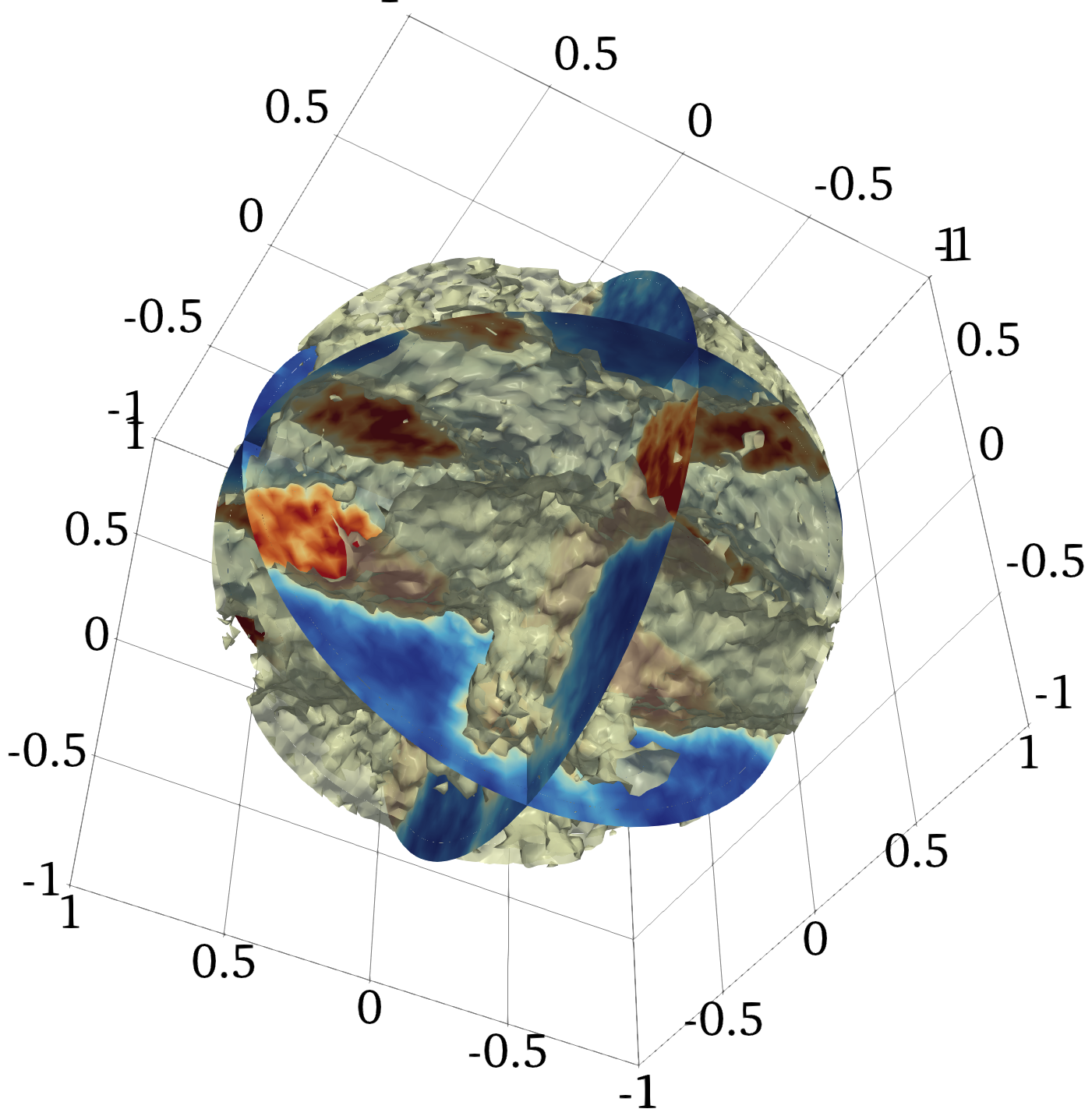}};
    \node at (15.5, -2.1) {\small{$c_3$}};
    \node at (14.6, -4.3) {\small{$c_2$}};
    \node at (11.7, -4.7) {\small{$c_1$}};

    \node[inner sep=0pt, anchor=north west] (img4) at (0,\dy)
      {\includegraphics[scale=0.1]{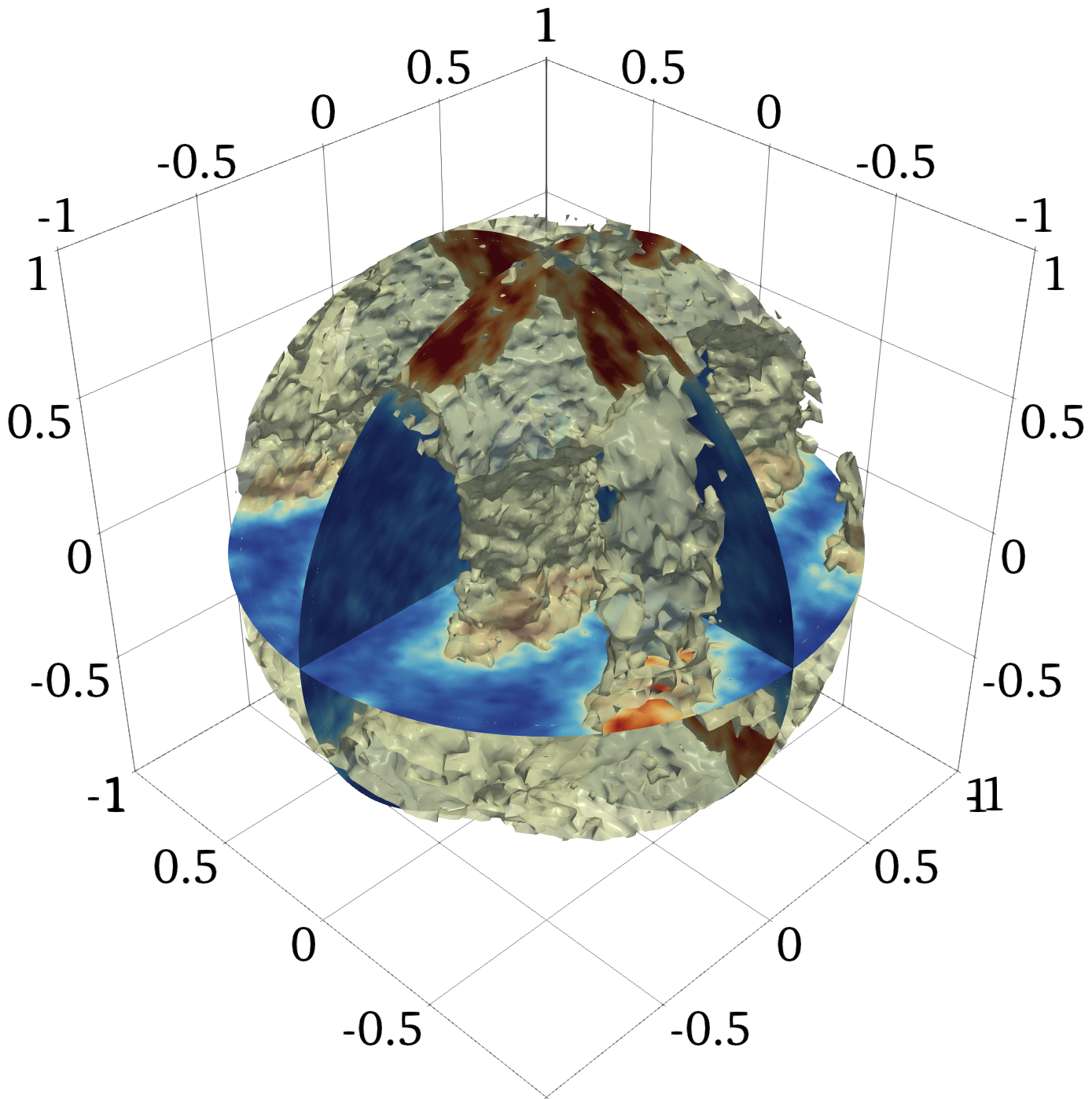}};
    \node at (4.3, {-9.7}) {\small{$c_1$}};
    \node at (0.6, {-9.7}) {\small{$c_2$}};

    \node[inner sep=0pt, anchor=north west] (img5) at (\dx,{-0.4+\dy})
      {\includegraphics[scale=0.10]{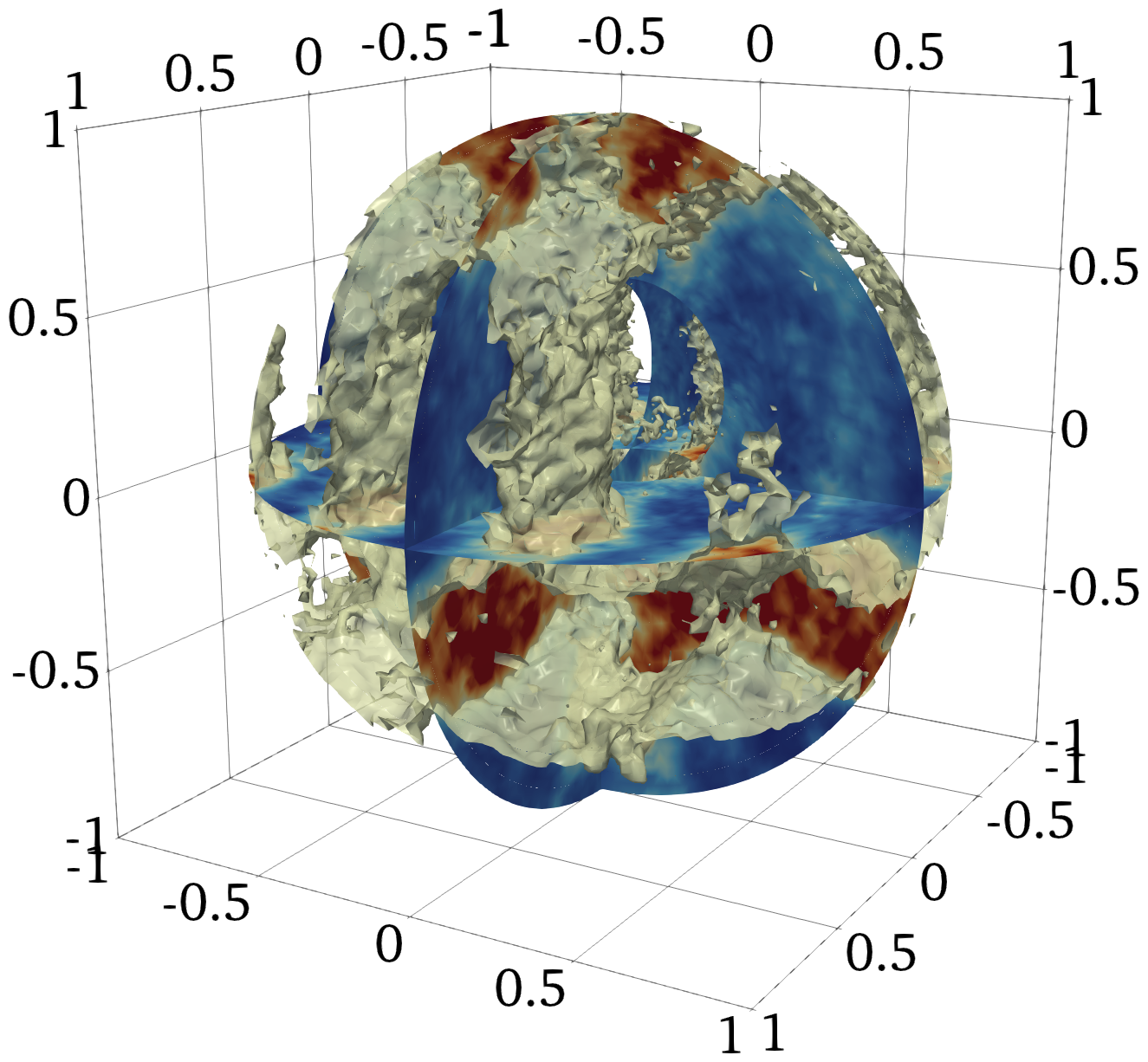}};
    \node at (5.3, {-7.7}) {\small{$c_3$}};
    \node at (6.7, {-9.7}) {\small{$c_2$}};
    \node at (9.3, {-9.7}) {\small{$c_1$}};

    \node[inner sep=0pt, anchor=north west] (img6) at ({1.75*\dx},{0.2+\dy})
      {\includegraphics[scale=0.10]{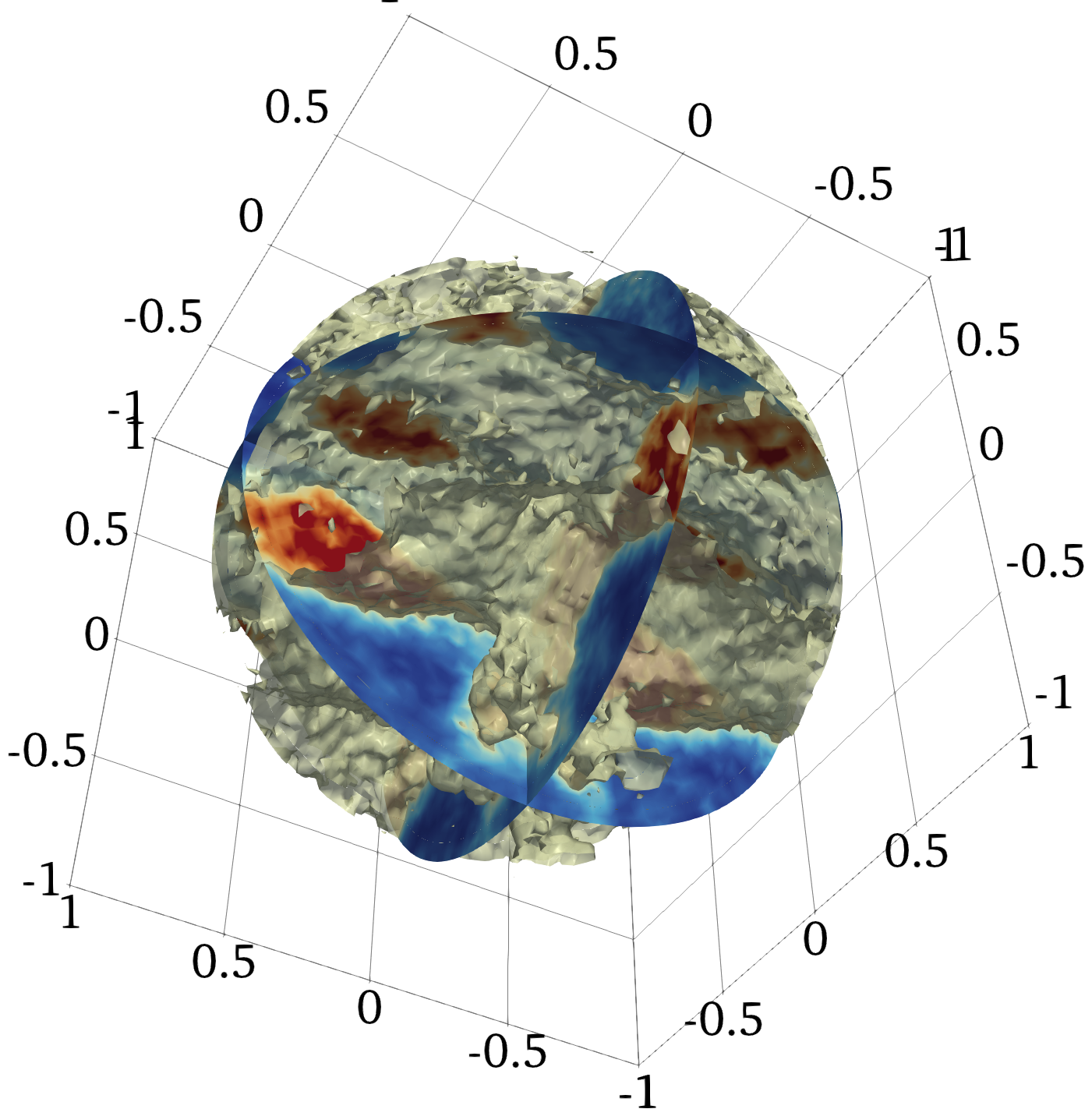}};
    \node at (15.5, {-7.4}) {\small{$c_3$}};
    \node at (14.6, {-9.9}) {\small{$c_2$}};
    \node at (11.7, {-10.2}) {\small{$c_1$}};

    \node[inner sep=0pt, anchor=north west] (img7) at (0,{2*\dy})
      {\includegraphics[scale=0.1]{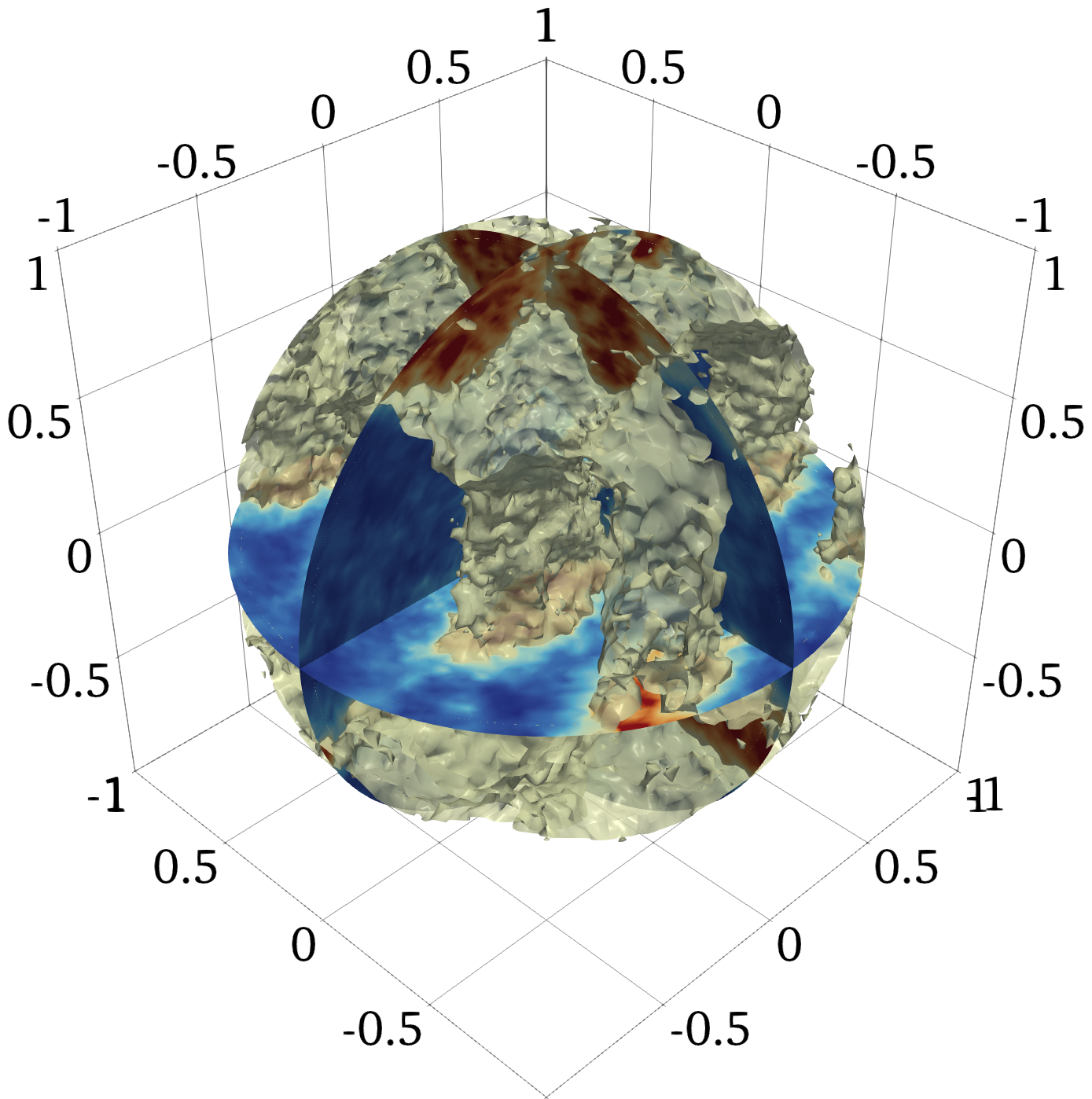}};
    \node at (4.3, {-15}) {\small{$c_1$}};
    \node at (0.6, {-15}) {\small{$c_2$}};

    \node[inner sep=0pt, anchor=north west] (img8) at (\dx,{-0.4+2*\dy})
      {\includegraphics[scale=0.10]{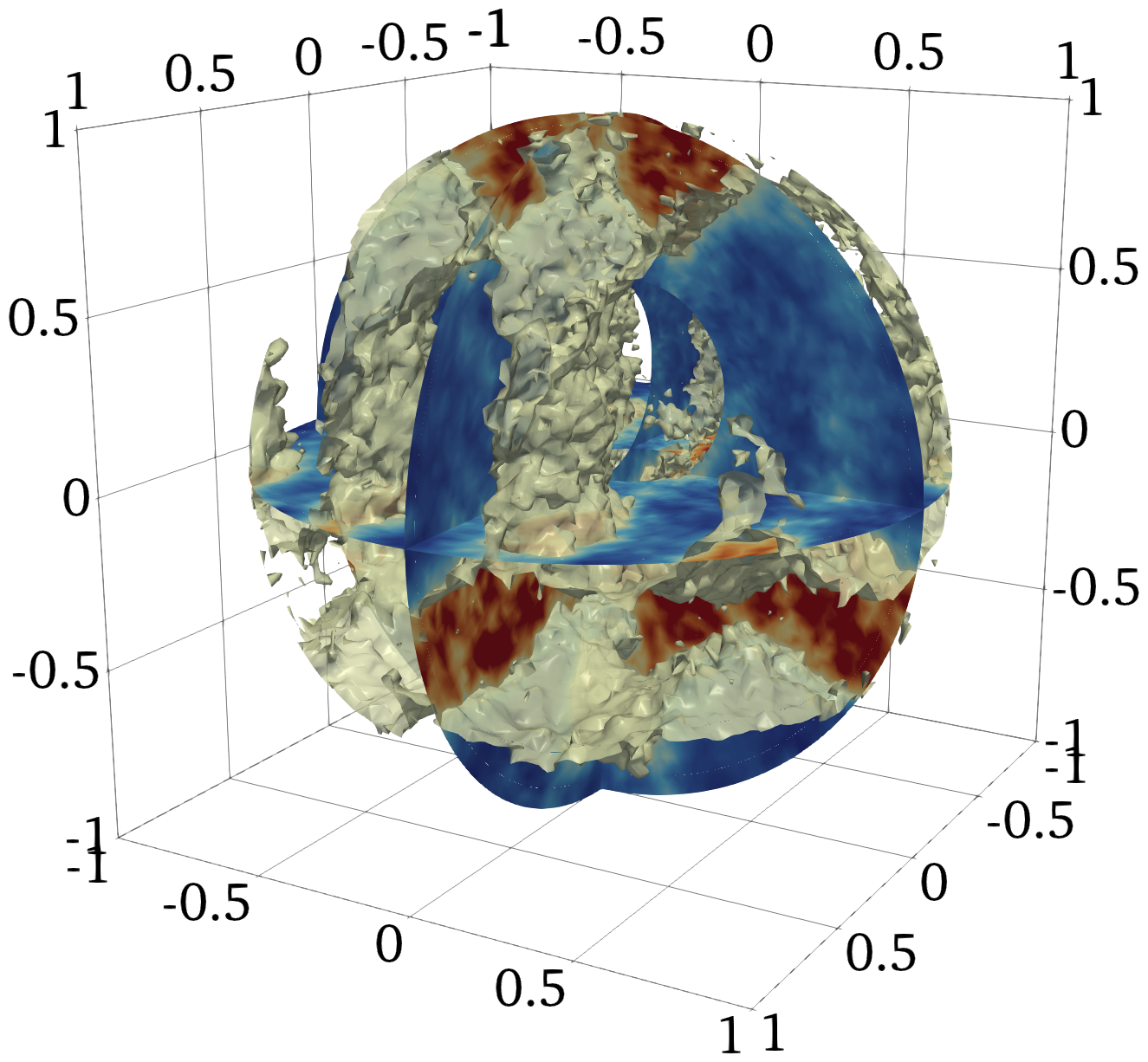}};
    \node at (5.3, {-13}) {\small{$c_3$}};
    \node at (6.7, {-15}) {\small{$c_2$}};
    \node at (9.3, {-15}) {\small{$c_1$}};

    \node[inner sep=0pt, anchor=north west] (img9) at ({1.75*\dx},{0.2+2*\dy})
      {\includegraphics[scale=0.10]{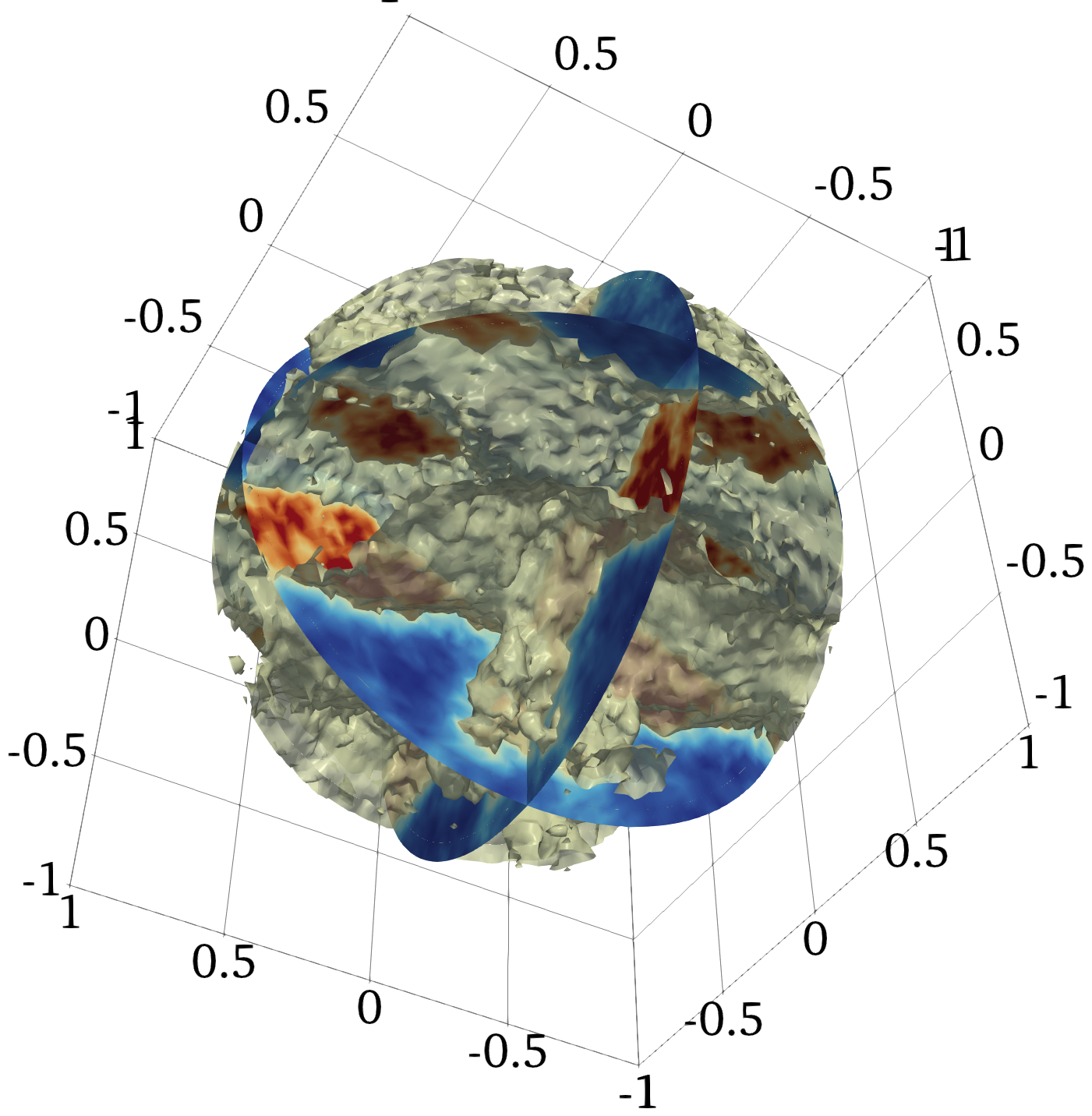}};
    \node at (15.5, {-12.5}) {\small{$c_3$}};
    \node at (14.6, {-15}) {\small{$c_2$}};
    \node at (11.7, {-15.5}) {\small{$c_1$}};

    \node[anchor=west, xshift=0.5cm] (colorbar) at ($(img3.north east)!0.5!(img9.south east)$) {
        \includegraphics[height=0.25\textwidth]{colorbar_standalone.png}
    };

    \node[rotate=90, anchor=south, font=\small] at ($(img1.north west)!0.5!(img1.south west)+(-0.6,0)$) {sample 1};
    \node[rotate=90, anchor=south, font=\small] at ($(img4.north west)!0.5!(img4.south west)+(-0.6,0)$) {sample 2};
    \node[rotate=90, anchor=south, font=\small] at ($(img7.north west)!0.5!(img7.south west)+(-0.6,0)$) {sample 3};
    \end{tikzpicture}}
\caption{Independent posterior samples drawn from $q(\bx|\bphi^*)$, visualized as 3D isosurfaces of the permeability field $k(\bx, \bc)$. Each row corresponds to one sample shown from three viewpoints (same as Figure~\ref{fig:gt_permeability}). All samples capture the dominant spatial structure of the ground truth, while fine-scale variations across samples reflect the remaining posterior uncertainty. Compare with the posterior mean in Figure~\ref{fig:posterior_mean_baseline} and the ground truth in Figure~\ref{fig:gt_permeability}.}
    \label{fig:posterior_samples_3d}
\end{figure}

Figure~\ref{fig:posterior_samples_3d} shows three independent samples drawn from the approximate posterior $q(\bx|\bphi^*)$ alongside the ground-truth field. Each row displays the isosurface representation of one posterior sample from the same three viewpoints used in Figure~\ref{fig:gt_permeability}. While the posterior mean (Figure~\ref{fig:posterior_mean_baseline}) recovers the dominant spatial features, individual samples reveal the remaining posterior uncertainty: fine-scale structures vary across draws, yet all samples reproduce the large-scale topology of the ground-truth isosurface.

\end{document}